\documentclass[pdflatex]{sn-jnl}

\usepackage{graphicx}\usepackage{multirow}\usepackage{amsmath,amssymb,amsfonts}\usepackage[dvipsnames]{xcolor}
\usepackage{booktabs}\usepackage{tabularx}

\usepackage{comment}
\usepackage{subcaption}
\usepackage{cleveref}
\usepackage{xspace}
\usepackage{url}
\usepackage{enumitem}

\usepackage[maxcitenames=2,mincitenames=1,backend=biber]{biblatex}
\newcommand{\referenceP}{\ensuremath{PF^{ref}}\xspace}
\newcommand{\computedP}{\ensuremath{PF^{c}}\xspace}
\newcommand{\independentRun}{30\xspace}
\newcommand{\nsga}{\textit{NSGA-II}\xspace}
\newcommand{\pesa}{\textit{PESA2}\xspace}
\newcommand{\spea}{\textit{SPEA2}\xspace}
\newcommand{\perfq}{\texttt{perfQ}\xspace}
\newcommand{\reliability}{\texttt{reliability}\xspace}
\newcommand{\achanges}{\texttt{\#changes}\xspace}
\newcommand{\pas}{\texttt{\#pas}\xspace}
\newcommand{\ttbs}{\texttt{TTBS}\xspace}
\newcommand{\ccm}{\texttt{CoCoME}\xspace}
\newcommand{\vda}{$\hat{A}_{12}$\xspace}
\newcommand{\mwu}{Mann--Whitney U\xspace}
\newcommand{\igdp}{IGD+\xspace}
\newcommand{\ebar}[2]{#1 & {\color{gray}\rule{#2cm}{5pt}}}

\usepackage{etoolbox}

\newboolean{showcomments}
\setboolean{showcomments}{false}

\ifthenelse{\boolean{showcomments}}
  {\newcommand{\nb}[2]{
    \fbox{\bfseries\sffamily\scriptsize#1}
    {\sf\small$\blacktriangleright$\textit{#2}$\blacktriangleleft$}
   }
    \newcommand{\rev}[1]{{\leavevmode\color{blue}#1}}
    \newenvironment{revfloatenv}
      {\par\noindent\color{blue}\rule{\linewidth}{1pt}\par\color{black}\vspace{.5ex}}
      {\vspace{-.5ex}\par\noindent\color{blue}\rule{\linewidth}{1pt}\par\color{black}}
  }
  {\newcommand{\nb}[2]{}
    \newcommand{\rev}[1]{{\leavevmode#1}}
    \newenvironment{revfloatenv}{}
}

\def\eg{\emph{e.g.},\xspace} 

\def\ie{\emph{i.e.},\xspace}

\newcommand{\SoBigDataITHack}{European Union - NextGenerationEU - National Recovery and Resilience Plan (Piano Nazionale di Ripresa e Resilienza, PNRR) - Project: ''SoBigData.it - Strengthening the Italian RI for Social Mining and Big Data Analytics`` - Prot. IR0000013 - Avviso n. 3264 del 28/12/2021\xspace}

\usepackage[most]{tcolorbox}
\usepackage{scalerel}
\newtcolorbox{rqbox}{
  enhanced,
  boxrule=0pt,frame hidden,
  borderline west={4pt}{0pt}{black!50},
  colback=gray!5,
  sharp corners
}

\begin{document}

\title{On the Role of Search Budgets in Model-Based Software Refactoring Optimization}

\author[1]{\fnm{J. Andres} \sur{Diaz-Pace}}\email{andres.diazpace@isistan.unicen.edu.ar}

\author*[2]{\fnm{Daniele} \sur{Di Pompeo}}\email{daniele.dipompeo@univaq.it}

\author[2]{\fnm{Michele} \sur{Tucci}}\email{michele.tucci@univaq.it}

\affil[1]{\orgname{ISISTAN, CONICET-UNICEN},
  \orgaddress{\city{Buenos Aires}, \country{Argentina}}}
  
\affil[2]{\orgname{University of L'Aquila},
            \orgaddress{\city{L'Aquila}, \country{Italy}}}

\abstract{
Software model optimization is a process that automatically generates design alternatives aimed at improving quantifiable non-functional properties of software systems, such as performance and reliability. Multi-objective evolutionary algorithms effectively help designers identify trade-offs among the desired non-functional properties.

To reduce the use of computational resources, this work examines the impact of implementing a search budget to limit the search for design alternatives. 
In particular, we analyze how time budgets affect the quality of Pareto fronts by utilizing quality indicators and exploring the structural features of the generated design alternatives.
This study identifies distinct behavioral differences among evolutionary algorithms when a search budget is implemented. 
It further reveals that design alternatives generated under a budget are structurally different from those produced without one. 
Additionally, we offer recommendations for designers on selecting algorithms in relation to time constraints, thereby facilitating the effective application of automated refactoring to improve non-functional properties.
}


\maketitle

\section{Introduction}\label{sec:intro}

Over the last decade, multi-objective optimization (MOO) techniques have been successfully applied to many software engineering problems~\cite{Ramirez-Romero-Ventura-2019,Bavota-Di-Penta-Oliveto-2014,Cortellessa-Di-Pompeo-2021}.
These techniques have proved effective on problems whose objectives can be expressed through quantifiable metrics. Problems related to non-functional properties (\eg performance and reliability) undoubtedly fit into this category, as witnessed by the literature in this domain~\cite{Aleti-Bjornander-Grunske-Meedeniya-2009,Mariani-Vergilio-2017,Ouni-Kessentini-Inoue-Cinneide-2017}.
Most of the approaches used evolutionary algorithms~\cite{Martens-Koziolek-Becker-Reussner-2010,Blum-Roli-2003} that allow exploration of the search space effectively.

One of the main drawbacks of applying MOO optimization techniques to improve non-functional properties is that the search for alternative solutions requires a considerable amount of computational resources, notably time.
Whenever a new solution is generated, the search algorithms have to evaluate it. 
This means computing quantifiable indices by solving non-functional models (\eg Queueing Network~\cite{DBLP:books/daglib/0076254} and PetriNet~\cite{DBLP:conf/ifip/Petri62} Models), either analytically or by simulating them. 
Due to the complexity of such models, it is difficult to further improve the efficiency of their evaluation.
Therefore, the time required to search for candidate solutions is negatively impacted by this evaluation phase.
When performed on realistic models, this type of optimization can even take days~\cite{Arcuri-Fraser-2011,Cortellessa-Di-Pompeo-Stoico-Tucci-2021,Cortellessa-Di-Pompeo-Stoico-Tucci-2023}, which poses an obstacle to its adoption in practical design and development scenarios.

To address the aforementioned challenge, the search for solutions can be constrained by search budgets of varying complexity~\cite{Arcuri-Fraser-2011, Zitzler_Künzli_2004}.
A simple strategy is to set a time budget that interrupts the search when the imposed time has expired~\cite{Arcuri-Fraser-2011}. 
However, choosing the right time budget is not straightforward. 
Time budgets that are too small heavily limit the exploration of the solution space, consequently hampering the quality of the computed Pareto fronts (\ie the set of non-dominated solutions obtained at the end of the optimization).
Conversely, larger time budgets may not be effective in saving enough optimization time, therefore, defeating their purpose.
Other complementary strategies to reduce evaluation cost and enhance search efficiency include the use of surrogate fitness functions that approximate objective values with lower computational effort, evaluation budgets that limit the time or resources allocated to assessing each individual solution, and parallel evaluation techniques that distribute the workload across computing nodes. While these strategies are not the focus of this study, they represent viable alternatives to be considered in future research on time-constrained software model optimization.

Specifically, the context of this work is a multi-objective optimization process that aims at improving software models through sequences of refactoring actions.
These actions are intended to alter an initial software model to maximize performance and reliability while minimizing the number of detected performance antipatterns\footnote{Performance antipatterns describe bad design practices that usually lead to performance degradation.} and the cost of the refactoring itself.

The type of budget-aware refactoring process we study could be applied in real-world model-based engineering workflows, where optimization and simulation techniques are increasingly integrated into tools like ArcheOpterix~\cite{Aleti-Bjornander-Grunske-Meedeniya-2009}, GATSE~\cite{GATSE:Procter2019}, and the Palladio Optimization Framework~\cite{Koziolek-Koziolek-Reussner-2011}. In such tools, designers and architects explore trade-offs across large design spaces involving performance, reliability, deployment cost, and other quality attributes.
Additionally, our work could be used in an interactive refactoring loop, where a designer runs constrained optimization procedures repeatedly to compare and refine architecture alternatives. This scenario is especially relevant in time-sensitive contexts, such as early-stage design analysis or tight iteration cycles, where full unconstrained searches are impractical. These ideas align with recent proposals to make multi-objective optimization more interactive and designer-friendly~\cite{cortellessaIntroducingInteractionsMultiObjective2025}.

In our previous work~\cite{Di-Pompeo-Tucci-2022}, we measured the impact of the time budget on the quality of Pareto fronts using the Hypervolume indicator~\cite{Li-Yao-2020}.
Since a quality indicator measures a specific property of a Pareto front, there is consensus that no single indicator can capture all the properties of a Pareto front~\cite{Cao-Smucker-Robinson-2015}.
Hence, here we enriched our analysis~\cite{Di-Pompeo-Tucci-2022} by using six quality indicators to measure the impact of the time budget on different quality properties of a Pareto front.
In addition, this extension also aims at helping the designer to find and evaluate a trade-off between the time spent on the search and the characteristics of the obtained software models. To achieve this, the analysis of the effects of time budgets is elaborated by investigating both the quality of the generated design alternatives and their structural features. 
Therefore, another novel aspect of this paper is that it analyzes and links the effects of time budgets on the search process to the structural features of the resulting software models, which is a rather unexplored topic in the literature.

\smallskip

To investigate the practical implications of imposing a time budget on the Model-Based Software Refactoring Optimization, we formulated RQ1, which focuses on the trade-offs between computational efficiency and solution quality.
\begin{itemize}[label=--]
 \item \textbf{RQ1}: How can we characterize the trade-offs between computational efficiency and solution quality in time-constrained multi-objective optimization for software refactoring?
 \begin{itemize}
	\item \textbf{RQ1.1}: Which algorithm completes the search process faster?
	\smallskip
	\item \textbf{RQ1.2}: Which algorithm performs better when limited by a time budget?
	\smallskip
	\item \textbf{RQ1.3}: To what extent does the time budget affect the quality of Pareto fronts?
 \end{itemize}
\end{itemize}

Prior work highlighted performance differences among algorithms, but did not evaluate their behavior under explicit time budgets~\cite{EPEW2023}. To address this, we decomposed RQ1 into three sub-questions. RQ1.1 examines runtime differences among algorithms, where our results show that \nsga consistently completes the search faster than \pesa and \spea. RQ1.2 evaluates algorithms’ ability to search the space under constrained time, revealing that \pesa generally achieves higher-quality results, even with limited time. RQ1.3 analyzes how increasing the time budget affects Pareto front quality. We found that while longer budgets benefit \pesa, other algorithms show limited improvement, and gains are often case-dependent. These findings are the result of analyzes that exploit experiments performed in our previous work~\cite{Di-Pompeo-Tucci-2022}, but that, in this extension, employ five additional quality indicators to provide a more comprehensive view of the problem and better inform the selection of algorithms and budget configurations.

\smallskip

While traditional quality indicators from multi-objective optimization are essential to compare the ability of algorithms and configurations to exhaustively search the solution space, they tend to be blind to the properties that the automatically generated solutions will exhibit in a specific domain (model-based software refactoring, in our case)~\cite{ECSA2024}.
Therefore, to understand the broader impact of time budgets on the outcomes of search-based refactoring, we formulated RQ2 and RQ3, which focus on how varying the time budget influences both the quality and structure of generated software models.

\begin{itemize}[label=--]
 \item \textbf{RQ2}: How does varying the time budget influence quality and structure of software models generated by search-based refactoring algorithms?
 \begin{itemize}
	\item \textbf{RQ2.1}: Do different time budgets significantly affect performance and reliability of the software models produced by search-based refactoring algorithms?
	\item \textbf{RQ2.2}: Do different time budgets significantly affect design properties of the software models produced by search-based refactoring algorithms?
 \end{itemize}	
 \smallskip
 \item \textbf{RQ3}: How are refactoring choices impacted by different time budgets?
\end{itemize}

RQ2.1 examines whether time budgets affect non-functional properties such as performance and reliability. 
We observed that, while certain configurations (notably with \pesa) yield improved performance under longer budgets, these effects vary across case studies. 
RQ2.2 investigates the influence of time budgets on design properties like the number of detected performance antipatterns and the cost of refactoring. Results show that design-level improvements are sensitive to both budget length and the algorithm used, with longer budgets sometimes leading to less converged, denser refactoring sequences. 
Finally, RQ3 explores how time budgets shape the use of refactoring actions. We found that constrained budgets lead to an overuse of certain actions (\eg moving an operation to a new component), while longer budgets promote more balanced and diverse sequence of refactoring actions (\eg inclusion of the cloning refactoring action), ultimately yielding structurally different models. 
These new research questions and findings represent the core of this extension and complement our investigation with domain-specific knowledge that was missing in our preliminary study.

In order to answer our research questions, we designed an experimental study with two model-based benchmarks, namely: Train Ticket Booking Service~\cite{Di-Pompeo-Tucci-Celi-Eramo-2019} and CoCoME~\cite{Herold-Klus-Welsch-Deiters-Rausch-Reussner-Krogmann-Koziolek-Mirandola-Hummel}.
In addition, we compared three genetic algorithms, \ie \nsga~\cite{Deb-Agarwal-Pratap-Meyarivan-2002}, \spea~\cite{Zitzler-Laumanns-Thiele-2001}, and \pesa~\cite{Corne-Jerram-Knowles-Oates-2001}, to identify whether any of them performs better when the search is limited by time budgets.

\smallskip
\noindent
In summary, the main contributions of this study are as follows:
\begin{itemize}[label=--]
	\item This study provides the first systematic investigation of how different time budgets affect the outcome of multi-objective optimization in model-based software refactoring. 
	\item We explore six complementary quality indicators to provide the designer with a comprehensive view on the behavior of genetic algorithms under constrained time when refactoring software models.
	\item Beyond quality metrics, this work analyzes how budget constraints influence the design and structure of refactored models, including action distribution and tree density, which is a novel representation of the structure of models.
    \item The findings are translated into actionable implications for practitioners, helping them select algorithms, tune budgets, and interpret search behavior in practical settings.
\end{itemize}

The remaining of the paper is structured as follows: \Cref{sec:related} reports related work, \Cref{sec:background} introduces background concepts, and \Cref{sec:approach} presents the design of this study. \Cref{sec:case-study} describes the two case studies employed in our analysis.
Research questions and results are presented and discussed in \Cref{sec:rqs}. 
\Cref{sec:implications} discusses the implications of our findings.
Threats to validity are covered in \Cref{sec:t2v}. 
Finally, \Cref{sec:conclusion} gives the conclusions and outlines future work.

\subsection*{Extension of Our Previous Study}
This article extends our prior work \cite{Di-Pompeo-Tucci-2022} in multiple dimensions. Specifically, we introduce substantial enhancements in research scope, experimental depth, and analytical granularity:
\begin{itemize}
    \item While our prior work used only the Hypervolume (HV) indicator to evaluate Pareto front quality, this study incorporates five additional quality indicators (IGD, IGD+, Epsilon, SPREAD, and GSPREAD) to provide a multi-faceted understanding of solution quality (see RQ1 and its sub-research questions).
    \item We introduce two entirely new research questions (RQ2 and RQ3) focused on the structural properties of the refactored models and how time budgets influence refactoring sequences and model features. These aspects were not addressed in \cite{Di-Pompeo-Tucci-2022}, nor in previous literaure.
    \item We perform an in-depth structural analysis, including: frequency analysis of refactoring actions, tree-based representations of refactoring sequences, and intersections of solution spaces with and without budgets. These analyses provide new insights into how and why budgeted searches lead to different structural outcomes.
    \item Based on the new findings, we deliver actionable recommendations for practitioners. These include algorithm selection strategies, time budget calibration, and heuristics for detecting convergence issues through tree density.
\end{itemize}

 \section{Related Work}\label{sec:related}

The idea of limiting the search using additional criteria has gotten attention within the search-based community~\cite{Ghoreishi-Clausen-Joergensen-2017}.
Often, it is unfeasible to use a ''formal`` stopping criterion in real-world multi-objective problems for which a mathematical formulation might be hard to define~\cite{Wagner-Trautmann-Marti-2011}.
To deal with this limitation, some proposals for stopping criteria are based on quality indicators~\cite{Wagner-Trautmann-Naujoks-2009,Guerrero-Marti-Berlanga-Garcia-Molina-2010}, while others are based on statistical testing of different metrics~\cite{Trautmann-Ligges-Mehnen-Preuss-2008,Marti-Garcia-Berlanga-Molina-2009}.
For example, \textcite{Arcuri-Fraser-2011} empirically studied the impact of imposing a time budget on the search process in the context of test case generation. 
\textcite{Luong-Nguyen-Gupta-Rana-Venkatesh-2021} introduced a cost-aware Bayesian optimization process that considers the cost of evaluating the objective function as the budget.
\textcite{Zitzler_Künzli_2004} introduced the IBEA algorithm that exploits a binary quality indicator, \eg hypervolume, to guide the search process.
The main idea was to guide the search through higher quality solutions, which are more likely to be part of the Pareto front.

To the best of our knowledge, there are no studies that investigate the usage of the aforementioned search budgets in refactoring optimization of model-based software.
In the following, we report on studies about multi-objective optimization of various non-functional properties of software models (\eg reliability, and energy \cite{Meedeniya-Buhnova-Aleti-Grunske-2010,Martens-Ardagna-Koziolek-Mirandola-Reussner-2010}), which have different degrees of freedom with respect to modifying the models (\eg service selection~\cite{Cardellini-Casalicchio-Grassi-Lo-Presti-Mirandola-2009}).

In the context of model-based non-functional properties evaluation, Palladio Component Model (PCM) is one the most popular Architecture Description Language (ADL)~\cite{Becker-Koziolek-Reussner-2009}.
In this context, \textcite{Koziolek-Koziolek-Reussner-2011} presented PerOpteryx, a performance-oriented multi-objective optimization problem.
PerOpteryx optimization process is guided by architectural tactics referring to component re-allocation and hardware.  
Furthermore, PerOpteryx was the first in investigating multi-objective optimization on software model.
However, PerOpteryx does not deal with model refactoring since it uses tactics that mainly change system configurations (\eg hardware settings, or operation demands).
\textcite{Rago-Vidal-Diaz-Pace-Frank-van-Hoorn-2017} proposed an extensible platform, called SQuAT, aimed at including flexibility in the definition of an architecture optimization problem. 
SQuAT exploits LQNs for performance evaluation and PerOpteryx tactics for architectural changes to optimize PCM  architectures.
\textcite{Ni-Du-Ye-Minku-Yao-Harman-Xiao-2021} compared the ability of two multi-objective optimization approaches to improve quality attributes where randomized search rules were applied to improve the software architecture.
\textcite{Ni-Du-Ye-Minku-Yao-Harman-Xiao-2021} used a fixed-length coding scheme for explainable solutions, which includes a do-nothing rule. This approach simplifies the management of evolutionary operators and incorporates a repair mechanism, improving efficiency and clarity of explanations. Effectiveness was assessed through experiments on six problem instances.

Several other ADLs have been explored in performance optimization. 
For example, \textcite{Cortellessa-Di-Pompeo-2021} previously studied the sensitivity of multi-objective software model refactoring to configuration characteristics, where models are defined in a performance-oriented ADL called \AE milia.
They also compared \nsga and \spea in terms of the quality of the solutions on the Pareto front. 
\textcite{Etemaadi-Chaudron-2015} presented an approach aimed at improving quality attributes of software architectures through genetic algorithms.
The multi-objective optimization considers component-based architectures described with an ADL called AQOSA-IR~\cite{Li-Etemaadi-Emmerich-Chaudron-2011}.
The architectures can be evaluated by means of several techniques, such as LQNs and Fault Trees.
The genetic algorithm considers the variation of designs (\eg number of hardware nodes) as objectives of the fitness function.
\textcite{Aleti-Bjornander-Grunske-Meedeniya-2009} proposed an approach for modeling and analyzing architectures expressed in the Architecture Analysis and Description Language (AADL). 
The authors also introduced a tool based on genetic algorithms for optimizing different quality attributes while varying the architecture deployment and the component redundancy. More recently, the GATSE project  \cite{GATSE:Procter2019} supported quality-attribute exploration of AADL configurations, enabling the designer to focus on certain regions of the space and narrow down the search.

Although non-functional analyses, such as performance evaluations, are inherently time-consuming, previous studies have not investigated search budgets in the context of software model optimization. 
To address this gap, we propose to analyze the impact of search budget on the multi-objective optimization of software model refactoring, as well as examine how the search budget affects the optimization of UML software models.
To the best of our knowledge, this paper presents the first comprehensive analysis of the impact of search budget on model refactoring optimization, through the use of three algorithms (\nsga, \spea, and \pesa), while also characterizing this impact using six quality indicators to assess various properties of the generated Pareto fronts.

 \section{The multi-objective optimization approach}\label{sec:background}

In this study, we analyze the impact of search budget on the refactoring of software models using three \emph{genetic algorithms}: \nsga~\cite{Deb-Agarwal-Pratap-Meyarivan-2002}, \spea~\cite{Zitzler-Laumanns-Thiele-2001}, and \pesa~\cite{Corne-Jerram-Knowles-Oates-2001}. 

We chose these algorithms because they have been already analyzed in prior studies, \eg \cite{liEvolutionaryMultiobjectiveOptimization2011,kingComparisonNSGAIISPEA22010,hiroyasuComparisonStudySPEA22005}, and on the basis of our prior work~\cite{cortellessaIntroducingInteractionsMultiObjective2025,cortellessaExploringSustainableAlternatives2024,cortellessaImpactPerformanceAntipatterns2021,Di-Pompeo-Tucci-2022}.
Furthermore, they use different policies when exploring the solution space. 
For example, \nsga uses the knowledge of non-dominated sorting to generate Pareto frontiers, \spea uses two archives to store computed Pareto fronts, and \pesa uses the hyper-grid concept to compute Pareto fronts.

\subsection{The Refactoring Engine}\label{sec:approach:refactoring}

The automated refactoring of UML models is a key point when evolutionary algorithms are employed in order to optimize non-functional properties of models.
To achieve such an automation, we have used the refactoring engine by \textcite{Arcelli-Cortellessa-D-Emidio-Di-Pompeo-2018,Cortellessa-Di-Pompeo-Stoico-Tucci-2021}, which can apply predefined refactoring actions on UML models.
Each solution produced by our evolutionary algorithm produces a sequence of refactoring actions that, once applied to an initial model, leads to a model alternative that shows different non-functional properties. 

We exploit our engine to verify in advance whether a sequence of refactoring actions is feasible or not~\cite{Arcelli-Cortellessa-Di-Pompeo-2019,Arcelli-Cortellessa-Pompeo-2018}.
In order to do so, we define pre- and post-conditions for each refactoring action, which are combined during the evolutionary approach. 
Thus, the refactoring engine checks the feasibility of the (partial) sequence of refactoring actions.
When the latest added action makes the sequence unfeasible, the engine discards that action and replaces it with a new one. 
Our engine reduces the number of invalid refactoring sequences, thus contributing to save computational time.
The refactoring actions employed in our study are briefly described below.
Furthermore, for each refactoring action we provide a graphical example of the initial and refactored model.

\subsubsection*{Clone a Node (Clon)}
This action is aimed at introducing a replica of a Node. Adding a replica means that every deployed artifact and every connection of the original Node has to be in turn cloned. Stereotypes and their tagged values are cloned as well. The rationale of this action is to introduce a replica of a platform device with the aim of reducing its utilization.

	\begin{figure}[ht]
    \begin{revfloatenv}
                \begin{subfigure}[b]{.46\textwidth}
                        \includegraphics[width=\textwidth]{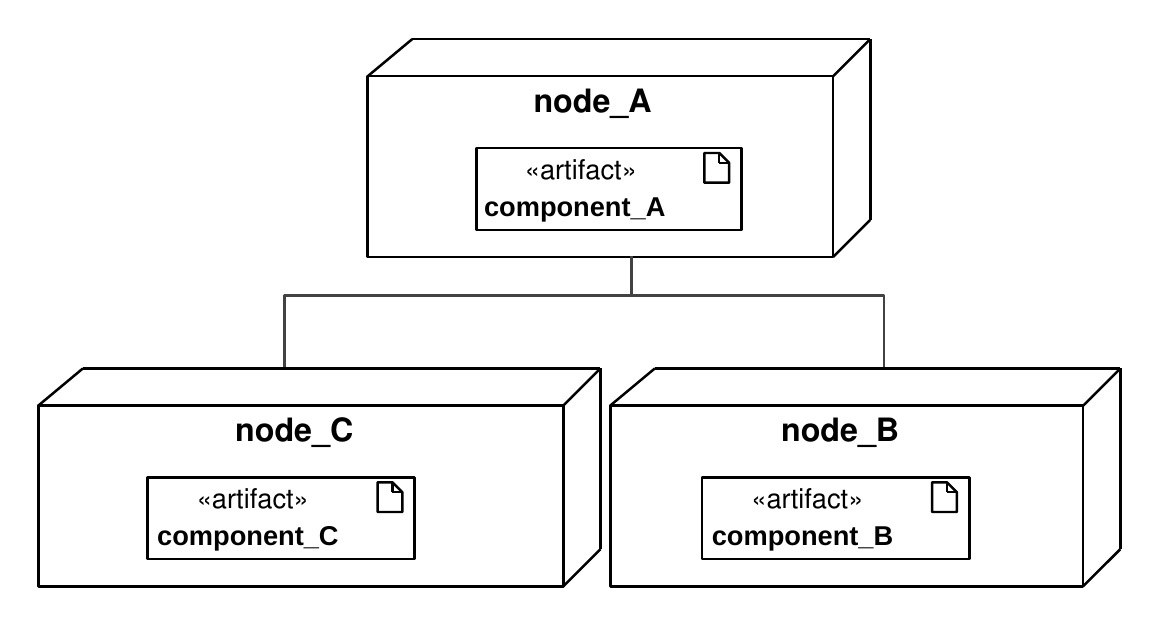}
                        \caption{Initial}\label{fig:clon-deploy}
                \end{subfigure}
                \begin{subfigure}[b]{.46\textwidth}
                        \includegraphics[width=.9\textwidth]{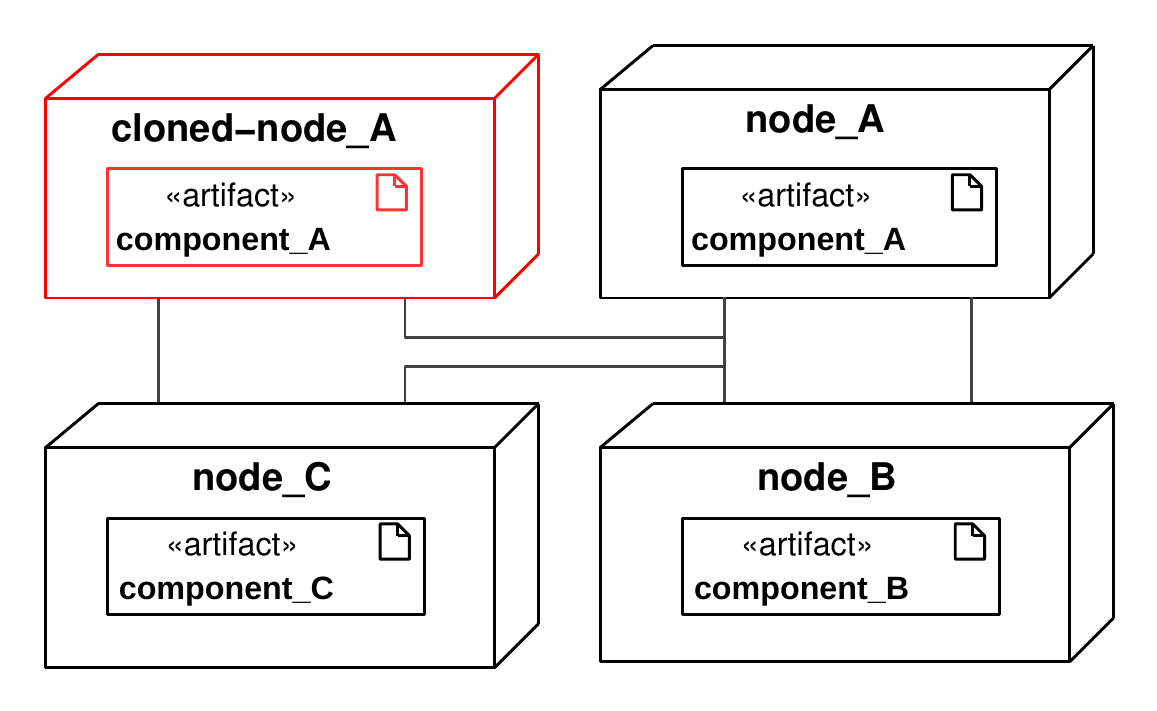}
                        \caption{Refactored}\label{fig:ref-clon-deploy}
                \end{subfigure}
		\caption{The \textit{Clon} refactoring action example on \emph{node\-A} through a UML Software Model}
                \label{fig:ref-clon-uml-diagrams}
    \end{revfloatenv}
        \end{figure}

\subsubsection*{Move an Operation to a new Component deployed on a new Node (MO2N)}
This action is in charge of randomly selecting an operation and moving it to a new Component. All the elements related to the moving operation (\eg links) will move as well. 
Since we adopt a multi-view model, and coherence among views has to be preserved, this action has to synchronize dynamic and deployment views. A lifeline for the newly created Component is added in the dynamic view, and messages related to the moved operation are forwarded to it. In the deployment view, instead, a new Node, a new artifact, and related links are created. The rationale of this action is to lighten the load of the original Component and Node.

	\begin{figure}[ht]
    \begin{revfloatenv}
                \begin{subfigure}[b]{.46\textwidth}
                \includegraphics[width=\textwidth]{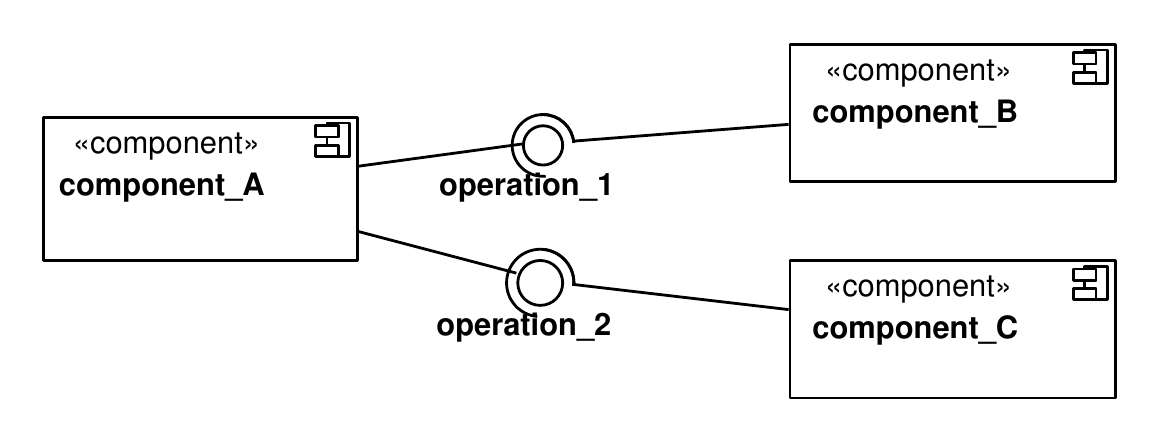}
                        \caption{Initial}\label{fig:mo2n-comp}
                \end{subfigure}
                \begin{subfigure}[b]{.46\textwidth}
                        \includegraphics[width=\textwidth]{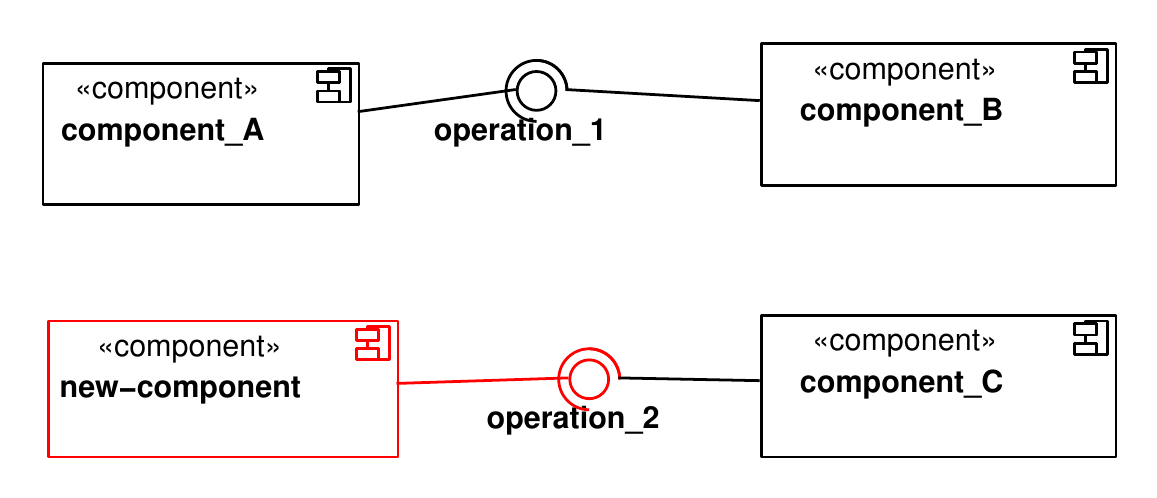}
                        \caption{Refactored}\label{fig:ref-mo2n-comp}
                \end{subfigure}
                \begin{subfigure}[b]{.46\textwidth}
                        \includegraphics[width=\textwidth]{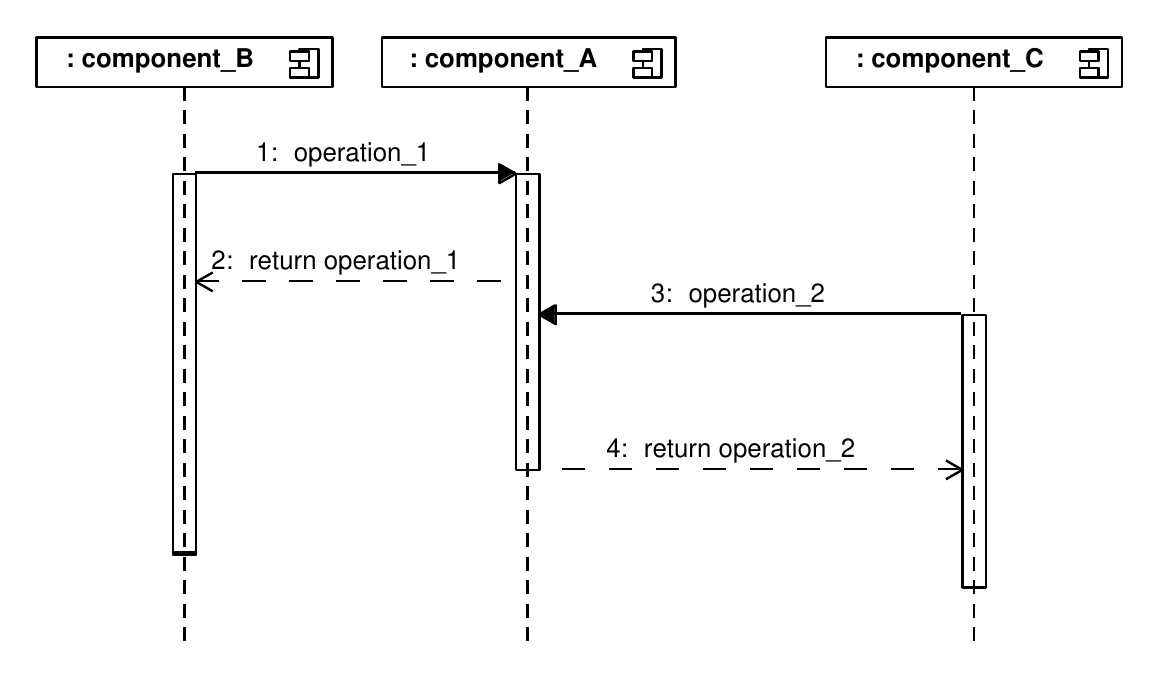}
                        \caption{Initial}\label{fig:mo2n-dynamic}
                \end{subfigure}
                \begin{subfigure}[b]{.46\textwidth}
                        \includegraphics[width=1.25\textwidth]{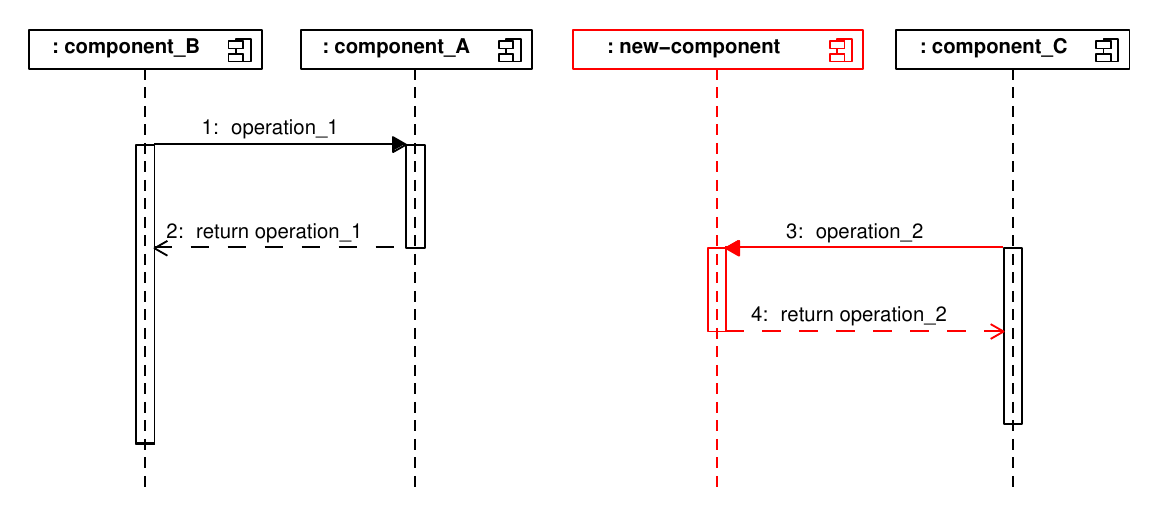}
                        \caption{Refactored}\label{fig:ref-mo2n-dynamic}
                \end{subfigure}
                \begin{subfigure}[b]{.46\textwidth}
                        \includegraphics[width=\textwidth]{fig/ref-actions/deployment}
                        \caption{Initial}\label{fig:mo2n-deploy}
                \end{subfigure}
                \begin{subfigure}[b]{.46\textwidth}
                        \includegraphics[width=\textwidth]{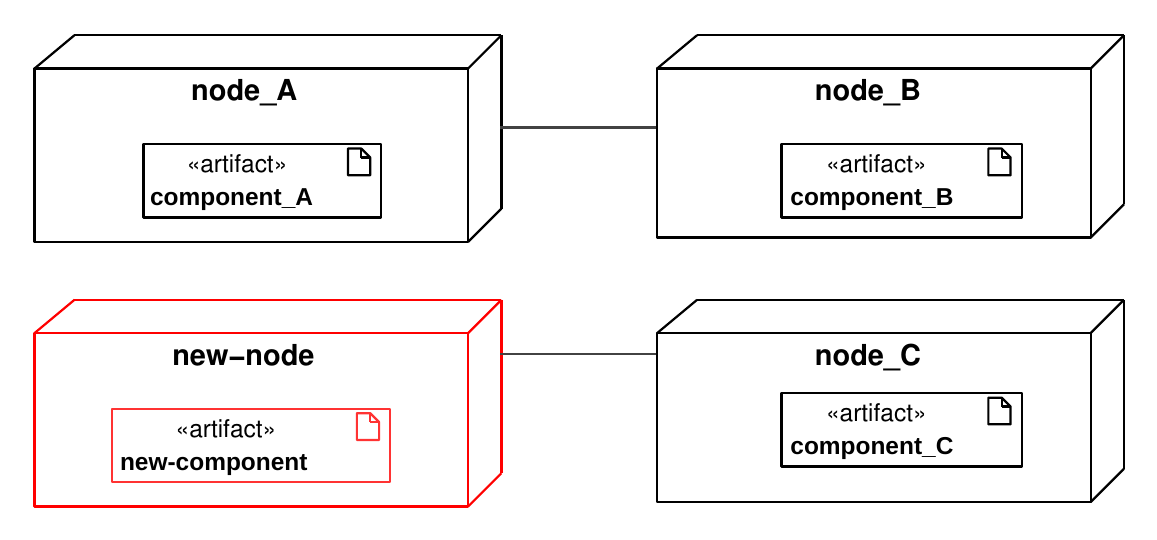}
                        \caption{Refactored}\label{fig:ref-mo2n-deploy}
                \end{subfigure}
                \caption{The \textit{MO2N} refactoring action example on \emph{operation\-2} through a UML Software Model}
                \label{fig:ref-mo2n-uml-diagrams}
    \end{revfloatenv}
    \end{figure}

\subsubsection*{Move an Operation to a Component (MO2C)}
This action is in charge of randomly selecting and transferring an Operation to an arbitrary existing target Component. The action consequently modifies each UML Use Case in which the Operation is involved. Sequence Diagrams are also updated to include a new lifeline representing the Component owning the Operation, but also to re-assign the messages invoking the operation to the newly created lifeline. The rationale of this action is quite similar to the previous refactoring action, but without adding a new UML Node to the model. 
\begin{figure}[ht]
    \begin{revfloatenv}
                \begin{subfigure}[b]{.46\textwidth}
                       \includegraphics[width=\textwidth]{fig/ref-actions/staticview}
                       \caption{Initial}\label{fig:mo2c-comp}
                \end{subfigure}
                \begin{subfigure}[b]{.46\textwidth}
                        \includegraphics[width=\textwidth]{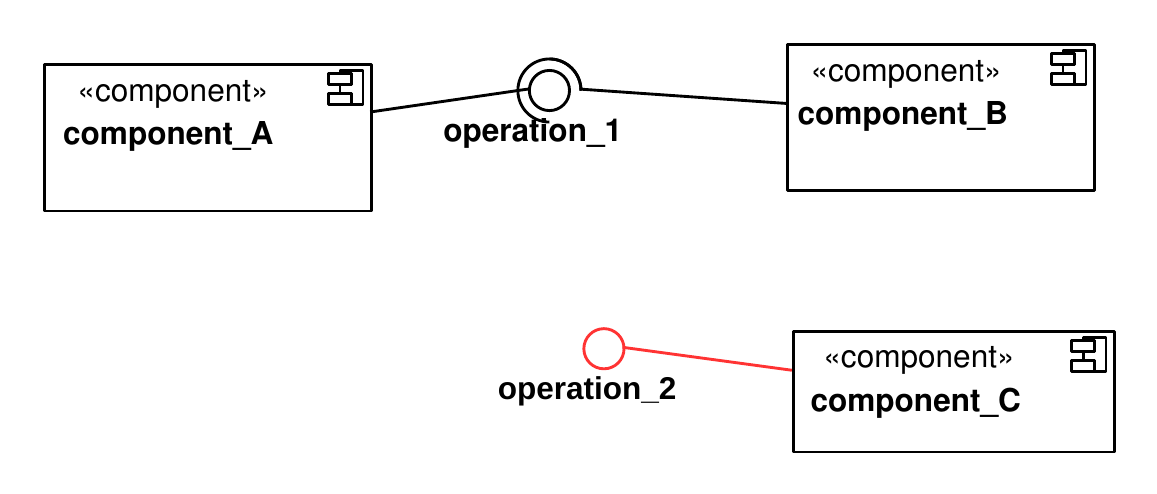}
                        \caption{Refactored}\label{fig:ref-mo2c-comp}
                \end{subfigure}
                \begin{subfigure}[b]{.46\textwidth}
                        \includegraphics[width=\textwidth]{fig/ref-actions/dynamic}
                        \caption{Initial}\label{fig:mo2c-dynamic}
                \end{subfigure}
                \begin{subfigure}[b]{.46\textwidth}
                        \includegraphics[width=1.07\textwidth]{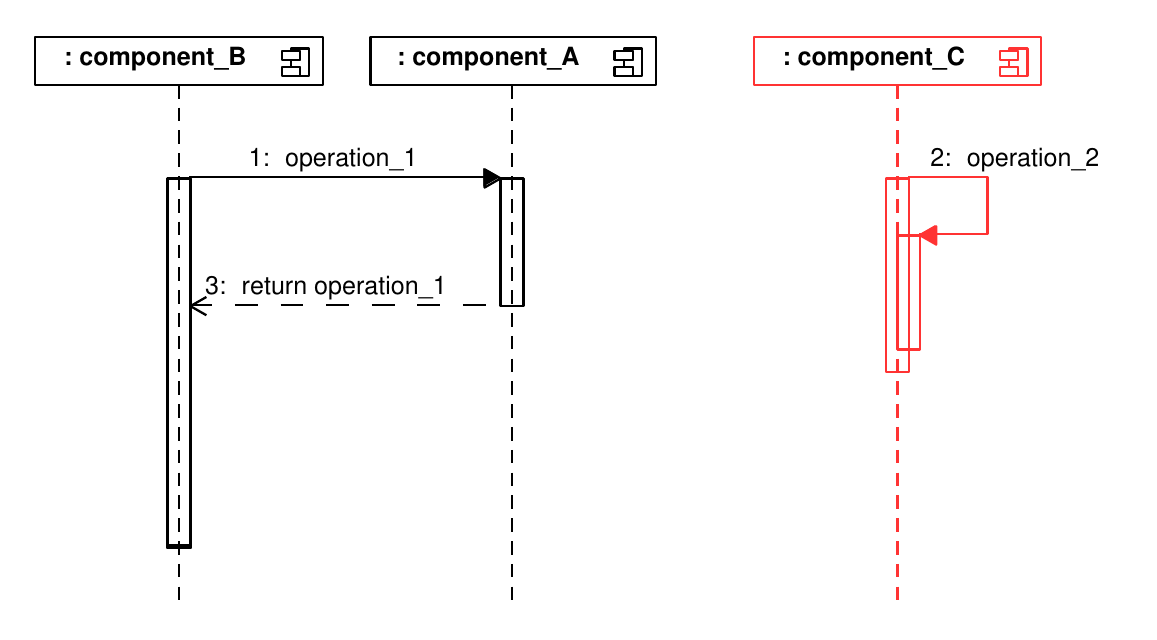}
                        \caption{Refactored}\label{fig:ref-mo2c-dynamic}
                \end{subfigure}
		\caption{The \textit{MO2C} refactoring action example on \emph{operation\-2} and \emph{component\-C} through a UML Software Model}
                \label{fig:ref-mo2c-uml-diagrams}
    \end{revfloatenv}
   \end{figure}

\subsubsection*{Deploy a Component on a new Node (ReDe)}
This action simply modifies the deployment view by redeploying a Component to a newly created Node.
In order to be consistent with the initial model, the new Node is connected with all other ones directly connected to the Node on which the target Component was originally deployed. The rationale of this action is to lighten the load of the original UML Node by transferring the load of the moving Component to a new UML Node. 

        \begin{figure}[ht]
        \begin{revfloatenv}
            
                \begin{subfigure}[b]{.46\textwidth}
                       \includegraphics[width=\textwidth]{fig/ref-actions/deployment}
                       \caption{Initial}\label{fig:rede-comp}
                \end{subfigure}
                \begin{subfigure}[b]{.46\textwidth}
                        \includegraphics[width=\textwidth]{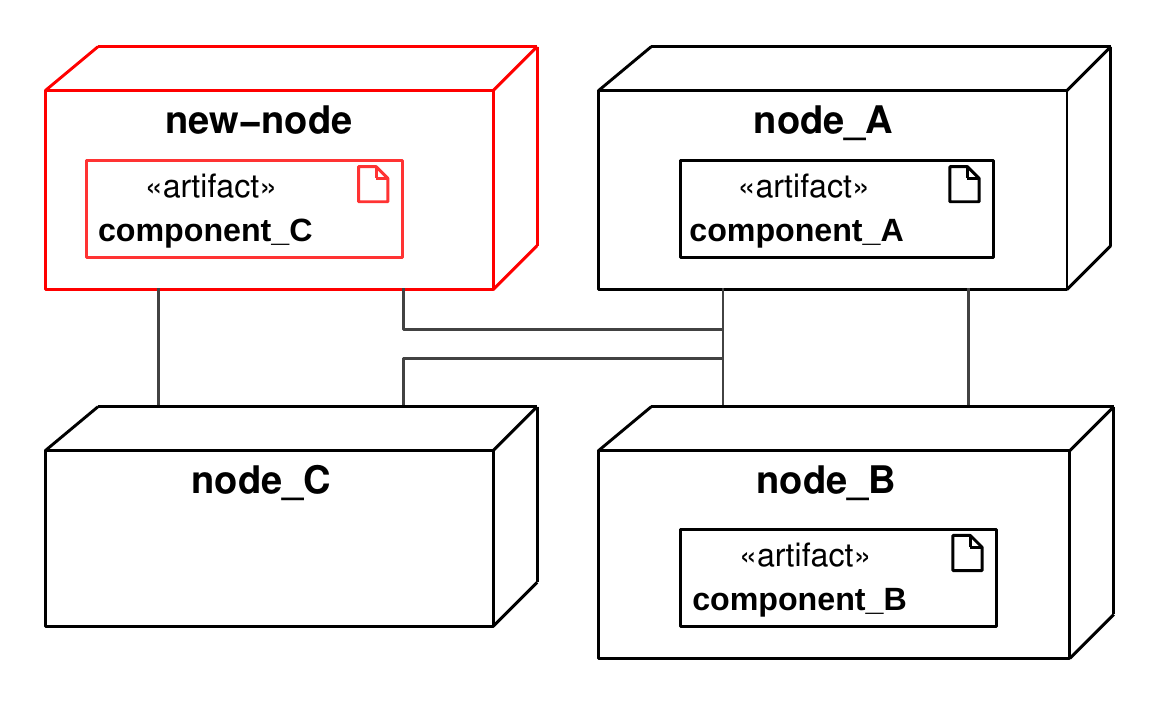}
                        \caption{Refactored}\label{fig:ref-rede-comp}
                \end{subfigure}
                \caption{The \textit{ReDe} refactoring action example on \emph{component\-C} through a UML Software Model}
		\label{fig:ref-rede-uml-diagrams}
        \end{revfloatenv}
        \end{figure}

\subsection{Objectives}
\begin{figure}[htbp]
	\centering
	\includegraphics[width=.7\linewidth]{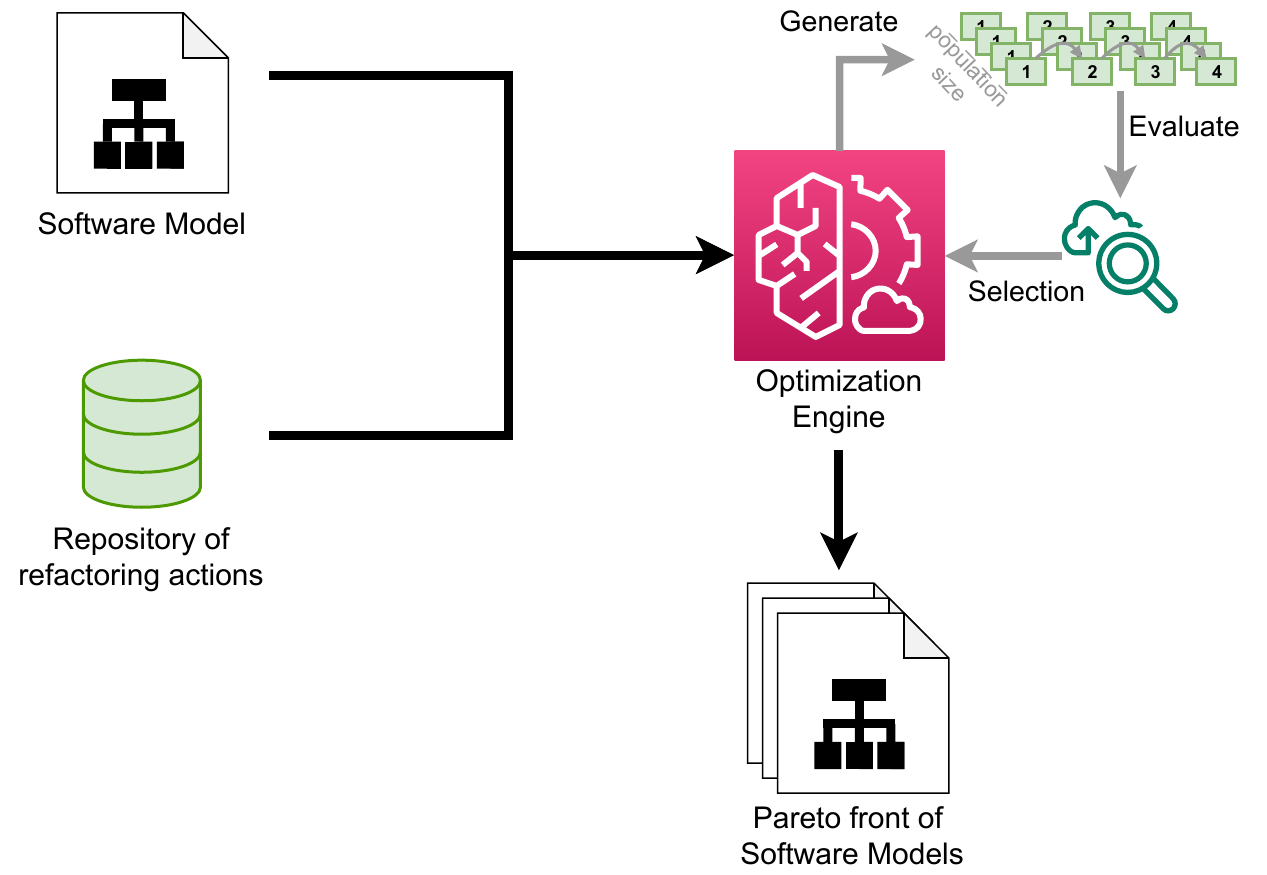}
\caption{\label{fig:approach}A graphical representation of the approach. It takes as input: the set of all the available refactoring actions (\emph{Repository of refactoring actions}), and an \emph{Software model} (\ie the subject model). The \emph{Optimization Engine} randomly selects and combines refactoring actions in order to generate a set of \emph{Model Alternatives}, which are \emph{Evaluate}d with respect to the objectives. Finally, the \emph{Optimization Engine} produces a \emph{Pareto front of Software Models}.
}
\end{figure}

Our process, as depicted in \Cref{fig:approach}, optimizes software models through refactoring, with respect to four conflicting objectives: the average system performance (\perfq)~\cite{Arcelli-Cortellessa-D-Emidio-Di-Pompeo-2018}, the reliability (\reliability) of the software model~\cite{Cortellessa-Singh-Cukic-2002}, the number of performance antipatterns (\pas) detected in the model, and the cost of the refactoring actions (\achanges) to generate a design alternative from the initial model~\cite{Cortellessa-Di-Pompeo-Stoico-Tucci-2021}.

\subsubsection*{Average System Performance (\perfq)}\label{sec:background:perfq}

With this objective, we quantify the performance improvement (or detriment) between two models.
\[\perfq(M)=\frac{1}{c}\sum\limits_{j=1}^{c} p_j\cdot \frac{F_j-I_j}{F_j+I_j}\]
\noindent where $M$ is a model obtained by applying a refactoring solution to the initial model, $F_j$ is the value of a performance index in $M$, and $I_j$ is the value of the same index on the initial model. 
$p\in\{-1,1\}$ is a multiplying factor that holds: i) $1$ if the $j$--th index has to be maximized (\ie, the higher the value, the better the performance), like the throughput; ii) $-1$ if the $j$--th index has to be minimized (\ie the smaller the value, the better the performance), like the response time. 
Furthermore, a single \perfq for each performance index is computed as the normalized ratio between the index value of a model alternative and the initial model.
Finally, the global \perfq is computed as the average across the number of performance indices considered in the performance analysis.

\subsubsection*{System Reliability (\reliability)}\label{sec:background:reliability}
The reliability analysis model that we adopt here to quantify the \reliability objective is based on~\cite{Cortellessa-Singh-Cukic-2002}. 
The mean failure probability $\theta_S$ of a software system $S$ is defined by the following equation:
\[ \theta_S = 1 - \sum\limits_{j=1}^K p_j \left( \prod\limits_{i=1}^N (1 - \theta_i)^{InvNr_{ij}} \cdot \prod\limits_{l=1}^L (1 - \psi_{l})^{MsgSize(l,j)} \right) \]
This model takes into account failure probabilities of components ($\theta_i$) and communication links ($\psi_{l}$), as well as the probability of a scenario to be executed ($p_j$). Such probabilities are combined to obtain the overall reliability on demand of the system ($\theta_S$), which represents how often the system is not expected to fail when its scenarios are invoked.
Such probabilities are combined to obtain the overall reliability on demand of the system, which represents how often the system is not expected to fail when its scenarios are invoked.

The model is considered to be composed of $N$ components and $L$ communication links, whereas its behavior is made of $K$ scenarios. The probability ($p_j$) of a scenario $j$ to be executed is multiplied by an expression that describes the probability that no component or link fails during the execution of the scenario. This expression is composed of two terms: $\prod_{i=1}^N (1 - \theta-i)^{InvNr_{ij}}$, which is the probability of the involved components not to fail raised to the power of their number of invocations in the scenario (denoted by $InvNr_{ij}$), and $\prod_{l=1}^L (1 - \psi_{l})^{MsgSize(l,j)}$, which is the probability of the involved links not to fail raised to the power of the size of messages traversing them in the scenario (denoted by $MsgSize(l,j)$). 

\subsubsection*{Performance Antipatterns (\pas)}\label{sec:background:pas}

\textcite{DBLP:books/sp/03/SmithW03} have introduced the concepts of performance antipatterns.  
A performance antipattern describes bad design practices that might lead to performance degradation in a system. 
These textual descriptions were later translated into first-order logic (FOL) equations~\cite{DBLP:journals/sosym/CortellessaMT14}.
The evolutionary approach exploits the FOL equations to detect the occurrence of performance antipatterns in the model.
\Cref{tab:supported-pas} lists the performance antipatterns detectable by the optimization engine.

\begin{table*}[htbp]
\begin{revfloatenv}
    
    \centering
   \begin{tabular}{p{.32\textwidth}p{.63\textwidth}}
    \toprule
        Performance antipattern & Description \\
    \midrule
        Pipe and Filter              & It occurs when the slowest filter in a ``pipe and filter'' causes the system to have unacceptable throughput. \\
	\midrule
        Blob                         & It occurs when a single component either (i) performs the greatest part of the work of a software system or (ii) holds the greatest part of the data of the software system. Either manifestation results in excessive message traffic that may degrade performance. \\
	\midrule
        Concurrent Processing System & It occurs when processing cannot make use of available processors. \\
	\midrule
        Extensive Processing         & It occurs when extensive processing in general impedes overall response time.\\ 
	\midrule
        Empty Semi-Truck             & It occurs when an excessive number of requests is required to perform a task. It may be due to inefficient use of available bandwidth, an inefficient interface, or both. \\
	\midrule
        The Tower of Babel               & It occurs when processes use different data formats and spend too much time converting them to an internal format. \\
    \bottomrule
    \end{tabular}
	\caption{Detectable performance antipatterns in our approach. Left column lists performance antipattern names, while the right column lists performance antipattern descriptions~\cite{DBLP:books/sp/03/SmithW03}.}   
    \label{tab:supported-pas}
\end{revfloatenv}
\end{table*}

\subsubsection*{Refactoring cost (\achanges)}\label{sec:background:distance}

The refactoring cost~\cite{Arcelli-Cortellessa-D-Emidio-Di-Pompeo-2018}, denoted as \achanges in this study, represents the effort required to transform the initial UML model into a modified version through the application of refactoring actions.

In a previous work~\cite{Cortellessa-Di-Pompeo-Stoico-Tucci-2021}, we introduced two key metrics to quantify this cost: the \emph{baseline refactoring factor} (\texttt{BRF}) and the \emph{architectural weight} (\texttt{AW}). 
The BRF is action-specific and captures the effort needed to perform a particular refactoring action, irrespective of the model element it targets. 
Conversely, the AW is element-specific and reflects the effort associated with applying a given action to a specific element.

Therefore, the overall refactoring cost (\achanges) is computed by summing the efforts of all refactoring actions within a given sequence, as expressed in the following equation:
\begin{revfloatenv}
\[
\achanges(\mathbb{A}) = \sum_{a_i(el_j) \in \mathbb{A}} BRF(a_i) \times AW(el_j)
\]
\end{revfloatenv}

\subsection*{Computational Cost of Objectives}

The cost of evaluating each objective function varies significantly and is a key factor in the overall runtime. 
The most expensive objective is the detection of performance antipatterns (\pas), which involves iteratively matching multiple antipattern definitions against all feasible positions in the UML model. This process requires deep model traversal and constraint evaluation, resulting in substantial computational effort (around $80\%$ of the time spent to evaluate a solution).
The second most expensive objective is the average system performance (\perfq), where although the final value is computed via a closed-form formula, the required input values are obtained through simulation of an LQN model transformed from the UML model. Simulation accuracy settings influence runtime, but even conservative settings require noticeable computation.
System reliability (\reliability) is comparatively inexpensive, as it only involves model navigation to read reliability values annotated on UML. The least expensive objective is the refactoring cost (\achanges), which is trivially computed as a factor of the number of refactoring actions applied to the model.

\subsection{Quality indicators}\label{sec:background:qi}

To estimate the quality of a computed Pareto front, different quality estimators have been introduced, such as the Hypervolume (HV)~\cite{Cao-Smucker-Robinson-2015,Beume-Naujoks-Emmerich-2007} and Inverse Generational Distance (IGD+)~\cite{Ishibuchi-Masuda-Tanigaki-Nojima-2015}.
Each estimator measures a different quality aspect of a Pareto front.
We adopt six commonly used multi-objective quality indicators to evaluate and compare the solution sets obtained: Hypervolume (HV), Inverted Generational Distance (IGD), IGD+, Epsilon, Spread, and Generalized Spread (GSPREAD). These were selected based on their popularity in the optimization literature and their complementary characteristics as analyzed by \textcite{Li-Yao-2020}. In their survey, quality indicators are assessed on four desirable properties of a solution set: convergence (closeness to the Pareto front), spread (region coverage), uniformity (even distribution), and cardinality (number of solutions). Each indicator is rated for its ability to reflect these aspects. HV, for instance, excels in convergence, spread, and cardinality; IGD performs well in convergence and spread, and moderately in the others; Epsilon focuses on convergence with emphasis on the worst-case solution. Spread and GSPREAD are more specialized, describing the solution distribution and handling higher-dimensional objective spaces, respectively.

While our study does not attempt to validate the sufficiency of this set in the specific context of model-based refactoring, we follow common practice in software engineering optimization to ensure broad and robust coverage of solution quality characteristics. To the best of our knowledge, this is the first such comparative use of these indicators in the model-based refactoring domain.

\subsubsection*{Hypervolume} 
The HV measures the amount of the volume of the solution space that a computed Pareto front (\computedP) covers with respect to a reference Pareto front (\referenceP), and it can assume values between $0$ and $1$.
When $HV = 0$, it means that the \computedP is fully dominated by the \referenceP, while $HV = 1$ means that each point within the \computedP is non-dominated by any points within the \referenceP.
Therefore, the closer to $1$ the HV, the higher the quality of the \computedP.
\subsubsection*{Inverse Generational Distance (IGD)}
The IGD measures the average distance from points on the reference Pareto front to the nearest points on the computed Pareto front. An IGD of $0$ indicates that all reference Pareto points are perfectly represented, and vice versa. 
Therefore, the closer to $0$ the IGD, the better the computed Pareto front in terms of both convergence and diversity.
\subsubsection*{Inverse Generational Distance plus (IGD+)}  
The Inverse Generational Distance plus (IGD+) indicator is an IGD enhancement that accounts for both convergence and diversity while ensuring strict Pareto compliance~\cite{Ishibuchi-Masuda-Tanigaki-Nojima-2015}. 
Unlike IGD, which measures the average distance to dominated solutions, IGD+ only considers the distances to solutions that dominate or are non-dominated by the reference Pareto front. 
This adjustment eliminates the possibility of penalizing solutions inappropriately and provides a more robust assessment of the Pareto front quality.
An IGD+ of $0$ indicates a perfect representation of the reference Pareto front by the computed Pareto front, while larger IGD+ values mean lower quality. 
Therefore, the closer to $0$ the IGD+, the better the balance between convergence and diversity in the obtained Pareto front.
\subsubsection*{Epsilon} 
The Epsilon ($\epsilon$-idicator) measures the maximum distance between two sets. 
For example, given \referenceP and \computedP, $\epsilon(\computedP,\referenceP) \leq 0$ implies that \computedP (weakly) dominates \referenceP. 
Therefore, the lower value of the Epsilon, the higher the quality of the Pareto front.
\subsubsection*{SPREAD}
The SPREAD evaluates the uniformity and extent of a distribution of solutions along the obtained Pareto front. A SPREAD of $0$ indicates that solutions are perfectly distributed across the front, while higher values denote an uneven distribution. Therefore, the lower the SPREAD, the better the diversity of the solutions.
\subsubsection*{GSPREAD}
The GSPREAD measures the average Euclidean distance between the computed Pareto front and the closest points on the reference Pareto front. A GSPREAD of $0$ indicates a perfect convergence to the reference front, while larger values mean increased divergence. Therefore, the closer to $0$ the GSPREAD, the higher the quality of the Pareto front in terms of convergence.

\bigskip

In our evaluation, we use the indicators above to estimate the quality of the \computedP obtained with a search budget when compared to a \referenceP computed without budgets but terminated after $100$ genetic evolutions.

\begin{table}[htbp]
	\centering
	\begin{tabular}{p{3.7cm}p{8.5cm}}
		\toprule
		\textbf{Quality Indicator} & \textbf{Description} \\
		\midrule
		\textbf{Hypervolume} (HV) & Measures the volume of the solution space covered by the computed Pareto front. \\
		\midrule
		\textbf{Inverse Generational Distance} (IGD)       & Evaluates the quality of representation of the reference Pareto front by the computed front. \\ 
		\midrule
		\textbf{Inverse Generational Distance plus} (IGD+) & Measures the distance from a solution in the reference Pareto front to the nearest solutions in the computed Pareto front. \\
		\midrule
		\textbf{Epsilon}   & Reflects the worst-case distance between the computed and reference Pareto fronts. \\
		\midrule
		\textbf{SPREAD}    & Measures the distribution and uniformity of the solutions along the Pareto front. \\
		\midrule
		\textbf{Generational SPREAD} (GSPREAD)   & Quantifies how closely the computed Pareto front approximates the reference front. \\
		\bottomrule
	\end{tabular}
	\caption{Quality indicators used in the evaluation}
	\label{tab:quality-indicators}
\end{table}
 \section{Study Design}\label{sec:approach}

The goal of the study is to determine whether the imposition of a time-based search budget can hamper the quality of the resulting Pareto fronts in the context of a model-based multi-objective optimization.
Additionally, we are interested in how different algorithms cope with the search budgets.
To this end, we selected two case studies and ran a number of optimization experiments with time-based budgets. 
We varied the budget limit between \texttt{15}, \texttt{30}, and \texttt{60} minutes, while considering the quality indicators described in~\Cref{sec:background}.

The selected budget values were informed by several considerations. First, in our previous work~\cite{EPEW2023}, we evaluated the same algorithms under unconstrained settings, which typically required between 3 and 6 hours of execution time depending on the algorithm and system under analysis. These extended runs provided valuable performance baselines, but they are impractical in design-time scenarios where timely feedback is essential. In contrast, the current study focuses on constrained optimization and aims to analyze algorithm behavior under increasingly tight runtime budgets. We intentionally selected substantially shorter durations to observe the degradation (or robustness) in optimization quality under limited computational resources.
Importantly, we designed these time budgets with practical use in mind. In an envisioned usage scenario, a software architect or designer would execute optimization runs as part of an iterative refactoring workflow. 
For such interactive use, waiting several hours for results would not be acceptable. Thus, budget values in the range of tens of minutes are far more realistic. Also, a time budget that is too generous would ultimately resemble unconstrained execution and undermine the purpose of studying budget-aware trade-offs.

Moreover, for each search budget, we ran three genetic algorithms: \nsga, \spea, and \pesa.
These algorithms were chosen due to their different search policies, as described in~\Cref{sec:background}.

To account for the random nature of genetic algorithms~\cite{Zitzler-Deb-Thiele-2000}, we ran the same experiment \textit{30} times and computed the QIs for each resulting Pareto front (\computedP).
Since the reference Pareto front (\referenceP) is unknown in our case studies\footnote{We remark that it is also challenging to extract an equation to compute it.}, we computed the quality indicator with respect to the best Pareto front obtained for each case study after running the algorithms for \textit{100} genetic evolutions (\ie without search budgets) across \textit{30} executions. 
The entire study consisted of \textbf{558} experiments that we performed on three AMD EPYC 7282, each with 64 cores and 512GB of RAM.\footnote{Replication package: \url{https://github.com/SpencerLabAQ/replication_package_search-budget.git}}

We followed the guidelines by \textcite{Arcuri-Briand-2014} to compare the experiments against each other.
Therefore, we applied the \mwu non-parametric statistical test (also referred to as Wilcoxon rank-sum test)~\cite{Mann-Whitney-1947} with a null hypothesis ($H_0$) stating that the experiments do not have a statistically significant difference.
Two experiments are considered to be significantly different on the basis of their quality indicator value if the test computes a p-value smaller than $\alpha=0.05$.
To assess the magnitude of the difference, we used the Vargha--Delaney \vda~\cite{Vargha-Delaney-2000}, a standardized non-parametric effect size measure.
\vda can take values between $0$ and $1$, and a value of $0.5$ indicates that the two experiments are equivalent.
The closer the \vda value gets to $0$ or $1$, the larger the effect size.
The interpretation of the magnitude as being negligible, small, medium, and large is performed according to the thresholds $0.147$, $0.33$, and $0.474$ respectively~\cite{Hess-Kromrey-2004}.

In addition to the quantitative analysis above, we conducted a qualitative analysis to assess differences in the software models when using different budgets. 
First, we looked at \achanges  and \pas as distinctive characteristics of the models, which were treated as optimization objectives in the experiments. 
Second, we relied on the types of refactoring actions and their arrangement in sequences (generated by the optimization) as proxies for the software models derived from those sequences. 
The sequences resulting from a given experiment (or search space) were represented as trees to facilitate comparisons between experiments.

\begin{table*}
	\centering
	\begin{tabular}{p{2.5cm}lp{4.5cm}}
		\toprule
		{} & Configuration & Eligible values  \\
		\midrule
		\multirow{6}{=}{\par{Common configuration}} & Number of genetic evolutions      &  100   \\
		                                      & Population Size                   & 16 \\
		                                      & Number of independent runs        & 30 \\
		                                      & $P_{crossover}$                   & 0.80 \\
		                                      & Crossover Operator                & Single Point \\
		                                      & $P_{mutation}$                    & 0.20 \\
		                                      & Mutation Operator                 & Simple Mutation \\
                \midrule
		\nsga                                 & Selection operator                & \par{Binary Tournament Selection with crowding distance} \\
                \midrule
		\multirow{3}{*}{\spea}                & Selection operator                & Binary Tournament Selection \\
		                                      & Archive population size           & 16 \\
						      & Distance to the k-th individual   & 1 \\
                \midrule
		\multirow{2}{*}{\pesa}                & Archive population size           & 16 \\
						      & Number of hyper-grids             & 5 \\
		\bottomrule
	\end{tabular}
	\caption{\label{tab:config_params} Configuration values for the evolutionary algorithms.}
\end{table*}
 \section{Case Studies}\label{sec:case-study}

We applied our approach to two case studies from the literature: i) the Train Ticket Booking Service (\ttbs)~\cite{Di-Pompeo-Tucci-Celi-Eramo-2019}, and ii) the well-established modeling case study \ccm, whose UML model has been derived by the specification in~\cite{Herold-Klus-Welsch-Deiters-Rausch-Reussner-Krogmann-Koziolek-Mirandola-Hummel}.\footnote{\url{https://github.com/SEALABQualityGroup/uml2lqn-casestudies}}
We note that we extracted the UML models and selected the performance-critical scenarios for analysis from the source code in the case of \ttbs and from the specification in the case of \ccm.

\subsubsection*{Train Ticket Booking Service}

\ttbs is a Web-based booking application whose architecture is based on the microservices paradigm.
The system is made up of 40 microservices, and it provides different scenarios through which users can perform realistic operations, \eg book a ticket or watch trip information.

We selected \emph{Login}, \emph{Update user details} and \emph{Rebook} as use cases because they commonly represent performance-critical scenarios in a ticketing booking service.
Also, the model defines two user categories: simple and admin users.

\subsubsection*{CoCoME}

\ccm describes a trading system containing several stores.
A store can have one or more cash desks for processing goods.
A cash desk is equipped with all the tools needed to serve a customer (\eg a Cash Box, Printer, Bar Code Scanner). 
\ccm covers possible use cases performed at a cash desk (\eg scanning products, paying by credit card, or ordering new goodies).

We focused on three scenarios: \emph{Process Sale}, \emph{Receive Ordered Products}, and \emph{Show stock reports} because they represent common activities in a trading system.
 \section{Research Questions}\label{sec:rqs}

The three research questions we intend to address in this study are presented below. 
Afterward, we describe the results for each question, and discuss the key findings and implications for the designer. 

\subsection{How can we characterize the trade-offs between computational efficiency and solution quality in time-constrained multi-objective optimization for software refactoring?}\label{sec:rq1}

This question targets the balance between computational efficiency and solution quality under limited execution time. RQ1 is broken down into three sub-questions, each addressing a specific aspect of this trade-off.

\subsubsection{RQ1.1: Which algorithm completes the search process faster?}\label{sec:rq1:rq11}

When a time constraint is imposed on the optimization, a designer might be interested in selecting the algorithm that provides the best quality solutions belonging to a computed Pareto front (\computedP) for the specific budget.
It is worth mentioning that the quality of \computedP can be estimated through several quality indicators (QIs)~\cite{Li-Yao-2020,Li-Chen-Yao-2022}.
Each quality indicator measures a specific characteristic of that \computedP, and none of them is a clear winner to estimate Pareto fronts~\cite{Li-Chen-Yao-2022}.
For this reason, we chose six quality indicators reported in \Cref{tab:quality-indicators} to assess several angles of \computedP in our experiments. 

\Cref{fig:QI-HV-timelines,fig:QI-IGDp-timelines,fig:QI-IGD-timelines,fig:QI-SPREAD-timelines,fig:QI-SPREAD-timelines,fig:QI-GSPREAD-timelines,fig:QI-EP-timelines} depict the timelines of how the quality indicators vary with different search budgets, and how many genetic evolutions were performed during the search. 
We can observe that \nsga was the fastest algorithm, \spea the slowest one.
Furthermore, the number of genetic evolutions of \ccm is consistent with that of \ttbs.

\subsubsection*{Hypervolume (HV)}

\begin{figure*}[!ht]
	\centering
   \begin{subfigure}{.99\linewidth}
      \includegraphics[width=.96\textwidth]{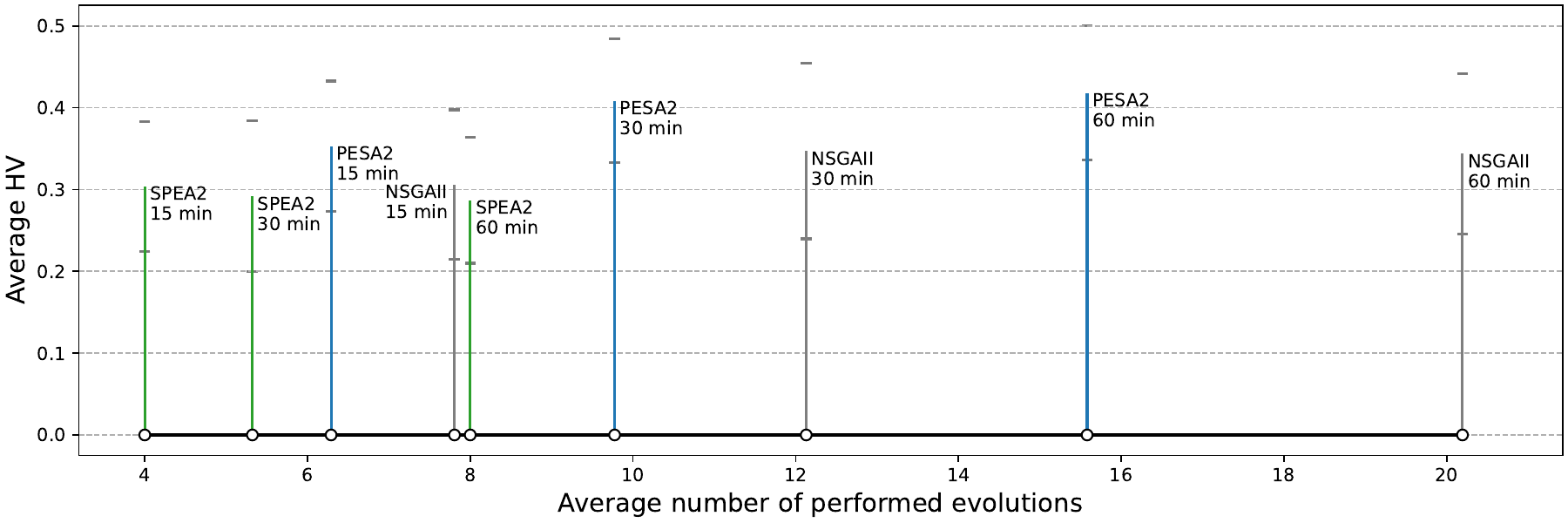}
      \caption{\ttbs}
      \label{fig:ttbs_HV_timeline}
   \end{subfigure}\hfill
\begin{subfigure}{.99\linewidth}
      \includegraphics[width=.96\textwidth]{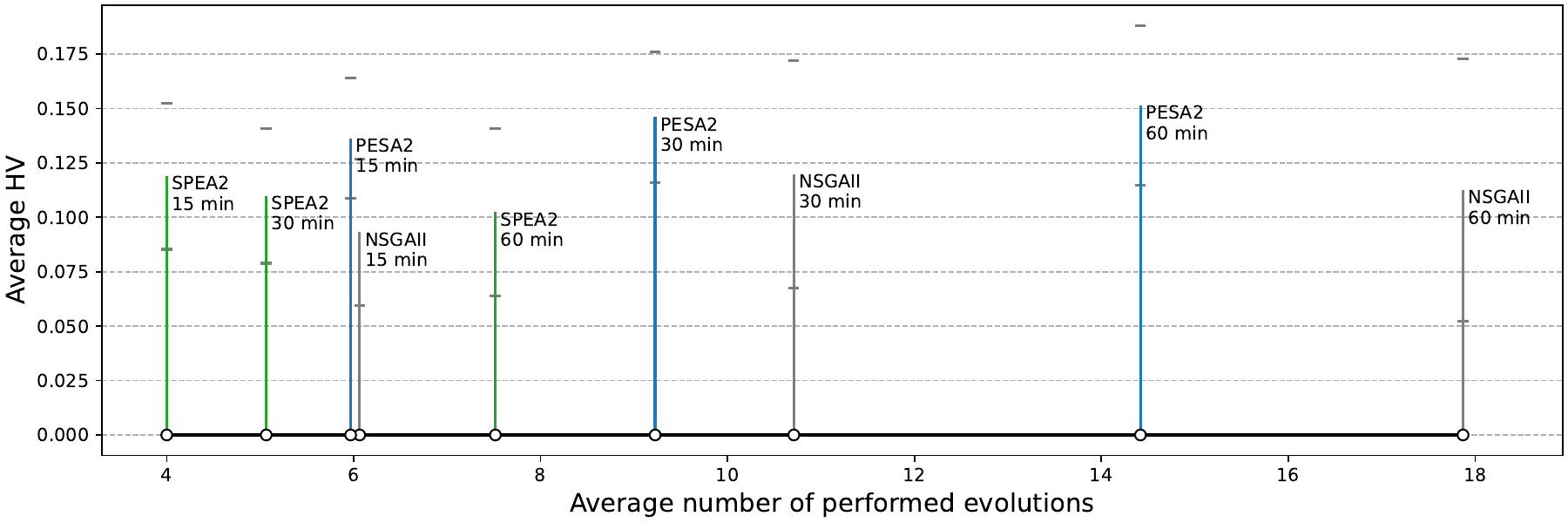}
      \caption{\ccm}
      \label{fig:simplified-cocome_HV_timeline}
   \end{subfigure}\hfill
\caption{Timelines of the number of evolutions performed by the algorithms for the different budget configurations, along with the achieved HV. Vertical bars show the average HV over 30 runs, while ticks represent the standard deviation from the mean.}
   \label{fig:QI-HV-timelines}
\end{figure*}

Analyzing the \ttbs results (see \Cref{fig:ttbs_HV_timeline}), we observe that the HV values of \spea lie almost close to $0.3$ for every search budget. 
For \pesa, in turn, the longer the search budget, the higher the HV values.
The HV values for \nsga increase between the $15$ and $30$ minutes budgets and then become mostly flat between $30$ and $60$ minutes.
In addition, the timelines of the two case studies seemed to resemble each other.

\subsubsection*{Inverse Generational Distance (IGD)}
\begin{figure*}[!ht]
        \centering
\begin{subfigure}{.99\linewidth}
      \includegraphics[width=.96\textwidth]{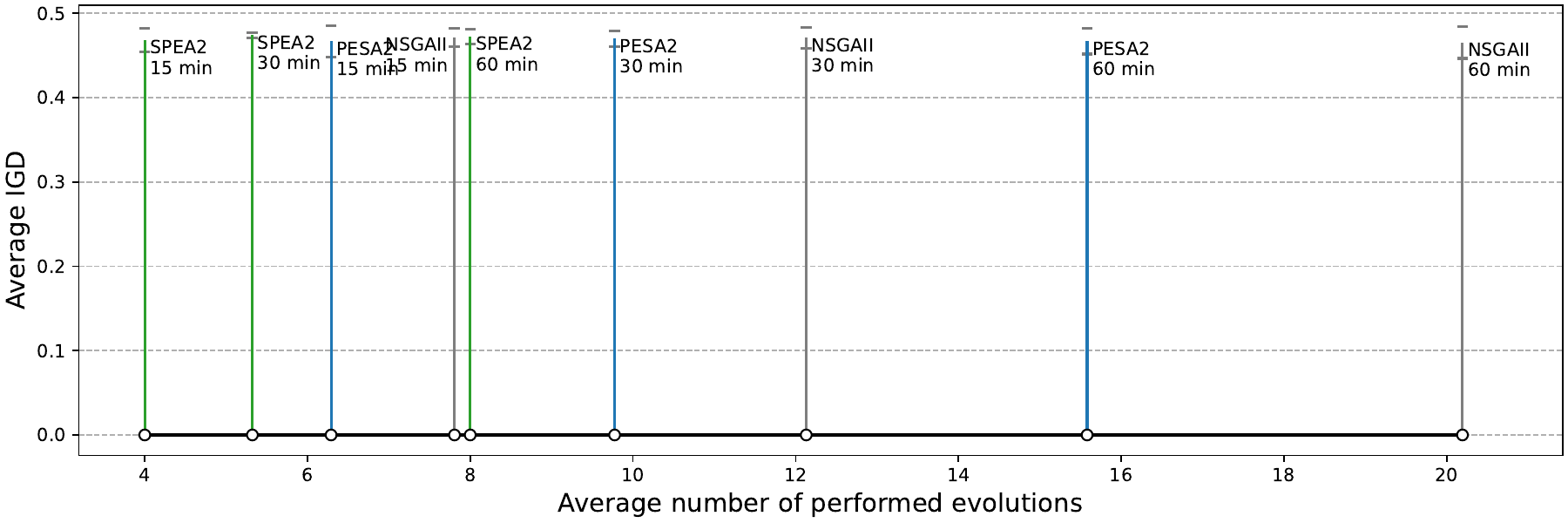}
      \caption{\ttbs}
      \label{fig:ttbs_IGD_timeline}
   \end{subfigure}\hfill
\begin{subfigure}{.99\linewidth}
      \includegraphics[width=.96\textwidth]{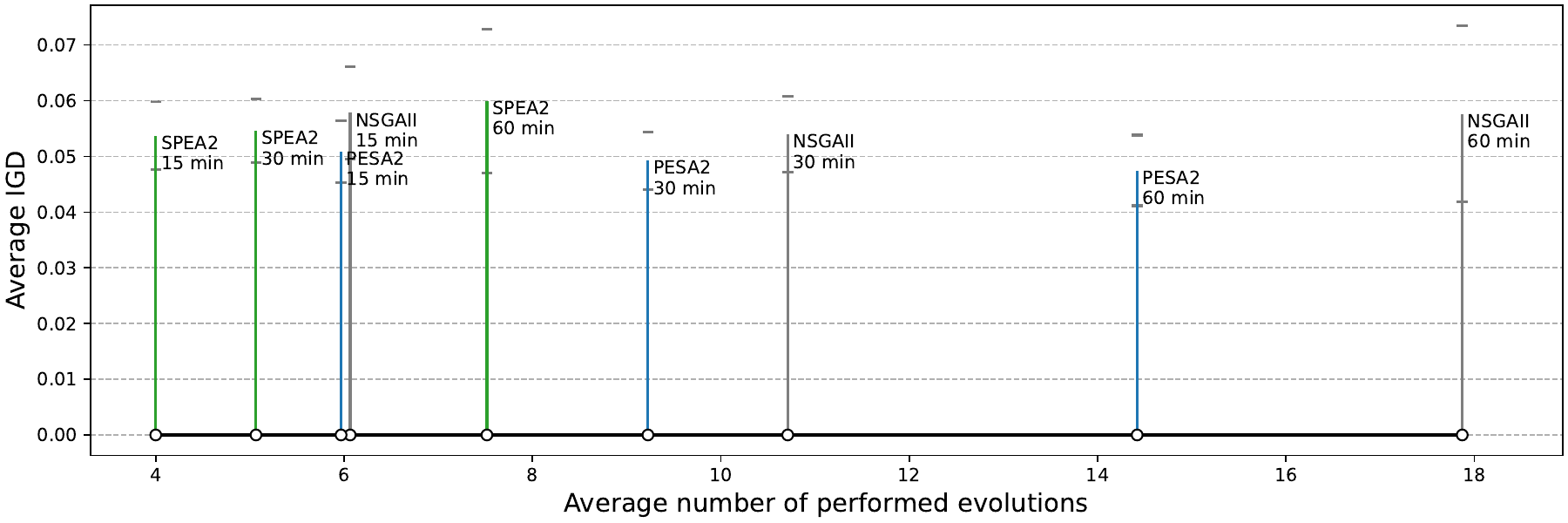}
      \caption{\ccm}
      \label{fig:simplified-cocome_IGD_timeline}
   \end{subfigure}\hfill
   \caption{Timelines of the number of evolutions performed by the algorithms for the different budget configurations, along with the achieved IGD. Vertical bars show the average IGD over 30 runs, while ticks represent the standard deviation from the mean.}
   \label{fig:QI-IGD-timelines}
\end{figure*}

When measuring the convergence and diversity of \computedP fronts for \ttbs case study, we observe that the three algorithms achieved similar IGD values, all close to $0.5$ (see \Cref{fig:ttbs_IGD_timeline}).
In contrast, for \ccm case study (see \Cref{fig:simplified-cocome_IGD_timeline}) the three algorithms exhibited different behaviors.
\spea showed a worsening trend with longer budget, \ie from $\sim 0.053$ for $15$-minute of budget to \rev{$\sim 0.06$} for $60$-minute of budget.  
\nsga showed a slight improvement between the $15$ and $30$ minutes of budget, and then a slight worsening between the $30$ and $60$ minutes of budget.
Contrary to \spea and \nsga, \pesa marginally improved the quality of the IGD values while increasing the budget, \ie from $\sim 0.05$ for $15$-minute of budget to \rev{$\sim 0.048$} for $60$-minute of budget.
 
\subsubsection*{Inverse Generational Distance plus (IGD+)}
\begin{figure*}[!ht]
	\centering
\begin{subfigure}{.99\linewidth}
      \includegraphics[width=.96\textwidth]{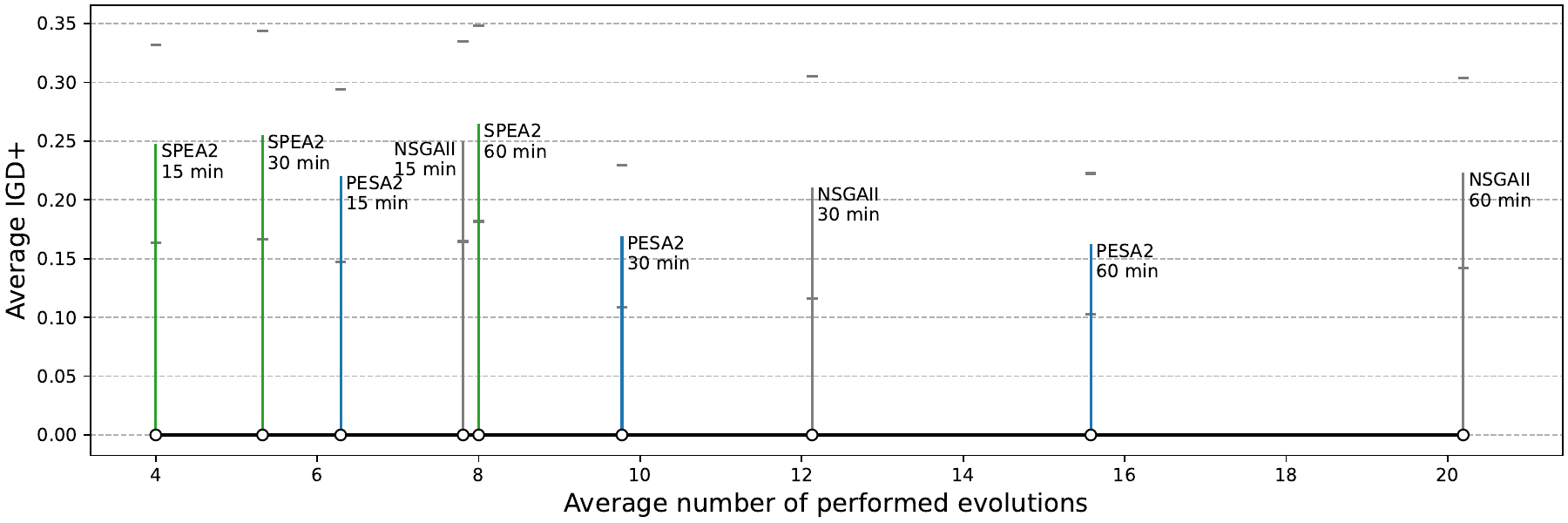}
      \caption{\ttbs}
      \label{fig:ttbs_IGDp_timeline}
   \end{subfigure}\hfill
\begin{subfigure}{.99\linewidth}
      \includegraphics[width=.96\textwidth]{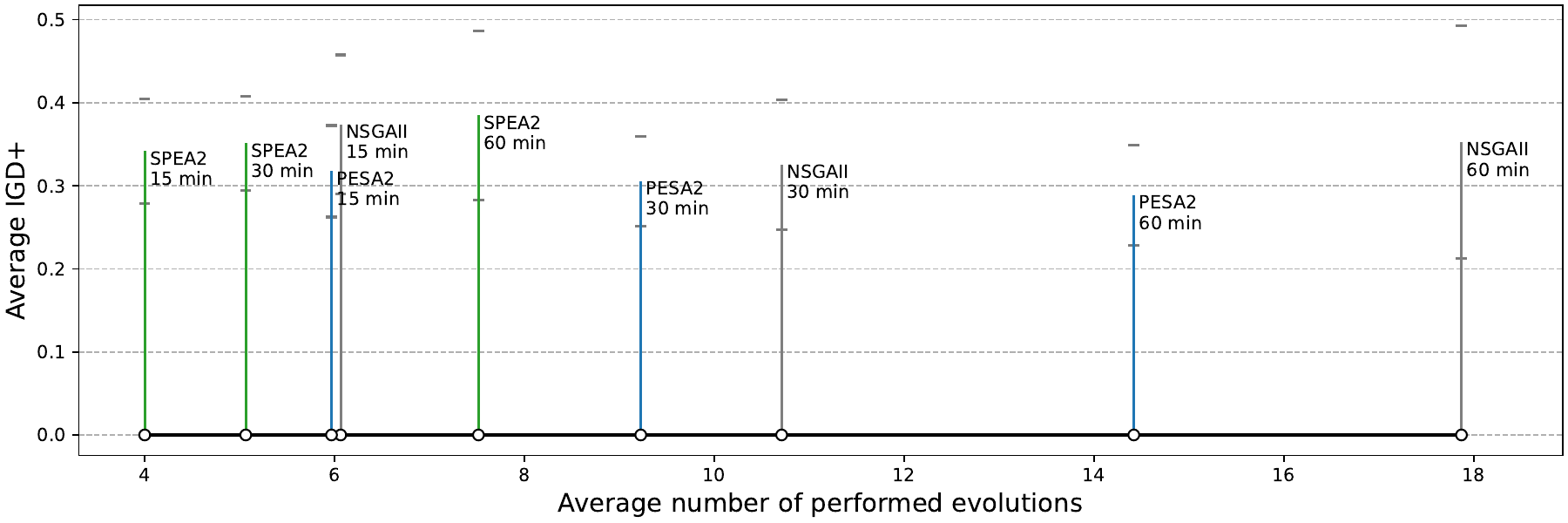}
      \caption{\ccm}
      \label{fig:simplified-cocome_IGDp_timeline}
   \end{subfigure}\hfill
   \caption{Timelines of the number of evolutions performed by the algorithms for the different budget configurations, along with the achieved \igdp. Vertical bars show the average \igdp over 30 runs, while ticks represent the standard deviation from the mean.}
   \label{fig:QI-IGDp-timelines}
\end{figure*}

Concerning the IGD+ (see \Cref{fig:QI-IGDp-timelines}), which measures the inverse euclidean distance between the \computedP and the \referenceP, we observed that \spea achieved the worst (\ie the highest) IGD+ values in all case studies. 
Furthermore, it showed an increasing trend with longer budgets for both case studies.
\pesa showed the best (\ie the lowest) IGD+ values in the two use cases and for the three budgets.
In addition, \pesa showed a larger improvement when moving from $15$ to $30$ minutes of budget for \ttbs case study, then the IGD+ values remained almost flat.
On the other hand, for \ccm case study, \pesa marginally improved the IGD+ values while increasing the budget, \ie from $\sim 0.3$ for $15$ minute of budget to $\sim 0.28$ for $60$ minutes of budget.
\nsga showed its best IGD+ values with $30$ minutes of budget, and then a slight worsening with $60$ minutes of budget for both case studies.

\subsubsection*{Epsilon}
\begin{figure*}[!ht]
	\centering
\begin{subfigure}{.99\linewidth}
      \includegraphics[width=.96\textwidth]{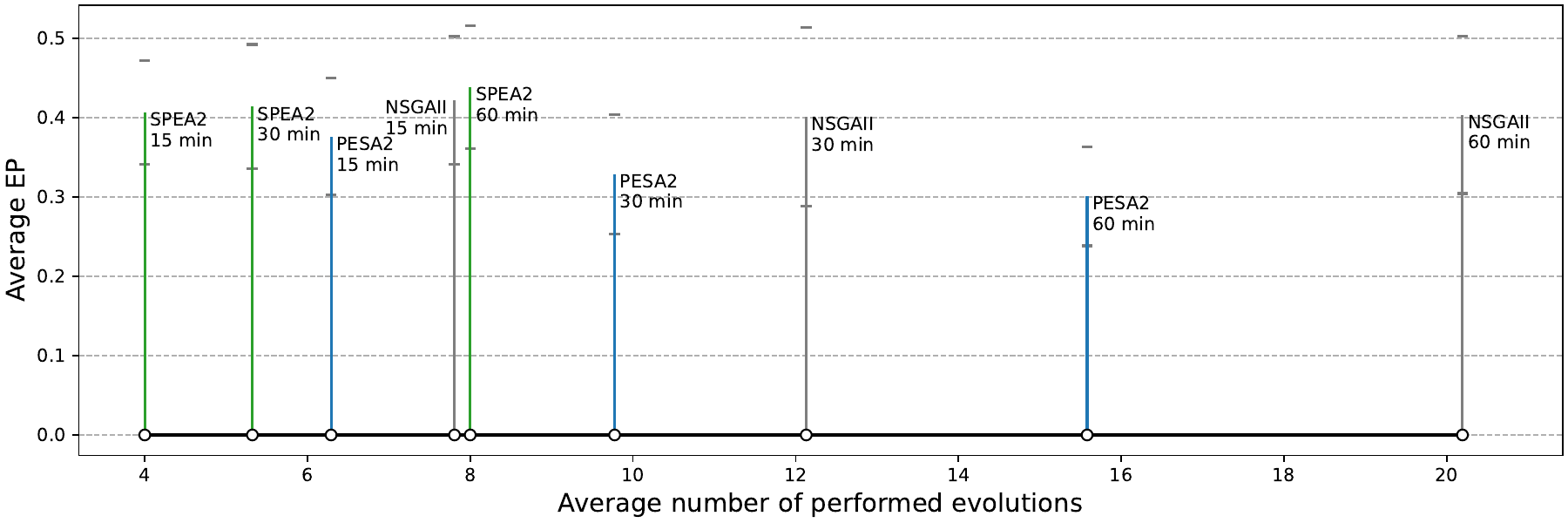}
      \caption{\ttbs}\label{fig:ttbs_EP_timelinep}
   \end{subfigure}\hfill
\begin{subfigure}{.99\linewidth}
      \includegraphics[width=.96\textwidth]{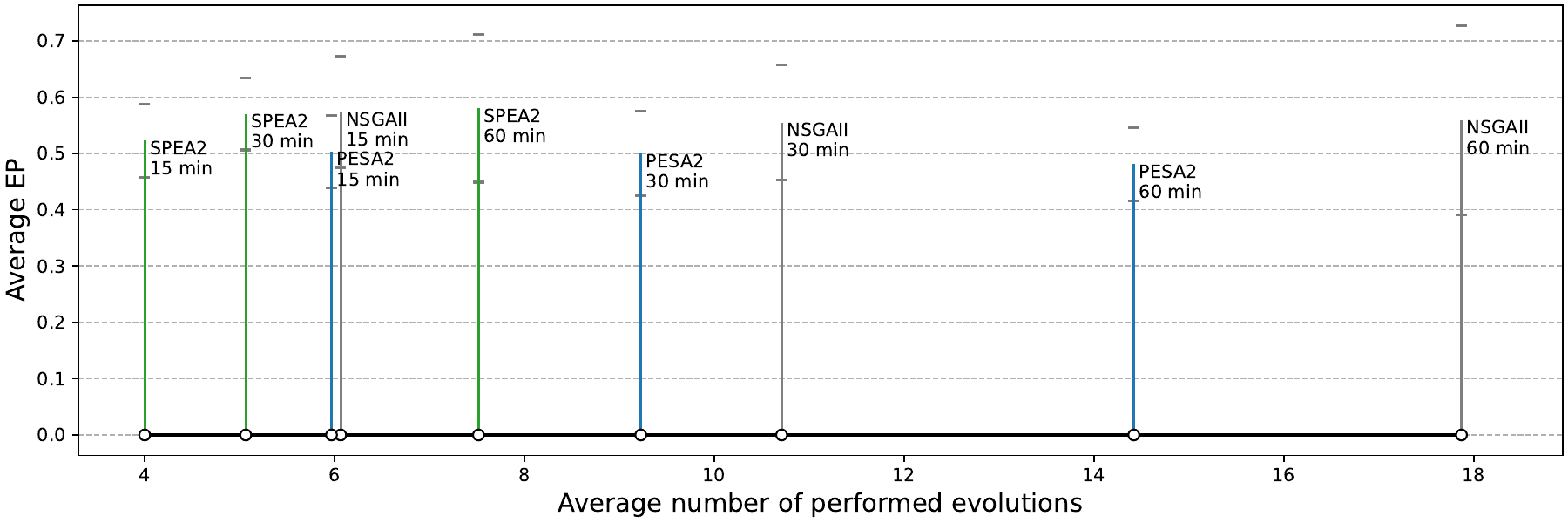}
      \caption{\ccm}\label{fig:simplified-cocome_EP_timeline}
   \end{subfigure}\hfill
   \caption{Timelines of the number of evolutions performed by the algorithms for the different budget configurations, along with the achieved Epsilon. Vertical bars show the average Epsilon over 30 runs, while ticks represent the standard deviation from the mean.}\label{fig:QI-EP-timelines}
\end{figure*}

Considering the Epsilon quality indicator ($\epsilon$-indicator), we can see from \Cref{fig:QI-EP-timelines} that \spea was the worst algorithm in all case studies for every budget.
\pesa, instead, showed the best $\epsilon$, \ie the lowest value, was $0.3$ for \ttbs and $\sim 0.48$ for \ccm with 60 minutes of budget.
We can notice that the longer the budget, the higher the Epsilon values for \spea, while the opposite trend was observed for \pesa.
\nsga showed a flat trajectory for the $\epsilon$-indicator values in the two case studies.
 
\subsubsection*{SPREAD}
\begin{figure*}[!ht]
	\centering
\begin{subfigure}{.99\linewidth}
      \includegraphics[width=.96\textwidth]{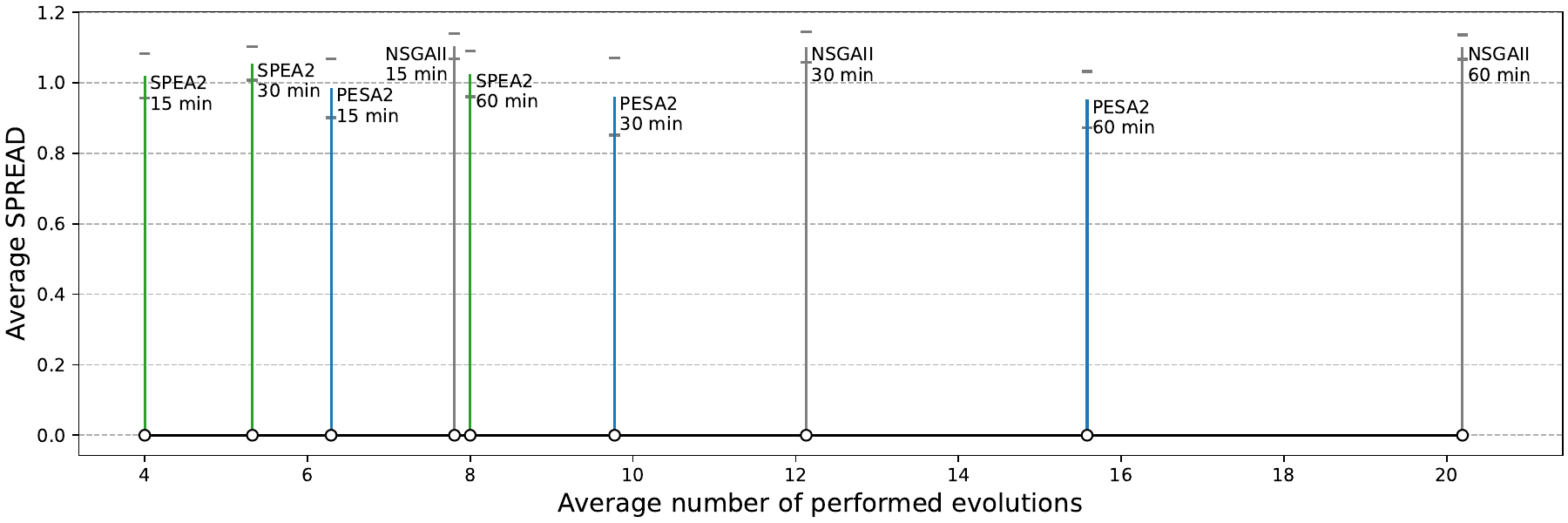}
      \caption{\ttbs}
      \label{fig:ttbs_SPREAD_timeline}
   \end{subfigure}\hfill
\begin{subfigure}{.99\linewidth}
      \includegraphics[width=.96\textwidth]{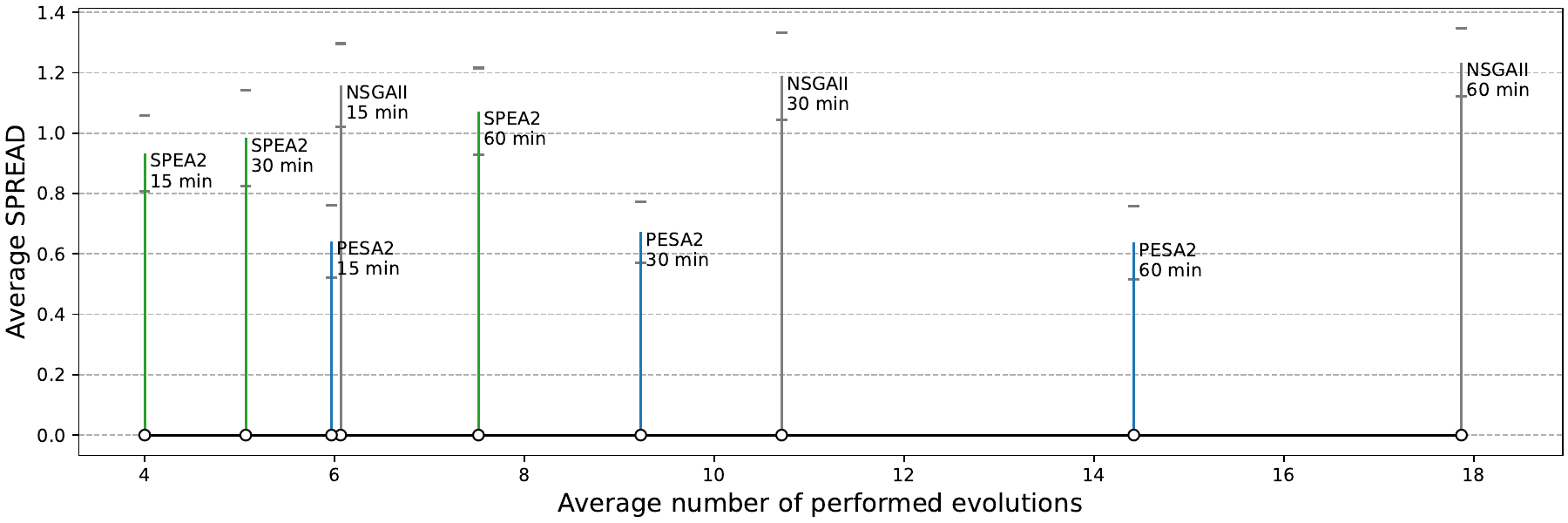}
      \caption{\ccm}
      \label{fig:simplified-cocome_SPREAD_timeline}
   \end{subfigure}\hfill
   \caption{Timelines of the number of evolutions performed by the algorithms for the different budget configurations, along with the achieved SPREAD. Vertical bars show the average SPREAD over 30 runs, while ticks represent the standard deviation from the mean.}
   \label{fig:QI-SPREAD-timelines}
\end{figure*}

Looking at the SPREAD indicator in \Cref{fig:QI-SPREAD-timelines}, the three algorithms showed different behaviors when varying the budget as well as the case study.
\nsga showed the worst SPREAD values for the two case studies and three budgets.
Furthermore, it performed worse for \ccm than \ttbs, and showed an increasing trend with longer budgets.
\spea showed an almost flat trajectory for \ttbs, while for \ccm it showed an increasing trend with longer budgets.
In addition, it performed better for \ccm than \ttbs for $15$ minutes and $30$ minutes of budget, conversely it performed worse for \ccm than \ttbs for $60$ minutes of budget.
\pesa was the best algorithm for the two case studies and for the three budgets by achieving the half values of \ttbs for \ccm.
Furthermore, it showed an almost flat trajectory in both case studies, with a slightly worse performance for \ccm with $30$ minutes of budget.

\subsubsection*{Generational SPREAD (GSPREAD)}
\begin{figure*}[!ht]
	\centering
\begin{subfigure}{.99\linewidth}
      \includegraphics[width=.96\textwidth]{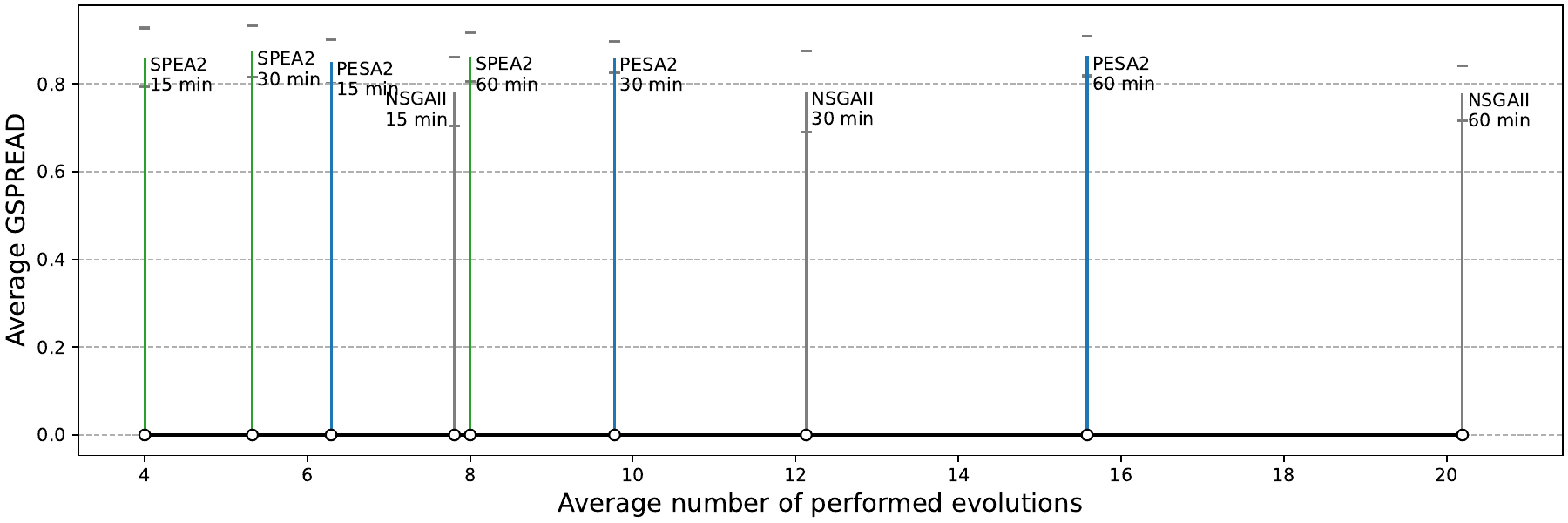}
      \caption{\ttbs}
      \label{fig:ttbs_GSPREAD_timeline}
   \end{subfigure}\hfill
\begin{subfigure}{.99\linewidth}
      \includegraphics[width=.96\textwidth]{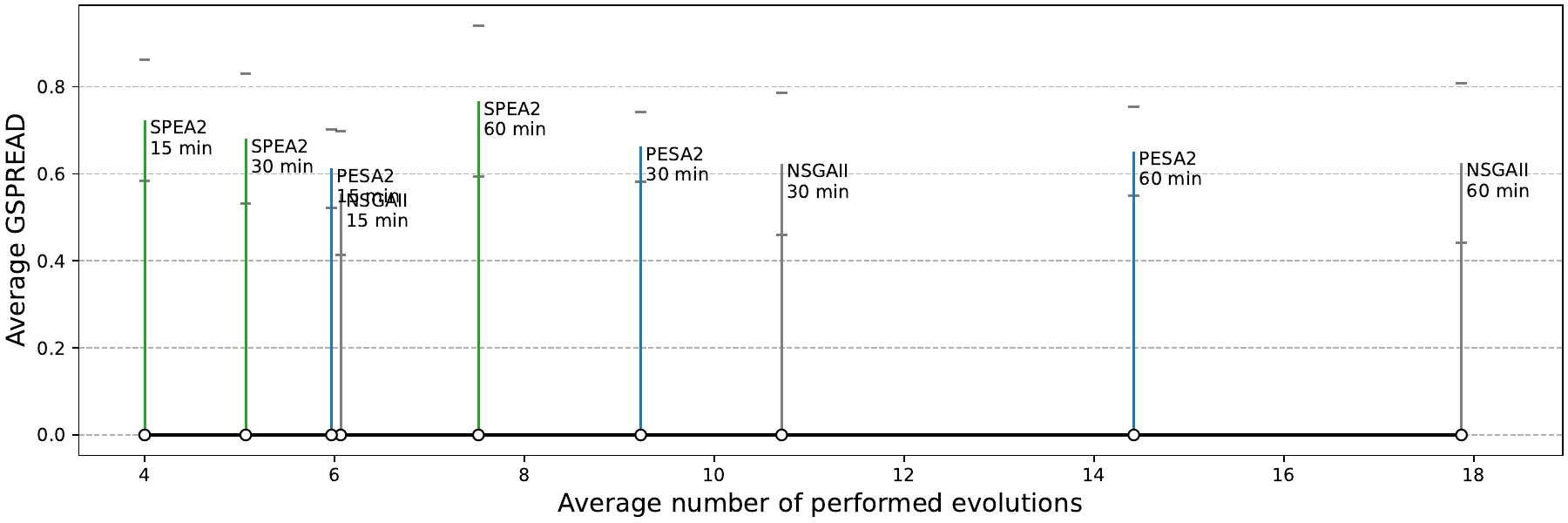}
      \caption{\ccm}
      \label{fig:simplified-cocome_GSPREAD_timeline}
   \end{subfigure}\hfill
   \caption{Timelines of the number of evolutions performed by the algorithms for the different budget configurations, along with the achieved GSPREAD. Vertical bars show the average GSPREAD over 30 runs, while ticks represent the standard deviation from the mean.}
   \label{fig:QI-GSPREAD-timelines}
\end{figure*}

Looking at the GSPREAD indicator in \Cref{fig:QI-GSPREAD-timelines}, we can see that the three algorithms found solutions showing better GSPREAD values for \ccm than \ttbs. 
Furthermore, the three algorithms showed a flat trajectory for the GSPREAD values in the \ttbs case study, while in the \ccm case study they showed different patterns.
Looking the \ccm case study, \spea increased the GSPREAD value between $15$ and $30$ minutes of budget and worsened the quality between $30$ and $60$ minutes of budget.
On the other hand, \nsga and \pesa worsened the GSPREAD values while increasing the budget. 

\bigskip

\subsubsection{RQ1.2: Which algorithm performs better when limited by a time budget?}

To understand the impact of the budget on the algorithms, we computed statistical test for each quality indicator, listed in \Cref{tab:HV_test_algo,tab:IGD+_test_algo,tab:IGD_test_algo,tab:SPREAD_test_algo,tab:GSPREAD_test_algo,tab:EP_test_algo}.
Furthermore, the tables report the results of the \mwu test, and the corresponding \vda effect sizes.
The name of the algorithm is underlined when i) the hypothesis test resulted in a significant difference, and ii) that algorithm yielded better QI values.

In the following, we discuss the results of the six QIs in detail by showing the timelines for each case study, and the hypothesis testing for the algorithms in the two case studies.

\subsubsection*{Hypervolume (HV)}

The hypothesis testing for the HV values is reported in \Cref{tab:HV_test_algo}, which reveals clear patterns in algorithm performance across different time budgets (highlighted in bold).
For the \ttbs case study, \pesa showed consistent superiority, significantly outperforming both \nsga and \spea at longer time budgets (\ie $30$ and $60$ minutes), with medium to large effect sizes.
\nsga also showed strengths, significantly outperforming \spea at both $30$ and $60$ minutes with small and medium effects. 
In the \ccm case study, the results showed interesting variations across time budgets. 
At $15$ minutes, both \pesa and \spea significantly outperformed \nsga with large and medium effects, respectively.
Also, \pesa also outperformed \spea with a small effect. 
As the time budget increased, \pesa emerged as the clear leader, significantly outperforming both competitors.
Overall, these results consistently demonstrate \pesa's superior performance in maximizing hypervolume, with its advantages becoming more pronounced with longer execution times.
\begin{table}[!ht]
\centering
\footnotesize
\begin{tabular}{llllll}
\toprule
Budget & Algor. 1 & Algor. 2 & MWU p & \multicolumn{2}{c}{Effect Size} \\

\midrule
\multicolumn{6}{c}{\textbf{\ttbs}} \\
\midrule
15 min & \underline{\pesa} & \nsga & \textbf{0.0487} & (S) \ebar{0.6462}{0.1462} \\
15 min & \underline{\pesa} & \spea & \textbf{0.0234} & (M) \ebar{0.6681}{0.1681} \\
15 min & \spea & \nsga & 0.8548 & (N) \ebar{0.4860}{0.0140} \\
30 min & \underline{\nsga} & \spea & \textbf{0.0385} & (S) \ebar{0.6535}{0.1535} \\
30 min & \underline{\pesa} & \nsga & \textbf{0.0167} & (M) \ebar{0.6774}{0.1774} \\
30 min & \underline{\pesa} & \spea & \textbf{3.9e-06} & (L) \ebar{0.8418}{0.3418} \\
60 min & \nsga & \underline{\pesa} & \textbf{0.0037} & (M) \ebar{0.2851}{0.2149} \\
60 min & \underline{\nsga} & \spea & \textbf{0.0202} & (M) \ebar{0.6722}{0.1722} \\
60 min & \spea & \underline{\pesa} & \textbf{5.8e-07} & (L) \ebar{0.1301}{0.3699} \\
\midrule
\multicolumn{6}{c}{\textbf{\ccm}} \\
\midrule
15 min & \nsga & \underline{\pesa} & \textbf{6.2e-06} & (L) \ebar{0.1655}{0.3345} \\
15 min & \nsga & \underline{\spea} & \textbf{0.0085} & (M) \ebar{0.3049}{0.1951} \\
15 min & \spea & \underline{\pesa} & \textbf{0.0487} & (S) \ebar{0.3538}{0.1462} \\
30 min & \underline{\pesa} & \nsga & \textbf{0.0066} & (M) \ebar{0.7014}{0.2014} \\
30 min & \underline{\pesa} & \spea & \textbf{1.1e-05} & (L) \ebar{0.8262}{0.3262} \\
30 min & \spea & \nsga & 0.5543 & (N) \ebar{0.4558}{0.0442} \\
60 min & \nsga & \spea & 0.3789 & (N) \ebar{0.5656}{0.0656} \\
60 min & \underline{\pesa} & \nsga & \textbf{0.0127} & (M) \ebar{0.6847}{0.1847} \\
60 min & \underline{\pesa} & \spea & \textbf{7.6e-06} & (L) \ebar{0.8314}{0.3314} \\

    \bottomrule
    \end{tabular}
    \caption{\mwu test and \vda effect sizes comparing the HV achieved by different algorithms in \independentRun runs. Magnitude interpretation: negligible (N), small (S), medium (M), large (L). The magnitude of the effect size is also represented by bars.}
    \label{tab:HV_test_algo}
\end{table}
    
 \subsubsection*{Inverse Generational Distance (IGD)}
\Cref{tab:IGD_test_algo} reports the hypothesis testing for the IGD values comparing pairs of algorithms for the two case studies.
IGD values showed few consistent patterns with few significant differences observed.
For the \ttbs case study, \pesa outperforms \spea at $30$ and $60$ minutes of budget with large effect sizes.
On the other hand, \pesa outperforms \nsga at $15$ and $60$ minutes with a medium effect size and at $30$ minutes with a large effect size.
The other comparisons are not statistically significant.
For the \ccm case study, \pesa still confirms its superiority over \spea and \nsga for all time budgets, with medium to large effect sizes.
\nsga and \spea do not show significant differences in the IGD values for the \ccm case study.
\begin{table}[!ht]
\centering
\footnotesize
\begin{tabular}{llllll}
\toprule
Budget & Algor. 1 & Algor. 2 & MWU p & \multicolumn{2}{c}{Effect Size} \\

\midrule
\multicolumn{6}{c}{\textbf{\ttbs}} \\
\midrule
15 min & \underline{\pesa} & \nsga & \textbf{0.0122} & (M) \ebar{0.6857}{0.1857} \\
15 min & \spea & \nsga & 0.3313 & (N) \ebar{0.5723}{0.0723} \\
15 min & \spea & \pesa & 0.0911 & (S) \ebar{0.3746}{0.1254} \\
30 min & \nsga & \underline{\pesa} & \textbf{9.1e-05} & (L) \ebar{0.2102}{0.2898} \\
30 min & \nsga & \spea & 0.8218 & (N) \ebar{0.5172}{0.0172} \\
30 min & \spea & \underline{\pesa} & \textbf{8.3e-07} & (L) \ebar{0.1353}{0.3647} \\
60 min & \nsga & \spea & 0.1181 & (S) \ebar{0.6160}{0.1160} \\
60 min & \underline{\pesa} & \nsga & \textbf{0.0024} & (M) \ebar{0.7253}{0.2253} \\
60 min & \underline{\pesa} & \spea & \textbf{1.0e-07} & (L) \ebar{0.8939}{0.3939} \\
\midrule
\multicolumn{6}{c}{\textbf{\ccm}} \\
\midrule
15 min & \underline{\pesa} & \nsga & \textbf{8.1e-04} & (L) \ebar{0.7482}{0.2482} \\
15 min & \spea & \nsga & 0.0995 & (S) \ebar{0.6223}{0.1223} \\
15 min & \spea & \pesa & 0.0738 & (S) \ebar{0.3673}{0.1327} \\
30 min & \nsga & \spea & 0.5927 & (N) \ebar{0.5401}{0.0401} \\
30 min & \underline{\pesa} & \nsga & \textbf{0.0075} & (M) \ebar{0.6982}{0.1982} \\
30 min & \underline{\pesa} & \spea & \textbf{6.9e-04} & (L) \ebar{0.7513}{0.2513} \\
60 min & \underline{\pesa} & \nsga & \textbf{1.6e-04} & (L) \ebar{0.7794}{0.2794} \\
60 min & \spea & \nsga & 0.1592 & (S) \ebar{0.3954}{0.1046} \\
60 min & \spea & \underline{\pesa} & \textbf{6.2e-07} & (L) \ebar{0.1311}{0.3689} \\

    \bottomrule
    \end{tabular}
    \caption{\mwu test and \vda effect sizes comparing the IGD achieved by different algorithms in \independentRun runs. Magnitude interpretation: negligible (N), small (S), medium (M), large (L). The magnitude of the effect size is also represented by bars.}
    \label{tab:IGD_test_algo}
    \end{table}
    
 \subsubsection*{Inverse Generational Distance plus (IGD+)}
\begin{table}[t]
\centering
\footnotesize
\begin{tabular}{llllll}
\toprule
Budget & Algor. 1 & Algor. 2 & MWU p & \multicolumn{2}{c}{Effect Size} \\

\midrule
\multicolumn{6}{c}{\textbf{\ttbs}} \\
\midrule
15 min & \nsga & \pesa & 0.2426 & (S) \ebar{0.4131}{0.0869} \\
15 min & \nsga & \spea & 0.8992 & (N) \ebar{0.5099}{0.0099} \\
15 min & \spea & \pesa & 0.1721 & (S) \ebar{0.3985}{0.1015} \\
30 min & \pesa & \nsga & 0.1431 & (S) \ebar{0.6087}{0.1087} \\
30 min & \underline{\pesa} & \spea & \textbf{1.2e-04} & (L) \ebar{0.7846}{0.2846} \\
30 min & \spea & \nsga & 0.0651 & (S) \ebar{0.3632}{0.1368} \\
60 min & \nsga & \underline{\pesa} & \textbf{0.0041} & (M) \ebar{0.2872}{0.2128} \\
60 min & \spea & \nsga & 0.0520 & (S) \ebar{0.3559}{0.1441} \\
60 min & \spea & \underline{\pesa} & \textbf{1.9e-06} & (L) \ebar{0.1478}{0.3522} \\
\midrule
\multicolumn{6}{c}{\textbf{\ccm}} \\
\midrule
15 min & \nsga & \underline{\pesa} & \textbf{0.0127} & (M) \ebar{0.3153}{0.1847} \\
15 min & \nsga & \spea & 0.1510 & (S) \ebar{0.3933}{0.1067} \\
15 min & \spea & \pesa & 0.2314 & (S) \ebar{0.4110}{0.0890} \\
30 min & \nsga & \spea & 0.1510 & (S) \ebar{0.6067}{0.1067} \\
30 min & \pesa & \nsga & 0.1510 & (S) \ebar{0.6067}{0.1067} \\
30 min & \underline{\pesa} & \spea & \textbf{0.0016} & (M) \ebar{0.7336}{0.2336} \\
60 min & \nsga & \pesa & 0.0555 & (S) \ebar{0.3580}{0.1420} \\
60 min & \underline{\nsga} & \spea & \textbf{0.0412} & (S) \ebar{0.6514}{0.1514} \\
60 min & \underline{\pesa} & \spea & \textbf{1.1e-05} & (L) \ebar{0.8252}{0.3252} \\

    \bottomrule
    \end{tabular}
    \caption{\mwu test and \vda effect sizes comparing the IGD+ achieved by different algorithms in \independentRun runs. Magnitude interpretation: negligible (N), small (S), medium (M), large (L). The magnitude of the effect size is also represented by bars.}
    \label{tab:IGD+_test_algo}
    \end{table}
     \Cref{tab:IGD+_test_algo} lists the hypothesis testing for the IGD+ values comparing pairs of algorithms for the two case studies.
For the \ttbs case study, \pesa demonstrated superior performance, particularly with longer execution times. 
At the $30$-minute budget, \pesa significantly outperformed \spea with a large effect size. 
This dominance became more pronounced at $60$-minute budget, where \pesa significantly outperformed both \nsga with medium effect and \spea with large effect. 
Similar patterns emerged in the \ccm scenario, where \pesa showed significant advantages over \nsga at $15$-minute budget with medium effect and over \spea at budgets of $30$-minute with medium effect and $60$-minute with a large effect.
These results suggest that \pesa generally provides better solution quality, with its advantages becoming more pronounced with longer execution times, while \spea typically yields worse results, particularly in extended time budgets.
To summarize, significant differences are primarily observed in the later stages ($30$ and $60$ minutes) for both problems. Some comparisons in early stages ($15$ minutes) are not statistically significant. 
\pesa frequently outperforms the other algorithms, particularly in \ccm, and \spea shows stronger performance in later stages for \ttbs.
\subsubsection*{Epsilon}

\Cref{tab:EP_test_algo} reports the hypothesis testing for the Epsilon values comparing pairs of algorithms for the two case studies.
For \ttbs and \ccm case studies, \pesa is statistically dominated by \nsga and \spea for all time budgets, with medium to large effect sizes.
\nsga and \spea do not show significant differences in the Epsilon values for the two case studies.
In particular, for the \ttbs case study, \spea outperforms \pesa with longer time budgets (\ie $30$ and $60$ minutes), while \nsga outperforms \pesa in all time budgets.
For the \ccm case case study, there are still differences between \pesa and the other two algorithms. 
However, the differences are less pronounced than in the \ttbs case study.
For example, for $15$ time budget, only \nsga statistically outperforms \pesa with a medium effect size, while the other differences are not statistically significant.
For longer time budgets, only \spea statistically outperforms \pesa with a large effect size.
From our results, we can conclude that \pesa is the algorithm that generates solutions being more distant from the \referenceP, while \spea is the algorithm that generates solutions closer to the \referenceP.
\begin{table}[!ht]
\centering
\footnotesize
\begin{tabular}{llllll}
\toprule
Budget & Algor. 1 & Algor. 2 & MWU p & \multicolumn{2}{c}{Effect Size} \\

\midrule
\multicolumn{6}{c}{\textbf{\ttbs}} \\
\midrule
15 min & \underline{\nsga} & \pesa & \textbf{0.0247} & (M) \ebar{0.6665}{0.1665} \\
15 min & \nsga & \spea & 0.5447 & (N) \ebar{0.5453}{0.0453} \\
15 min & \spea & \pesa & 0.0807 & (S) \ebar{0.6296}{0.1296} \\
30 min & \pesa & \underline{\nsga} & \textbf{0.0201} & (M) \ebar{0.3278}{0.1722} \\
30 min & \spea & \nsga & 0.3526 & (N) \ebar{0.5692}{0.0692} \\
30 min & \underline{\spea} & \pesa & \textbf{4.5e-04} & (L) \ebar{0.7596}{0.2596} \\
60 min & \underline{\nsga} & \pesa & \textbf{3.4e-05} & (L) \ebar{0.8070}{0.3070} \\
60 min & \nsga & \spea & 0.1084 & (S) \ebar{0.3809}{0.1191} \\
60 min & \underline{\spea} & \pesa & \textbf{3.8e-09} & (L) \ebar{0.9360}{0.4360} \\
\midrule
\multicolumn{6}{c}{\textbf{\ccm}} \\
\midrule
15 min & \underline{\nsga} & \pesa & \textbf{0.0102} & (M) \ebar{0.6904}{0.1904} \\
15 min & \spea & \nsga & 0.0693 & (S) \ebar{0.3652}{0.1348} \\
15 min & \spea & \pesa & 0.3007 & (S) \ebar{0.5770}{0.0770} \\
30 min & \nsga & \spea & 0.2127 & (S) \ebar{0.4074}{0.0926} \\
30 min & \pesa & \nsga & 0.0536 & (S) \ebar{0.3569}{0.1431} \\
30 min & \pesa & \underline{\spea} & \textbf{3.2e-04} & (L) \ebar{0.2336}{0.2664} \\
60 min & \pesa & \nsga & 0.1148 & (S) \ebar{0.3829}{0.1171} \\
60 min & \pesa & \underline{\spea} & \textbf{1.2e-04} & (L) \ebar{0.2149}{0.2851} \\
60 min & \spea & \nsga & 0.3074 & (S) \ebar{0.5760}{0.0760} \\

    \bottomrule
    \end{tabular}
    \caption{\mwu test and \vda effect sizes comparing the EP achieved by different algorithms in \independentRun runs. Magnitude interpretation: negligible (N), small (S), medium (M), large (L). The magnitude of the effect size is also represented by bars.}
    \label{tab:EP_test_algo}
    \end{table}

\subsubsection*{SPREAD}
\Cref{tab:SPREAD_test_algo} reports the hypothesis testing for the SPREAD values comparing pairs of algorithms for the two case studies.
The SPREAD values are statistically significant for both case studies.
For \ttbs case study, \nsga is dominated by \pesa and \spea for all time budgets, with large effect sizes.
Furthermore, \pesa outperforms \spea at $30$ and $60$ minutes with large effect sizes, while there is no significant difference between them for a $15$ minutes budget.
For the \ccm case study, \pesa outperforms \nsga and \spea for all time budgets, with large effect sizes.
Also, \spea outperforms \nsga for all time budgets, with large effect sizes.
From these results, we can conclude that \pesa can find Pareto frontiers with higher diversity. On the other hand, \spea solutions are more diverse than \nsga solutions.
\begin{table}[!ht]
\centering
\footnotesize
\begin{tabular}{llllll}
\toprule
Budget & Algor. 1 & Algor. 2 & MWU p & \multicolumn{2}{c}{Effect Size} \\

\midrule
\multicolumn{6}{c}{\textbf{\ttbs}} \\
\midrule
15 min & \nsga & \underline{\pesa} & \textbf{2.3e-08} & (L) \ebar{0.0864}{0.4136} \\
15 min & \underline{\spea} & \nsga & \textbf{1.5e-07} & (L) \ebar{0.8887}{0.3887} \\
15 min & \spea & \pesa & 0.0833 & (S) \ebar{0.3715}{0.1285} \\
30 min & \nsga & \underline{\pesa} & \textbf{1.9e-08} & (L) \ebar{0.0843}{0.4157} \\
30 min & \nsga & \underline{\spea} & \textbf{1.5e-04} & (L) \ebar{0.2196}{0.2804} \\
30 min & \underline{\pesa} & \spea & \textbf{1.8e-04} & (L) \ebar{0.7773}{0.2773} \\
60 min & \nsga & \underline{\pesa} & \textbf{9.4e-11} & (L) \ebar{0.0208}{0.4792} \\
60 min & \nsga & \underline{\spea} & \textbf{9.0e-07} & (L) \ebar{0.1363}{0.3637} \\
60 min & \spea & \underline{\pesa} & \textbf{5.1e-04} & (L) \ebar{0.2425}{0.2575} \\
\midrule
\multicolumn{6}{c}{\textbf{\ccm}} \\
\midrule
15 min & \nsga & \underline{\spea} & \textbf{2.4e-07} & (L) \ebar{0.1176}{0.3824} \\
15 min & \underline{\pesa} & \nsga & \textbf{1.9e-11} & (L) \ebar{0.9969}{0.4969} \\
15 min & \underline{\pesa} & \spea & \textbf{2.0e-09} & (L) \ebar{0.9438}{0.4438} \\
30 min & \nsga & \underline{\pesa} & \textbf{1.7e-11} & (L) \ebar{0.0021}{0.4979} \\
30 min & \underline{\spea} & \nsga & \textbf{1.8e-05} & (L) \ebar{0.8179}{0.3179} \\
30 min & \spea & \underline{\pesa} & \textbf{1.3e-09} & (L) \ebar{0.0510}{0.4490} \\
60 min & \nsga & \underline{\pesa} & \textbf{1.4e-11} & (L) \ebar{0.0000}{0.5000} \\
60 min & \underline{\spea} & \nsga & \textbf{1.3e-05} & (L) \ebar{0.8231}{0.3231} \\
60 min & \spea & \underline{\pesa} & \textbf{1.2e-10} & (L) \ebar{0.0239}{0.4761} \\

    \bottomrule
    \end{tabular}
    \caption{\mwu test and \vda effect sizes comparing the SPREAD achieved by different algorithms in \independentRun runs. Magnitude interpretation: negligible (N), small (S), medium (M), large (L). The magnitude of the effect size is also represented by bars.}
    \label{tab:SPREAD_test_algo}
    \end{table}
    
 \subsubsection*{Generational SPREAD (GSPREAD)}

\Cref{tab:GSPREAD_test_algo} reports the hypothesis testing for the GSPREAD values comparing pairs of algorithms for the two case studies.
For the \ttbs case study, \nsga outperforms the other two algorithms for all time budgets, with large effect sizes, while \pesa and \spea do not show significant differences.
For the \ccm case study, \spea is dominated by \nsga and \pesa for all time budgets, with medium and large effect sizes.
Furthermore, \nsga and \pesa do not show significant differences for all time budgets.
From our results,  we can conclude that \nsga found Pareto frontiers with the highest convergence, while \pesa solutions showed higher convergence in terms of GSPREAD than \spea ones. 
\begin{table}[!ht]
\centering
\footnotesize
\begin{tabular}{llllll}
\toprule
Budget & Algor. 1 & Algor. 2 & MWU p & \multicolumn{2}{c}{Effect Size} \\

\midrule
\multicolumn{6}{c}{\textbf{\ttbs}} \\
\midrule
15 min & \underline{\nsga} & \spea & \textbf{1.4e-04} & (L) \ebar{0.7825}{0.2825} \\
15 min & \pesa & \underline{\nsga} & \textbf{3.9e-04} & (L) \ebar{0.2373}{0.2627} \\
15 min & \pesa & \spea & 0.4062 & (N) \ebar{0.5619}{0.0619} \\
30 min & \underline{\nsga} & \pesa & \textbf{3.1e-05} & (L) \ebar{0.8085}{0.3085} \\
30 min & \underline{\nsga} & \spea & \textbf{2.6e-05} & (L) \ebar{0.8117}{0.3117} \\
30 min & \spea & \pesa & 0.6831 & (N) \ebar{0.4693}{0.0307} \\
60 min & \pesa & \underline{\nsga} & \textbf{5.8e-07} & (L) \ebar{0.1301}{0.3699} \\
60 min & \pesa & \spea & 0.9663 & (N) \ebar{0.5036}{0.0036} \\
60 min & \spea & \underline{\nsga} & \textbf{1.7e-06} & (L) \ebar{0.1457}{0.3543} \\
\midrule
\multicolumn{6}{c}{\textbf{\ccm}} \\
\midrule
15 min & \pesa & \nsga & 0.0715 & (S) \ebar{0.3663}{0.1337} \\
15 min & \spea & \underline{\nsga} & \textbf{5.7e-05} & (L) \ebar{0.2019}{0.2981} \\
15 min & \spea & \underline{\pesa} & \textbf{0.0021} & (M) \ebar{0.2726}{0.2274} \\
30 min & \nsga & \pesa & 0.1857 & (S) \ebar{0.5983}{0.0983} \\
30 min & \spea & \nsga & 0.1215 & (S) \ebar{0.3850}{0.1150} \\
30 min & \spea & \pesa & 0.7354 & (N) \ebar{0.4745}{0.0255} \\
60 min & \nsga & \pesa & 0.4223 & (N) \ebar{0.5598}{0.0598} \\
60 min & \underline{\nsga} & \spea & \textbf{0.0031} & (M) \ebar{0.7190}{0.2190} \\
60 min & \underline{\pesa} & \spea & \textbf{0.0043} & (M) \ebar{0.7118}{0.2118} \\

    \bottomrule
    \end{tabular}
    \caption{\mwu test and \vda effect sizes comparing the GSPREAD achieved by different algorithms in \independentRun runs. Magnitude interpretation: negligible (N), small (S), medium (M), large (L). The magnitude of the effect size is also represented by bars.}
    \label{tab:GSPREAD_test_algo}
    \end{table}

\bigskip

Our analysis strengthens the idea that the quality of Pareto frontiers cannot be measured with a single QI, and that the choice of the QI should be tailored to the specific problem at hand.
Hence, from our analysis, designers can choose the algorithm that best fits their needs, depending on the QI they are interested in.
For example, if the designer is interested in finding a diverse set of solutions, \pesa is the best choice, while if the designer is interested in inspecting the solution space quickly, \nsga is the recommended choice.

\subsubsection{RQ1.3: To what extent does the time budget affect the quality of Pareto fronts?}\label{sec:rq1:rq13}

In order to answer RQ1.3, we analyze the quality of the Pareto fronts computed by the algorithms under different time budgets through several quality indicators.
Intuitively, each indicator gives an idea of the impact of the search budget on specific properties of \computedP compared to the \referenceP.

To assess whether doubling or quadrupling the time budget makes a significant difference in the quality indicator value of the \computedP, we compare the results obtained with different budgets but with the same algorithm.
\Cref{tab:HV_test_time,tab:IGD_test_time,tab:IGD+_test_time,tab:EP_test_time,tab:SPREAD_test_time,tab:GSPREAD_test_time} show the results of the \mwu test and the corresponding \vda effect size.
The p-value is highlighted in bold when the detected difference is statistically significant.
The time budget is underlined when i) the test showed a significant difference, and ii) the experiment running on that time budget led to better quality indicator values.
\subsubsection*{Hypervolume (HV)}
Regarding the HV, the search budget had a different impact on the two case studies.
In \ttbs, the search was able to achieve a higher HV in all cases, when compared to that of \ccm.
This is probably due to the difference in size and complexity between the two case studies.
\ccm permits a larger number of possible refactoring candidates, and its model defines a more complex behavior.
Therefore, on average, the longer it takes to complete a single evolution, the fewer evolutions will be performed for a given time budget.

\Cref{tab:HV_test_time} shows the results of the \mwu test and the corresponding \vda effect size.
Only the \ttbs case study showed a statistical difference with a medium effect size of HV in two cases.
Furthermore, in both cases this trend was detected for \pesa.
This situation suggests that, except for \pesa, the main difference in the HV values might be attributed to a difference in the algorithm being used rather than to a difference in the budget.
We further investigate this aspect in the next section.
\begin{table}[t]
\centering
\footnotesize
\setlength{\tabcolsep}{3pt}
\begin{minipage}{.5\textwidth}
\begin{tabular}{llllrl}
\toprule
Algor & B 1 & B 2 & MWU p & \multicolumn{2}{c}{Effect Size} \\
        
\midrule
\multicolumn{6}{c}{\textbf{\ccm}} \\
\midrule
\nsga & 15  & 30  & 0.0574 & (S) \ebar{0.3590}{0.1410} \\
\nsga & 15  & 60  & 0.1054 & (S) \ebar{0.3798}{0.1202} \\
\nsga & 30  & 60  & 0.8769 & (N) \ebar{0.5120}{0.0120} \\
\pesa & 30  & 15  & 0.1721 & (S) \ebar{0.6015}{0.1015} \\
\pesa & 60  & 15  & 0.0592 & (S) \ebar{0.6400}{0.1400} \\
\pesa & 60  & 30  & 0.6024 & (N) \ebar{0.5390}{0.0390} \\
\spea & 15  & 60  & 0.1249 & (S) \ebar{0.6139}{0.1139} \\
\spea & 30  & 15  & 0.2483 & (S) \ebar{0.4142}{0.0858} \\
\spea & 30  & 60  & 0.6123 & (N) \ebar{0.5380}{0.0380} \\
\midrule
\multicolumn{6}{c}{\textbf{\ttbs}} \\
\midrule
\nsga & 15  & 30  & 0.1677 & (S) \ebar{0.3975}{0.1025} \\
\nsga & 60  & 15  & 0.1677 & (S) \ebar{0.6025}{0.1025} \\
\nsga & 60  & 30  & 0.9327 & (N) \ebar{0.4932}{0.0068} \\
\pesa & 15  & \underline{30} & \textbf{0.0180} & (M) \ebar{0.3247}{0.1753} \\
\pesa & 15  & \underline{60} & \textbf{0.0031} & (M) \ebar{0.2810}{0.2190} \\
\pesa & 30  & 60  & 0.4223 & (N) \ebar{0.4402}{0.0598} \\
\spea & 15  & 30  & 0.4556 & (N) \ebar{0.5557}{0.0557} \\
\spea & 60  & 15  & 0.4992 & (N) \ebar{0.4495}{0.0505} \\
\spea & 60  & 30  & 0.7999 & (N) \ebar{0.5193}{0.0193} \\

        \bottomrule
        \end{tabular}
        \caption{\mwu test and \vda effect sizes comparing the HV achieved with different time budgets in \independentRun runs. Magnitude interpretation: negligible (N), small (S), medium (M), large (L). The magnitude of the effect size is also represented by bars.}
        \label{tab:HV_test_time}
\end{minipage}
\hspace{1em}
\begin{minipage}{.5\textwidth}
\begin{tabular}{llllll}
\toprule
Algor & B 1 & B 2 & MWU p & \multicolumn{2}{c}{Effect Size} \\

\midrule
\multicolumn{6}{c}{\textbf{\ccm}} \\
\midrule
\nsga & 30  & 15  & 0.1181 & (S) \ebar{0.6160}{0.1160} \\
\nsga & 60  & 15  & 0.2660 & (S) \ebar{0.5827}{0.0827} \\
\nsga & 60  & 30  & 0.6728 & (N) \ebar{0.4683}{0.0317} \\
\pesa & 30  & 15  & 0.1765 & (S) \ebar{0.6004}{0.1004} \\
\pesa & 30  & 60  & 0.1431 & (S) \ebar{0.3913}{0.1087} \\
\pesa & \underline{60} & 15 & \textbf{0.0085} & (M) \ebar{0.6951}{0.1951} \\
\spea & 15  & 30 & 0.3905 & (N) \ebar{0.5640}{0.0640} \\
\spea & \underline{15} & 60 & \textbf{0.0194} & (M) \ebar{0.6733}{0.1733} \\
\spea & 60 & 30 & 0.0784 & (S) \ebar{0.3694}{0.1306} \\
\midrule
\multicolumn{6}{c}{\textbf{\ttbs}} \\
\midrule
\nsga & 30  & 15  & 0.8108 & (N) \ebar{0.5182}{0.0182} \\
\nsga & 60  & 15  & 0.1634 & (S) \ebar{0.6035}{0.1035} \\
\nsga & 60  & 30  & 0.2483 & (S) \ebar{0.5858}{0.0858} \\
\pesa & 15  & 30  & 0.3983 & (N) \ebar{0.4370}{0.0630} \\
\pesa & 60  & 15  & 0.0574 & (S) \ebar{0.6410}{0.1410} \\
\pesa & 60  & 30  & 0.2846 & (S) \ebar{0.5796}{0.0796} \\
\spea & 15  & 30  & 0.1765 & (S) \ebar{0.6004}{0.1004} \\
\spea & 15  & 60  & 0.1116 & (S) \ebar{0.6181}{0.1181} \\
\spea & 30  & 60  & 0.9103 & (N) \ebar{0.5088}{0.0088} \\

        \bottomrule
        \end{tabular}
        \caption{\mwu test and \vda effect sizes comparing the IGD achieved with different time budgets in \independentRun runs. Magnitude interpretation: negligible (N), small (S), medium (M), large (L). The magnitude of the effect size is also represented by bars.}
        \label{tab:IGD_test_time}
\end{minipage}
        \end{table}
        
 \subsubsection*{Inverse Generational Distance (IGD)}
\Cref{tab:IGD_test_time} shows the results of the \mwu test and the corresponding \vda effect size for the IGD indicator.
The results show that the time budget had a statistically significant impact, with a medium effect size, on the IGD values in two cases for the \ccm case study only for \pesa and \spea algorithms, respectively.
The two cases are different and interesting to analyze.
Looking at the \pesa algorithm, we can see that quadrupling the time budget led to a significant improvement in the IGD value. 
This result is not surprising and confirms the intuition that a longer time budget allows the algorithm to explore more of the solution space.

On the other hand, quadrupling the time budget for the \spea algorithm led to a significant decrease in the IGD value.
In this case, the algorithm was unable to exploit the additional time to improve the quality of the Pareto front.
There are several possible reasons for this behavior, such as the algorithm's inability to explore the solution space effectively or the presence of a local optimum that the algorithm was unable to escape from.
We remark that we repeated each experiment $30$ times to reduce the probability of being stuck in a local optimum.

\subsubsection*{Inverse Generational Distance plus (IGD+)}

\Cref{tab:IGD+_test_time} shows the results of the \mwu test and the corresponding \vda effect size for the IGD+ indicator.
The results show that the time budget had a statistically significant impact for \pesa algorithm, with a medium effect size in two cases for the \ccm case study and in one case with a small effect size for the \ttbs case study.
Regarding the \ccm case study, we can see that only quadrupling the time budget led to a significant improvement in the IGD+ value.

On the other hand, looking at the \ttbs case study, we can see that doubling and quadrupling the time budget led to a significant improvement of the quality indicator value.
A marginal consideration is that \nsga and \spea did not show any significant difference in the IGD+ values when changing the time budget.
This suggests that the algorithms could not exploit the additional time to improve the quality of the Pareto front, due to several reasons. 

We suspect that the \nsga and \spea algorithms in our setting stuck at a local optimum or that they were unable to explore the solution space effectively.
Such a behavior could be due to the properties of the two case studies or the specific algorithm configurations. 
To confirm this hypothesis, we will further investigate this aspect in future work.

\begin{table}[t]
\centering
\footnotesize
\setlength{\tabcolsep}{3.5pt}
\begin{minipage}{.5\textwidth}
\begin{tabular}{llllll}
\toprule
Algor & B 1 & B 2 & MWU p & \multicolumn{2}{c}{Effect Size} \\
\midrule
\multicolumn{6}{c}{\textbf{\ccm}} \\
\midrule
\nsga & 15  & 30  & 0.0672 & (S) \ebar{0.3642}{0.1358} \\
\nsga & 60  & 15  & 0.0761 & (S) \ebar{0.6316}{0.1316} \\
\nsga & 60  & 30  & 0.8769 & (N) \ebar{0.5120}{0.0120} \\
\pesa & 30  & 15  & 0.3107 & (S) \ebar{0.5754}{0.0754} \\
\pesa & 30  & 60  & 0.1857 & (S) \ebar{0.4017}{0.0983} \\
\pesa & \underline{60} & 15 & \textbf{0.0291} & (S) \ebar{0.6618}{0.1618} \\
\spea & 30  & 15  & 0.2783 & (S) \ebar{0.4194}{0.0806} \\
\spea & 60  & 15  & 0.0784 & (S) \ebar{0.3694}{0.1306} \\
\spea & 60  & 30  & 0.2483 & (S) \ebar{0.4142}{0.0858} \\
\midrule
\multicolumn{6}{c}{\textbf{\ttbs}} \\
\midrule
\nsga & 15  & 30  & 0.0651 & (S) \ebar{0.3632}{0.1368} \\
\nsga & 15  & 60  & 0.2154 & (S) \ebar{0.4079}{0.0921} \\
\nsga & 60  & 30  & 0.5830 & (N) \ebar{0.4589}{0.0411} \\
\pesa & 15  & \underline{30} & \textbf{0.0060} & (M) \ebar{0.2966}{0.2034} \\
\pesa & 15  & \underline{60} & \textbf{0.0028} & (M) \ebar{0.2789}{0.2211} \\
\pesa & 30  & 60  & 0.4815 & (N) \ebar{0.4475}{0.0525} \\
\spea & 15  & 30  & 0.6831 & (N) \ebar{0.5307}{0.0307} \\
\spea & 15  & 60  & 0.5449 & (N) \ebar{0.5453}{0.0453} \\
\spea & 60  & 30 & 0.9775 & (N) \ebar{0.4974}{0.0026} \\
\bottomrule
\end{tabular}
\caption{\mwu test and \vda effect sizes comparing the IGD+ achieved with different time budgets in \independentRun runs. Magnitude interpretation: negligible (N), small (S), medium (M), large (L). The magnitude of the effect size is also represented by bars.}
\label{tab:IGD+_test_time}
\end{minipage}
\hspace{1em}
\begin{minipage}{.5\textwidth}
\begin{tabular}{llllll}
\toprule
Algor & $B_1$ & $B_2$ & MWU p & \multicolumn{2}{c}{Effect Size} \\
\midrule
\multicolumn{6}{c}{\textbf{\ccm}} \\
\midrule
\nsga & 15  & 30  & 0.5172 & (N) \ebar{0.5484}{0.0484} \\
\nsga & 15  & 60  & 0.2127 & (S) \ebar{0.5926}{0.0926} \\
\nsga & 30  & 60  & 0.4991 & (N) \ebar{0.5505}{0.0505} \\
\pesa & 30  & 15  & 0.8108 & (N) \ebar{0.4818}{0.0182} \\
\pesa & 30  & 60  & 0.2939 & (S) \ebar{0.5780}{0.0780} \\
\pesa & 60  & 15  & 0.1976 & (S) \ebar{0.4043}{0.0957} \\
\spea & 15  & \underline{30} & \textbf{0.0132} & (M) \ebar{0.3163}{0.1837} \\
\spea & 15  & \underline{60} & \textbf{0.0412} & (S) \ebar{0.3486}{0.1514} \\
\spea & 60  & 30 & 0.8713 & (N) \ebar{0.5125}{0.0125} \\
\midrule
\multicolumn{6}{c}{\textbf{\ttbs}} \\
\midrule
\nsga & 15  & 30  & 0.2599 & (S) \ebar{0.5838}{0.0838} \\
\nsga & 60  & 15  & 0.4101 & (N) \ebar{0.4386}{0.0614} \\
\nsga & 60  & 30  & 0.8713 & (N) \ebar{0.5125}{0.0125} \\
\pesa & 30  & 15  & 0.0640 & (S) \ebar{0.3626}{0.1374} \\
\pesa & 60  & \underline{15} & \textbf{1.2e-04} & (L) \ebar{0.2149}{0.2851} \\
\pesa & 60  & 30  & 0.0844 & (S) \ebar{0.3720}{0.1280} \\
\spea & 30  & 15  & 0.6984 & (N) \ebar{0.5291}{0.0291} \\
\spea & 30  & 60  & 0.1856 & (S) \ebar{0.4017}{0.0983} \\
\spea & 60  & 15  & 0.0691 & (S) \ebar{0.6348}{0.1348} \\
\bottomrule
\end{tabular}
\caption{\mwu test and \vda effect sizes comparing the EP achieved with different time budgets in \independentRun runs. Magnitude interpretation: negligible (N), small (S), medium (M), large (L). The magnitude of the effect size is also represented by bars.}
\label{tab:EP_test_time}
\end{minipage}
\end{table}
        
 \subsubsection*{Epsilon}
\Cref{tab:EP_test_time} lists the results of the \mwu test and the corresponding \vda effect size for the Epsilon quality indicator.
At a glance there are limited differences in the Epsilon values when changing the time budget.
Looking at the \ccm case study, we can see that only the \spea algorithm showed a significant difference in the Epsilon values when doubling or quadrupling the time budget.
This suggests that \nsga and \pesa were not able to exploit the additional time to improve the quality of the Pareto front in our setting for the \ccm case study.

On the other hand, looking at the \ttbs case study, only \pesa algorithm showed a significant difference in the Epsilon value. 
Furthermore, the statistical test proved that quadrupling the time budget led to a significant detriment in the Epsilon value.
This suggests that \pesa was not able to exploit the additional time to improve the quality of the Pareto front in our setting for the \ttbs case study, for example, due to the presence of a local optimum that the algorithm was unable to escape from or the structure of the problem.
To generalize these results, we need to further investigate this aspect in future work by considering more case studies of different sizes and complexities.

\subsubsection*{SPREAD}
\Cref{tab:SPREAD_test_time} shows the results of the \mwu test and the corresponding \vda effect size for the SPREAD quality indicator.
Again, the results show a different impact of the time budget on the two case studies.
In the \ccm case study, the time budget did not have a significant impact on the SPREAD values for the \pesa algorithm, which means that the algorithm cannot find varied solutions in the \computedP when the time budget is increased.
On the other hand, a shorter time budget led to statistically significant better SPREAD values for both \nsga and \spea algorithms. For example, \spea found more varied solutions in the \computedP when the time budget was set to $15$ and $30$ minutes against $60$ minutes.

Looking at the \ttbs case study, we can see that only \spea showed a significant difference in the SPREAD values with the $15$ minutes time budget against $30$ minutes. 
This suggests that the structure of the case study hindered the search process of all algorithms, which were unable to find varied solutions in the \computedP when the time budget was increased.

\subsubsection*{GSPREAD}
\begin{table}[t]
\centering
\footnotesize
\setlength{\tabcolsep}{3.5pt}
\begin{minipage}{.5\textwidth}
\begin{tabular}{llllll}
\toprule
Algor & B 1 & B 2 & MWU p & \multicolumn{2}{c}{Effect Size} \\
\midrule
\multicolumn{6}{c}{\textbf{\ccm}} \\
\midrule
\nsga & 15  & 30  & 0.3983 & (N) \ebar{0.5630}{0.0630} \\
\nsga & 60  & \underline{15} & \textbf{0.0347} & (S) \ebar{0.3434}{0.1566} \\
\nsga & 60  & 30  & 0.2206 & (S) \ebar{0.4089}{0.0911} \\
\pesa & 15  & 30  & 0.2154 & (S) \ebar{0.5921}{0.0921} \\
\pesa & 15  & 60  & 0.7143 & (N) \ebar{0.5276}{0.0276} \\
\pesa & 30  & 60  & 0.3751 & (N) \ebar{0.4339}{0.0661} \\
\spea & 30  & 15  & 0.2314 & (S) \ebar{0.4110}{0.0890} \\
\spea & \underline{30} & 60 & \textbf{0.0173} & (M) \ebar{0.6764}{0.1764} \\
\spea & 60  & \underline{15} & \textbf{3.9e-05} & (L) \ebar{0.1956}{0.3044} \\
\midrule
\multicolumn{6}{c}{\textbf{\ttbs}} \\
\midrule
\nsga & 30  & 15  & 0.9663 & (N) \ebar{0.4964}{0.0036} \\
\nsga & 30  & 60  & 0.8880 & (N) \ebar{0.4891}{0.0109} \\
\nsga & 60  & 15  & 0.8327 & (N) \ebar{0.5161}{0.0161} \\
\pesa & 30  & 15  & 0.5356 & (N) \ebar{0.5463}{0.0463} \\
\pesa & 60  & 15  & 0.1320 & (S) \ebar{0.6119}{0.1119} \\
\pesa & 60  & 30  & 0.3827 & (N) \ebar{0.5650}{0.0650} \\
\spea & \underline{15} & 30  & \textbf{0.0226} & (M) \ebar{0.6691}{0.1691} \\
\spea & 15  & 60  & 0.6728 & (N) \ebar{0.5317}{0.0317} \\
\spea & 60  & 30  & 0.0911 & (S) \ebar{0.6254}{0.1254} \\
\bottomrule
\end{tabular}
\caption{\mwu test and \vda effect sizes comparing the SPREAD achieved with different time budgets in \independentRun runs. Magnitude interpretation: negligible (N), small (S), medium (M), large (L). The magnitude of the effect size is also represented by bars.}
\label{tab:SPREAD_test_time}
\end{minipage}
\hspace{1em}
\begin{minipage}{.5\textwidth}
\begin{tabular}{llllll}
\toprule
Algor & B 1 & B 2 & MWU p & \multicolumn{2}{c}{Effect Size} \\
\midrule
\multicolumn{6}{c}{\textbf{\ccm}} \\
\midrule
\nsga & 30  & 15  & 0.1356 & (S) \ebar{0.3892}{0.1108} \\
\nsga & 60  & 15  & 0.1284 & (S) \ebar{0.3871}{0.1129} \\
\nsga & 60  & 30  & 1.0000 & (N) \ebar{0.5005}{0.0005} \\
\pesa & \underline{15} & 30 & \textbf{0.0372} & (S) \ebar{0.6545}{0.1545} \\
\pesa & 60  & 15  & 0.1249 & (S) \ebar{0.3861}{0.1139} \\
\pesa & 60  & 30  & 0.8108 & (N) \ebar{0.5182}{0.0182} \\
\spea & 30  & 15  & 0.2660 & (S) \ebar{0.5827}{0.0827} \\
\spea & 30  & 60  & 0.0555 & (S) \ebar{0.6420}{0.1420} \\
\spea & 60  & 15  & 0.2783 & (S) \ebar{0.4194}{0.0806} \\
\midrule
\multicolumn{6}{c}{\textbf{\ttbs}} \\
\midrule
\nsga & 15  & 60  & 0.8437 & (N) \ebar{0.4849}{0.0151} \\
\nsga & 30  & 15  & 0.6934 & (N) \ebar{0.4703}{0.0297} \\
\nsga & 30  & 60  & 0.4815 & (N) \ebar{0.4475}{0.0525} \\
\pesa & 15  & 30  & 0.3751 & (N) \ebar{0.5661}{0.0661} \\
\pesa & 60  & 15  & 0.3751 & (N) \ebar{0.4339}{0.0661} \\
\pesa & 60  & 30  & 0.7675 & (N) \ebar{0.4776}{0.0224} \\
\spea & 15  & 60  & 0.9327 & (N) \ebar{0.4932}{0.0068} \\
\spea & 30  & 15  & 0.7568 & (N) \ebar{0.4766}{0.0234} \\
\spea & 30  & 60  & 0.7783 & (N) \ebar{0.4787}{0.0213} \\
\bottomrule
\end{tabular}
\caption{\mwu test and \vda effect sizes comparing the GSPREAD achieved with different time budgets in \independentRun runs. Magnitude interpretation: negligible (N), small (S), medium (M), large (L). The magnitude of the effect size is also represented by bars.}
\label{tab:GSPREAD_test_time}
\end{minipage}
\end{table}
        
 \Cref{tab:GSPREAD_test_time} shows the results of the \mwu test and the corresponding \vda effect size for the GSPREAD quality indicator.
The results for both case studies showed that the time budget did not have a statistically significant impact on the GSPREAD values.
Therefore, the three studied algorithms were not able to find varied solutions in the \computedP with respect to the \referenceP when the time budget was increased.
In only one case \pesa showed a statistically difference with a small effect size in the \ccm case study when comparing the $15$ minutes time budget against $30$ minutes.
\bigskip 

\begin{rqbox}
In summary, we can answer \textbf{RQ1} by saying that \spea was the slowest algorithm in our experiments, whereas \nsga was the fastest one, and \pesa was the algorithm that benefits from longer budgets.

In addition, our results suggest that the effectiveness of time budgets and algorithmic behavior may depend on the complexity of the system under analysis, which, in our context, refers to factors such as the number of software components, exposed interfaces, interactions, and use cases. While we do not explicitly quantify these aspects in our study, the differing behaviors observed between \ttbs and \ccm hint at their potential influence.
Predicting the impact of the time budget on the quality of the \computedP without considering the specific characteristics of the system under analysis and the specific quality indicator is challenging.

\end{rqbox}  \subsection{RQ2: How does varying the time budget influence quality and design of software models generated by search-based refactoring algorithms?}\label{sec:rq2}

In order to answer RQ2, we divided the analysis of the influence of search budgets into two parts: non-functional properties (\ie \perfq and \reliability), and structural design properties (\ie \pas and \achanges). We separately compared the non-functional and design properties of \computedP by each algorithm.

\subsubsection*{RQ2.1: Do different time budgets significantly affect performance and reliability of the software models produced by search-based refactoring algorithms?}

\begin{figure*}
        \begin{revfloatenv}
   \begin{subfigure}{\dimexpr0.28\textwidth+20pt\relax}\centering
     \makebox[20pt]{\raisebox{40pt}{\rotatebox[origin=l]{90}{\ttbs}}}\includegraphics[width=\dimexpr\linewidth-23pt\relax]{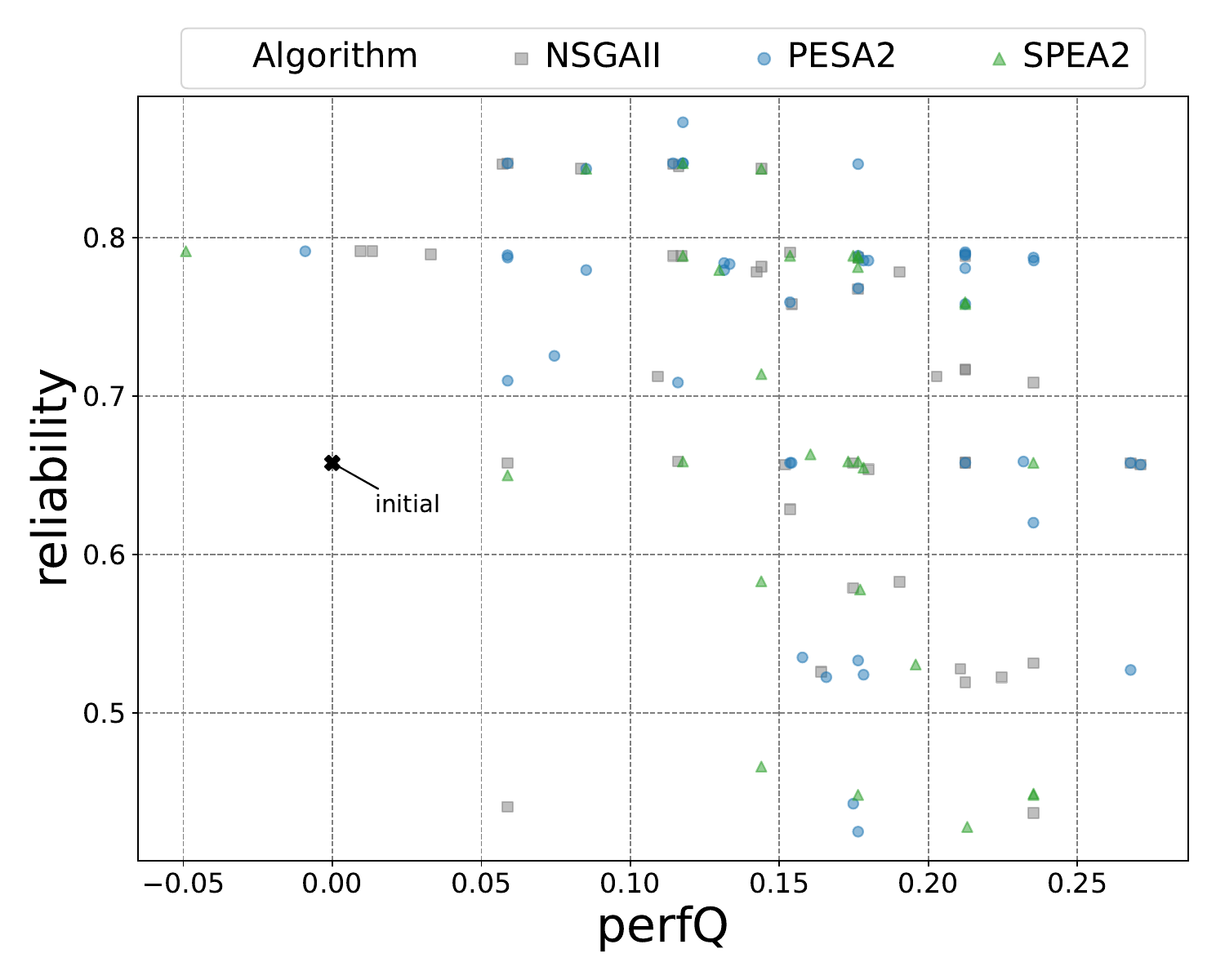}
     \makebox[20pt]{\raisebox{40pt}{\rotatebox[origin=l]{90}{\ccm}}}\includegraphics[width=\dimexpr\linewidth-23pt\relax]{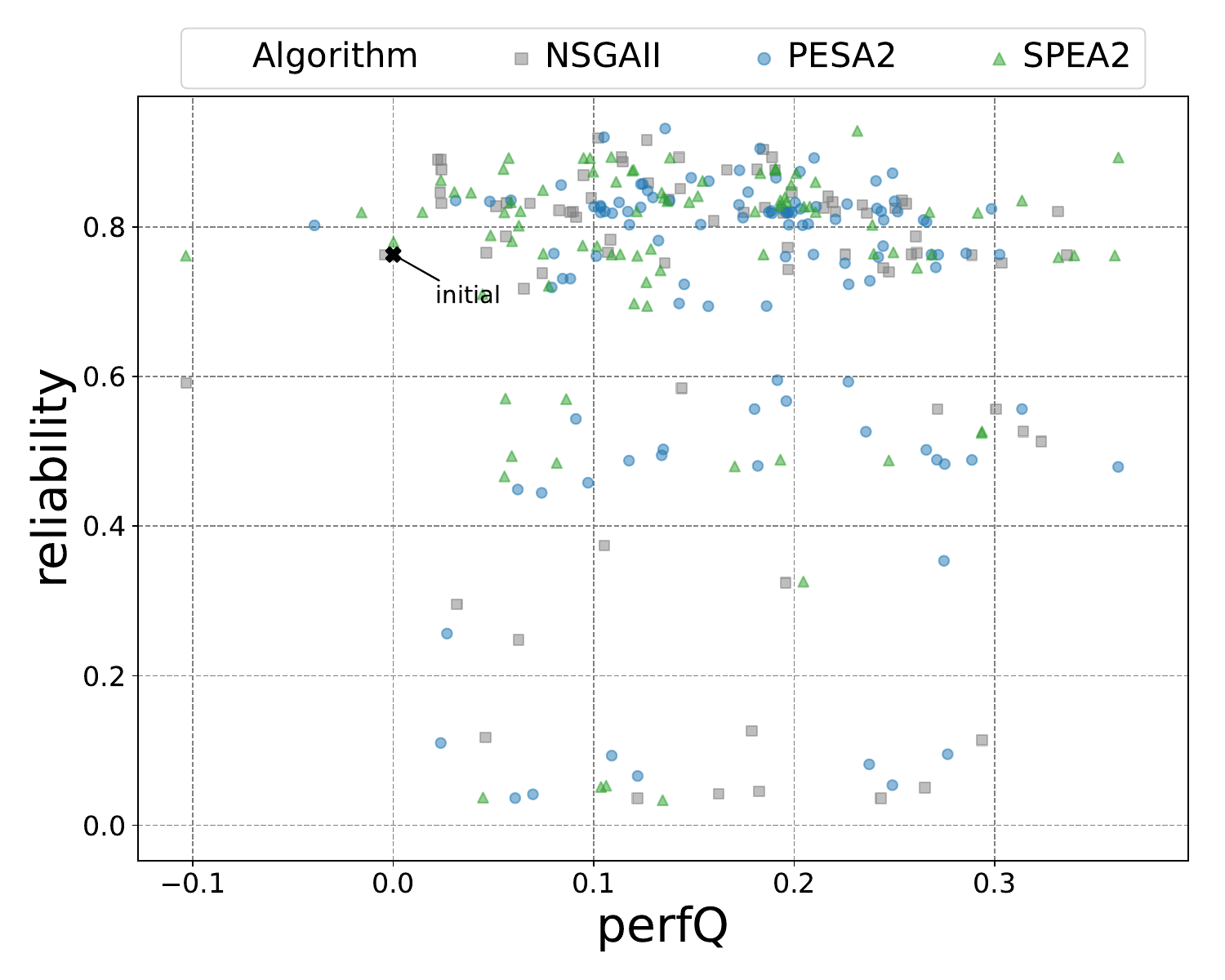}
     \caption{15 min budget}
     \label{fig:byalgo_15_min}
   \end{subfigure}\hfill \begin{subfigure}{\dimexpr0.28\textwidth+17pt\relax}\centering
     \includegraphics[width=\dimexpr\linewidth-20pt\relax]{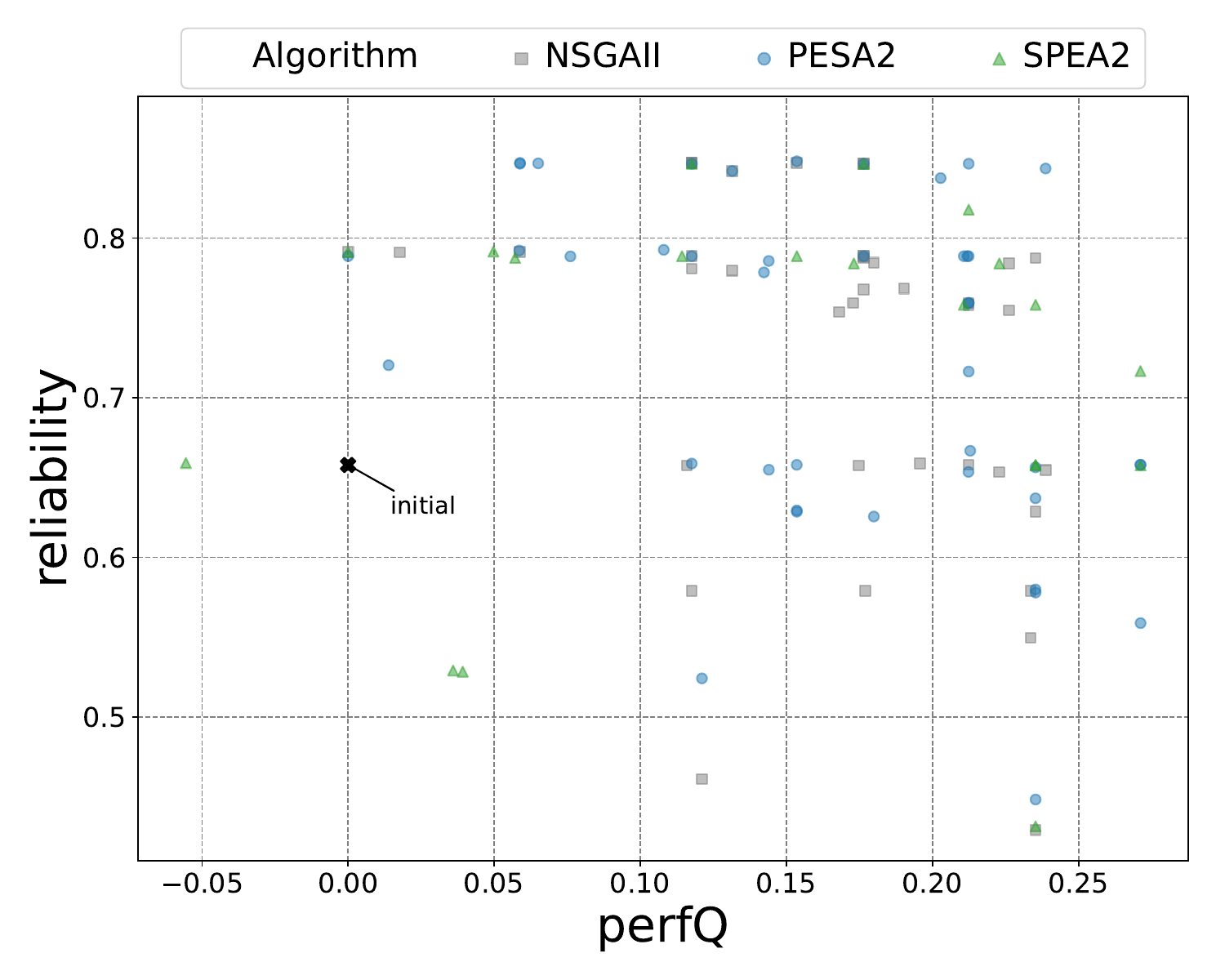}
     \includegraphics[width=\dimexpr\linewidth-20pt\relax]{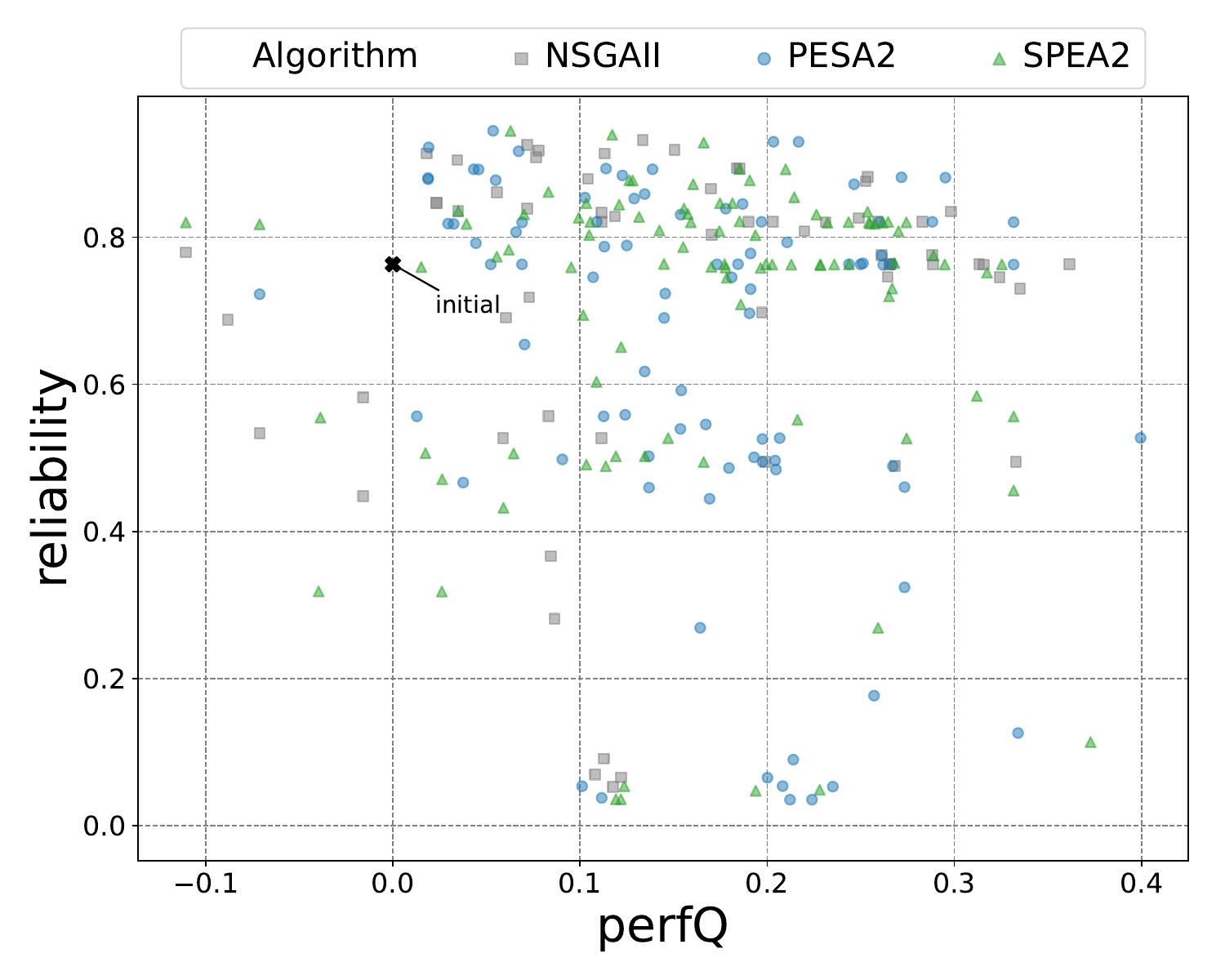}
     \caption{30 min budget}
     \label{fig:byalgo_30_min}
   \end{subfigure}\hfill \begin{subfigure}{\dimexpr0.28\textwidth+17pt\relax}\centering
     \includegraphics[width=\dimexpr\linewidth-20pt\relax]{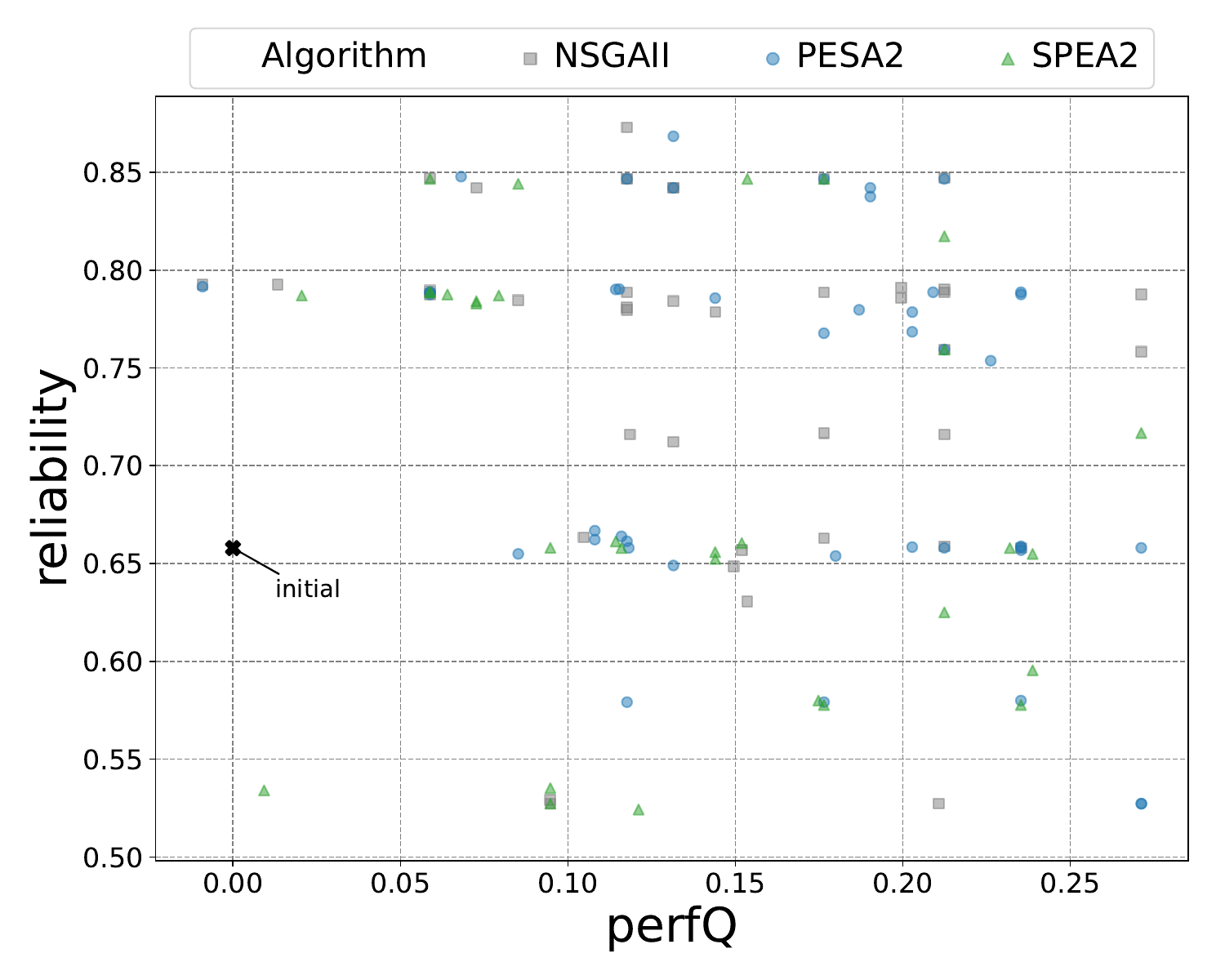}
     \includegraphics[width=\dimexpr\linewidth-20pt\relax]{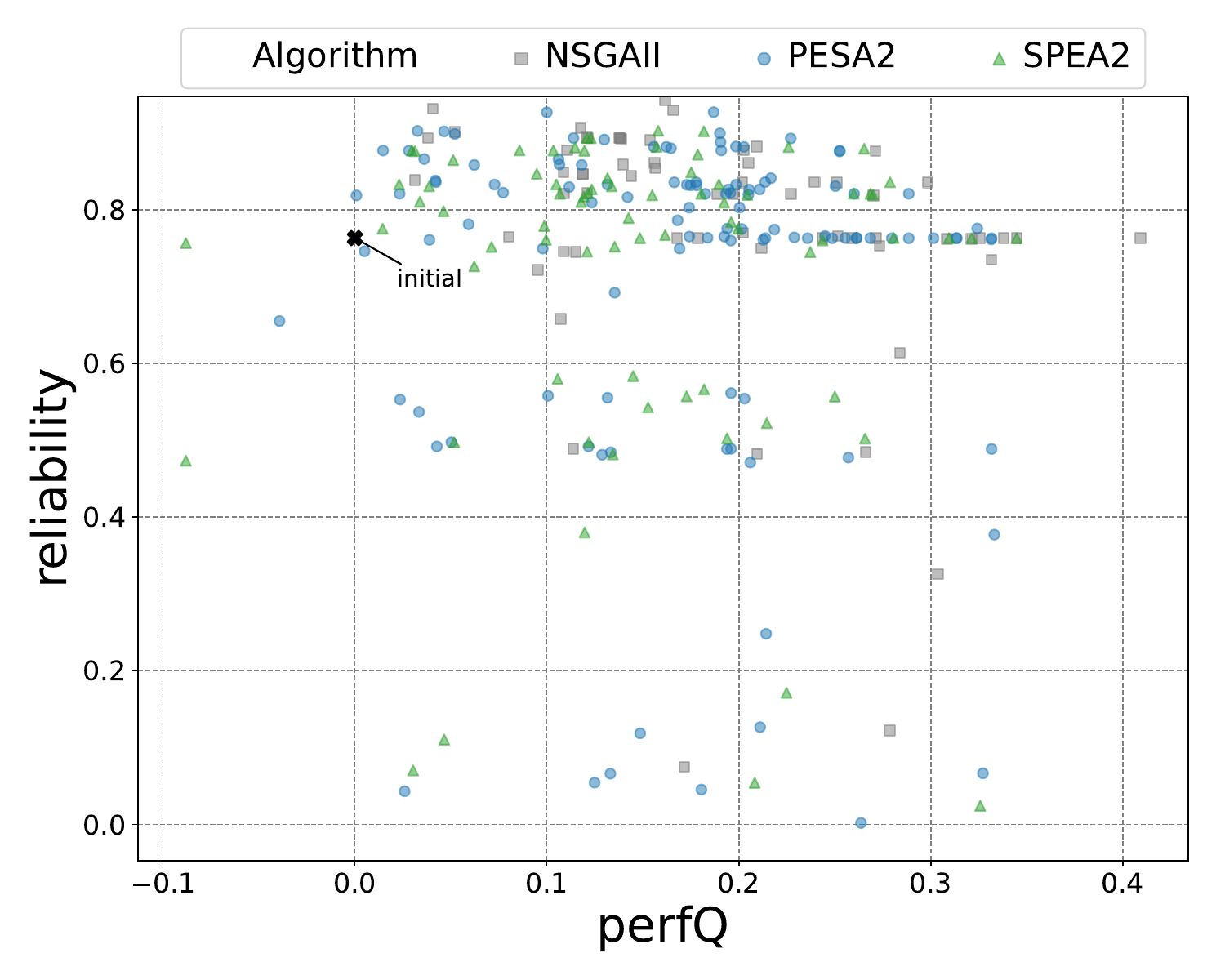}
     \caption{60 min budget}
     \label{fig:byalgo_60_min}
   \end{subfigure}\hfill \caption{\ttbs and \ccm Pareto frontiers for \perfq and \reliability obtained by the three algorithms when varying the time budget between 15, 30, and 60 minutes. The top-right corner is the optimal point, whereas the bottom-left corner is the worst one. Filled symbols correspond to the results of each algorithm \label{fig:byalgo_scatter}}
  \end{revfloatenv}
\end{figure*}

To visually assess differences among the algorithms in the quality (non-functional properties.) of the computed solutions, we relied on scatter plots comparing \perfq and \reliability. Along this line, \Cref{fig:byalgo_scatter} depicts the three \computedP when varying the time budget of all three genetic algorithms for both case studies.
At a glance, we can observe a more densely populated \computedP for \ccm than for \ttbs, while \ttbs showed a more evident trend towards the top-right corner (the objective optimization direction).
Regarding the \computedP for \ccm, a horizontal clustering (with three bands) was observed for the three search budgets.
The cluster that lies around $0.8$ for \reliability is always more populated than the other two clusters: one between $0.4$ and $0.6$, and the other between $0.0$ and $0.2$, approximately. Also, the \ccm solution space seems to be less homogeneous than \ttbs. There is no evident motivation for the clustering pattern of \ccm. 
We conjecture that the characteristics of the \ccm model, which has a more complex behavior than \ttbs (in terms of numbers of components and interfaces), affect the behavior of the algorithms.

\subsubsection*{RQ2.2: Do different time budgets significantly affect design properties of the
software models produced by search-based refactoring algorithms?}

\begin{figure*}
\begin{revfloatenv}
   \begin{subfigure}{\dimexpr0.28\textwidth+20pt\relax}\centering
     \makebox[20pt]{\raisebox{40pt}{\rotatebox[origin=l]{90}{\ttbs}}}\includegraphics[width=\dimexpr\linewidth-23pt\relax]{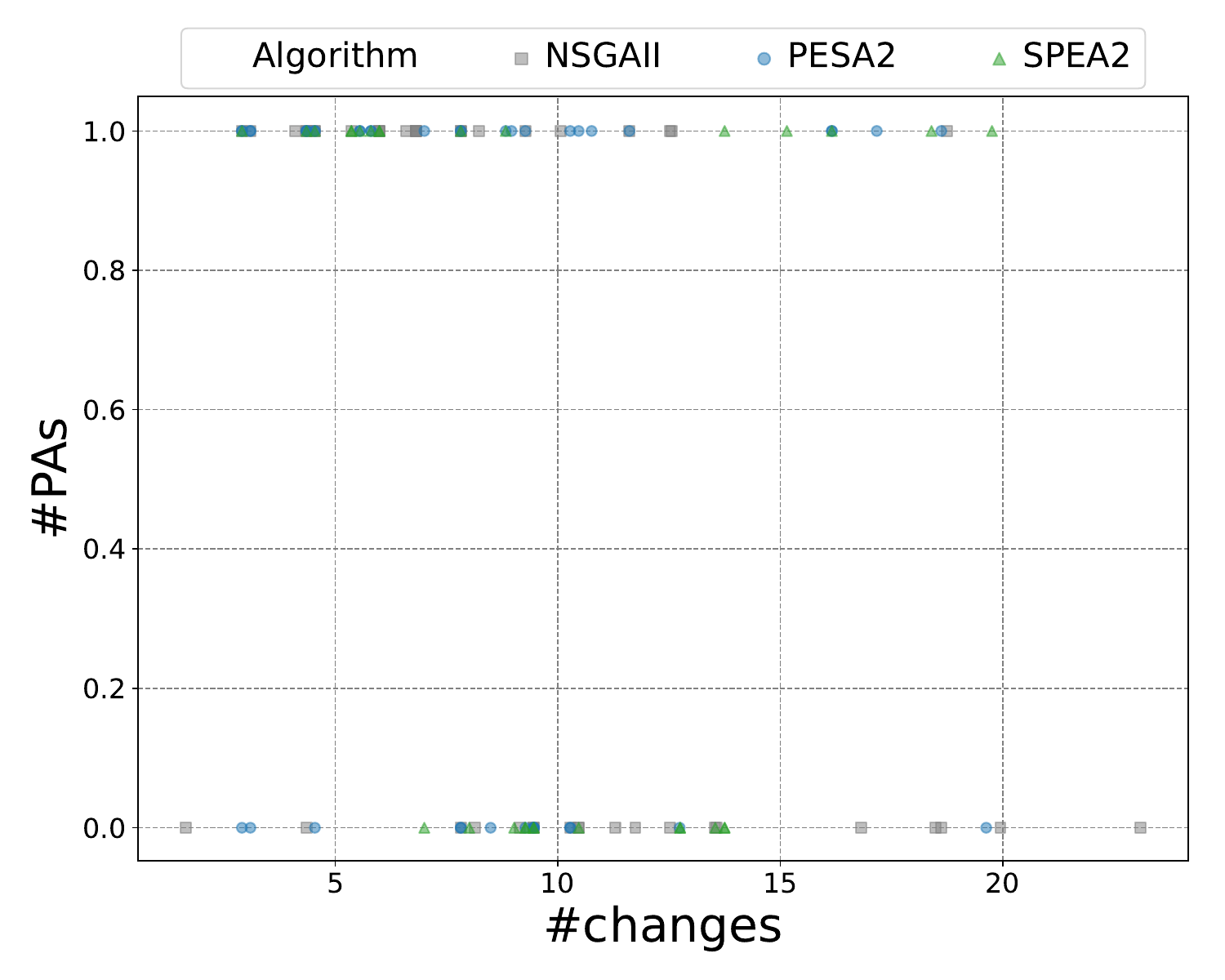}
     \makebox[20pt]{\raisebox{40pt}{\rotatebox[origin=l]{90}{\ccm}}}\includegraphics[width=\dimexpr\linewidth-23pt\relax]{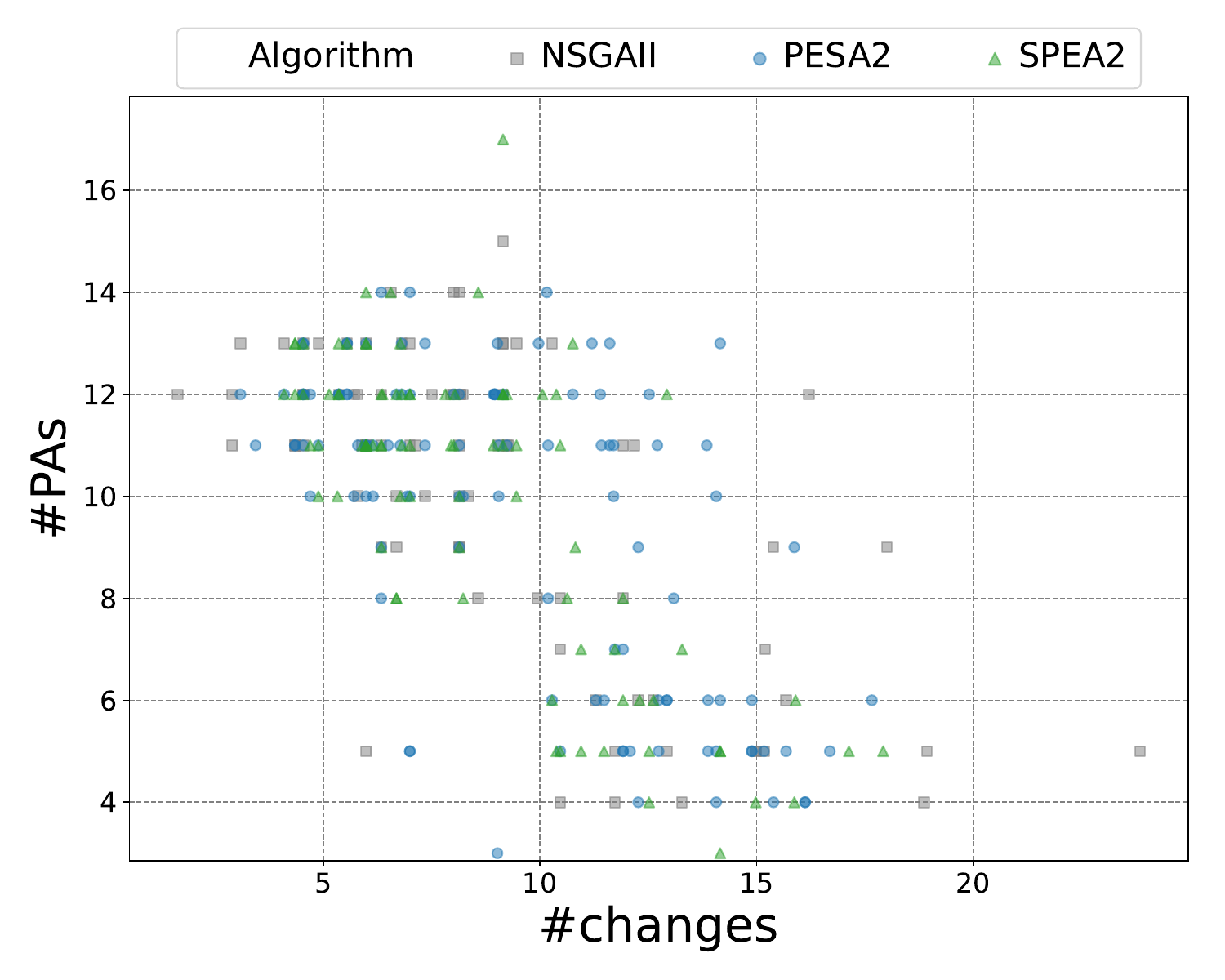}
     \caption{15 min budget}
     \label{fig:byalgo_15_min_changespas}
   \end{subfigure}\hfill \begin{subfigure}{\dimexpr0.28\textwidth+17pt\relax}\centering
     \includegraphics[width=\dimexpr\linewidth-20pt\relax]{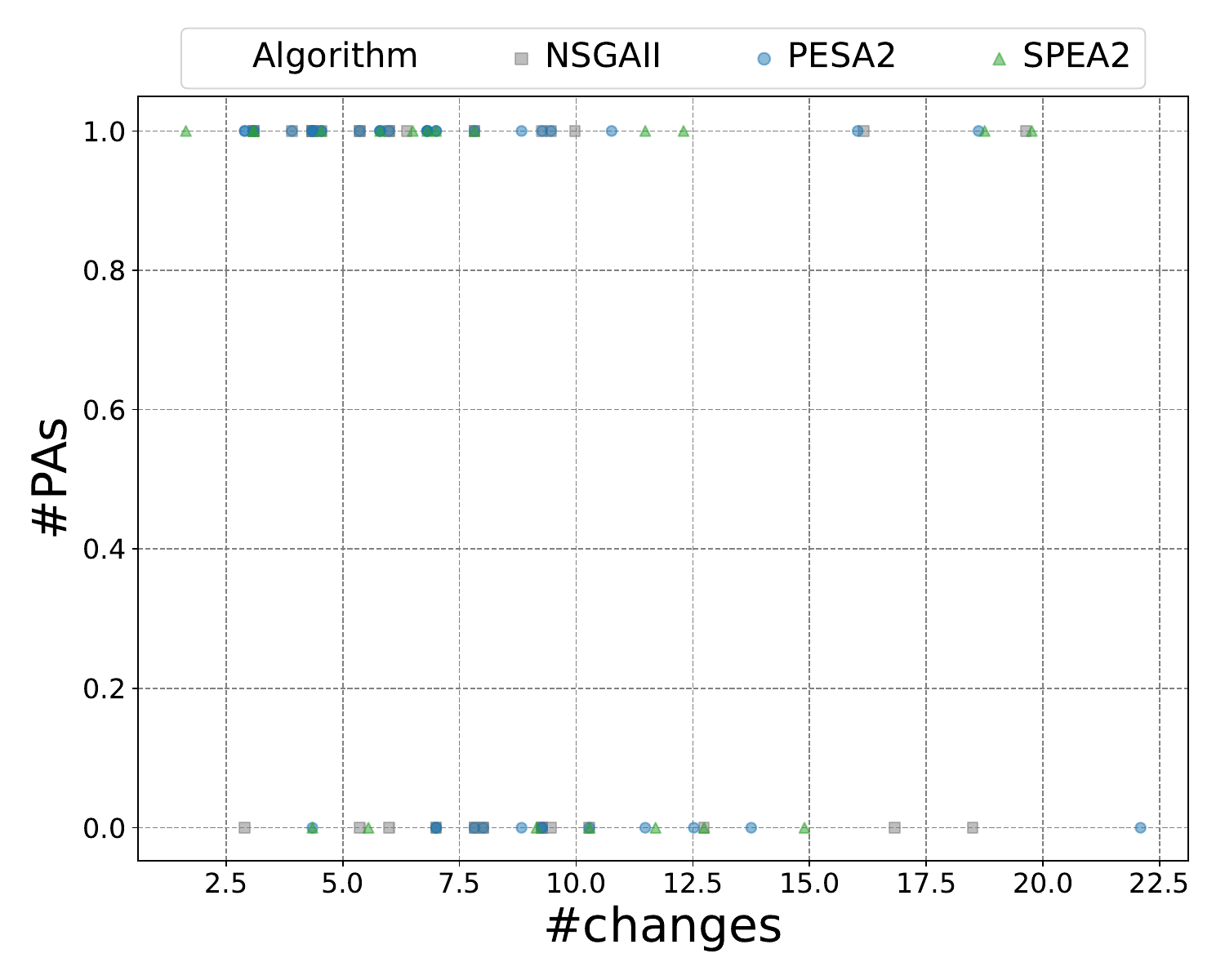}
     \includegraphics[width=\dimexpr\linewidth-20pt\relax]{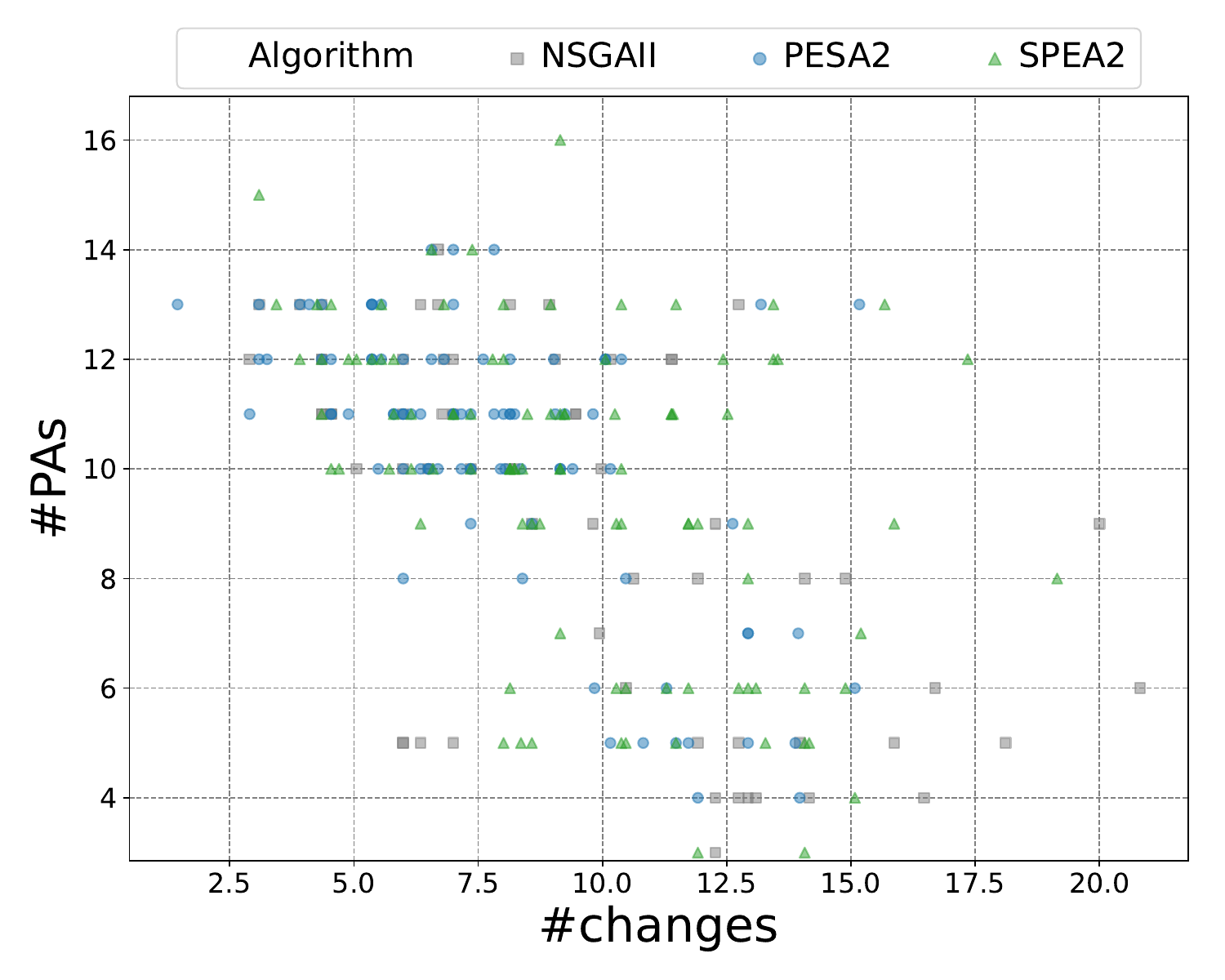}
     \caption{30 min budget}
     \label{fig:byalgo_30_min_changespas}
   \end{subfigure}\hfill \begin{subfigure}{\dimexpr0.28\textwidth+17pt\relax}\centering
     \includegraphics[width=\dimexpr\linewidth-20pt\relax]{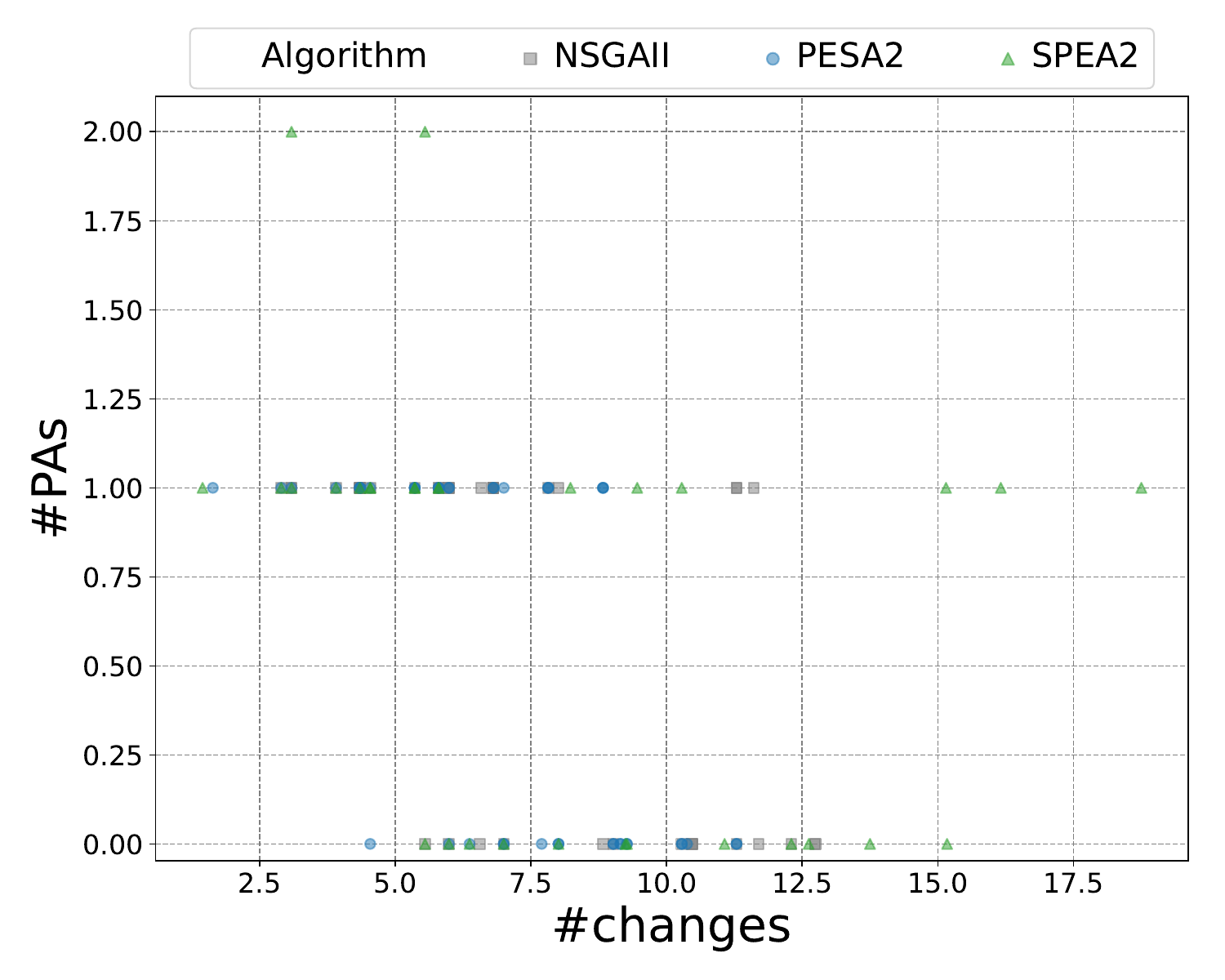}
     \includegraphics[width=\dimexpr\linewidth-20pt\relax]{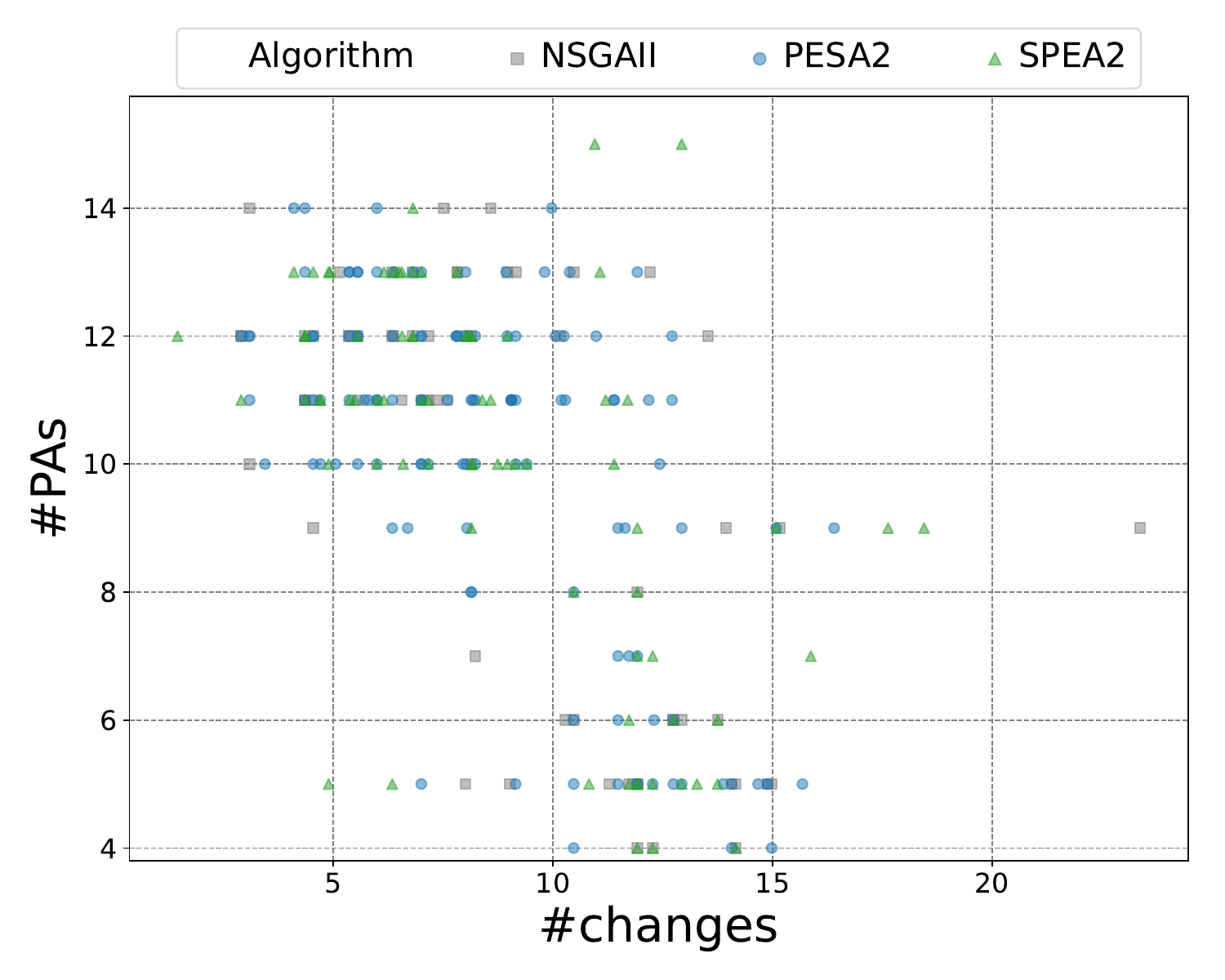}
     \caption{60 min budget}
     \label{fig:byalgo_60_min_changespas}
   \end{subfigure}\hfill \caption{\ttbs and \ccm Pareto frontiers for \achanges and \pas obtained by the three algorithms when varying the time budget between 15, 30, and 60 minutes. The bottom-left corner is the optimal point, whereas the top-right corner is the worst one. Filled symbols correspond to the results of each algorithm. \label{fig:byalgo_scatter_changespas}}
\end{revfloatenv}
\end{figure*}

Like we did for the non-functional properties, we initially generated scatter plots for \achanges and \pas objectives to assess the kinds of software models resulting from the budgets, as depicted in \Cref{fig:byalgo_scatter_changespas}.
We observed that the solutions were confined to compact, well-defined regions of the design space, in contrast to the variety of solutions offered by the reference Pareto front (\referenceP) in the previous section. 
In both case studies, two main clusters of solutions were identified. 
The clusters were very clear (like stripes) in \ttbs, with the majority of the models having at most one antipattern and their refactoring costs in a mid-range ($[3-20]$). 
For \ccm, the clusters resulting from the three budgets shared the same boundaries. 
The refactoring cost was around the same range as for \ttbs, but the number of antipatterns covered had more variability ($[2-15]$). We think the complexity of the \ccm models must have played a role in the higher number of detected antipatterns.

The patterns for the clusters were similar, regardless of the algorithm being used. 
Although there were slight differences in the \ccm results, increasing the time budget did not affect the general cluster patterns. 
This means that even when imposing a time budget, the designer has chances of finding a number of (Pareto) optimal solutions for the refactoring problem. 
Certainly, the corresponding (alternative) models will be fewer (in terms of \achanges and \pas) than those obtainable when running the algorithms without budgets.

It should be noticed that \achanges and \pas provide a limited characterization of the underlying models, as other structural properties of the models are not captured. 
For example, two models having one antipattern and a refactoring cost of $10$ might still differ in their design structure. 
Thus, a finer-grained characterization of the models can help to expose additional differences.

\begin{rqbox}
In summary, we can answer \textbf{RQ2} by saying that, if we only look at the non-functional properties, the shape of the \computedP and the explored design space do not differ much from those covered by the \referenceP, regardless of the algorithm. A factor affecting the shape of the design space seems to be the model complexity.
Furthermore, using time budgets leads to a restricted set of model alternatives, but some of them fall into the \referenceP. 
\end{rqbox}
 \subsection{RQ3: How do the sequences of refactoring actions look like when using different budgets?}\label{sec:rq3}
 
From a constructive (or structural) perspective, the software models are obtained by applying (sequences of) refactoring actions on an initial model. 
Altogether, these actions constitute the search space explored by a given algorithm. 
In this context, we can take all the sequences used in a given experiment and arrange them as a \textit{prefix tree}, in which the leaves correspond to models and the inner nodes capture actions shared by the different sequences. 
This tree representation is useful for identifying unique sequences in a given search space, but also for computing sequence intersections between the trees coming from different algorithms or budgets. 

\begin{figure*}[t]
    \centering
	\begin{subfigure}{.5\textwidth}
	    \includegraphics[width=0.95\textwidth]{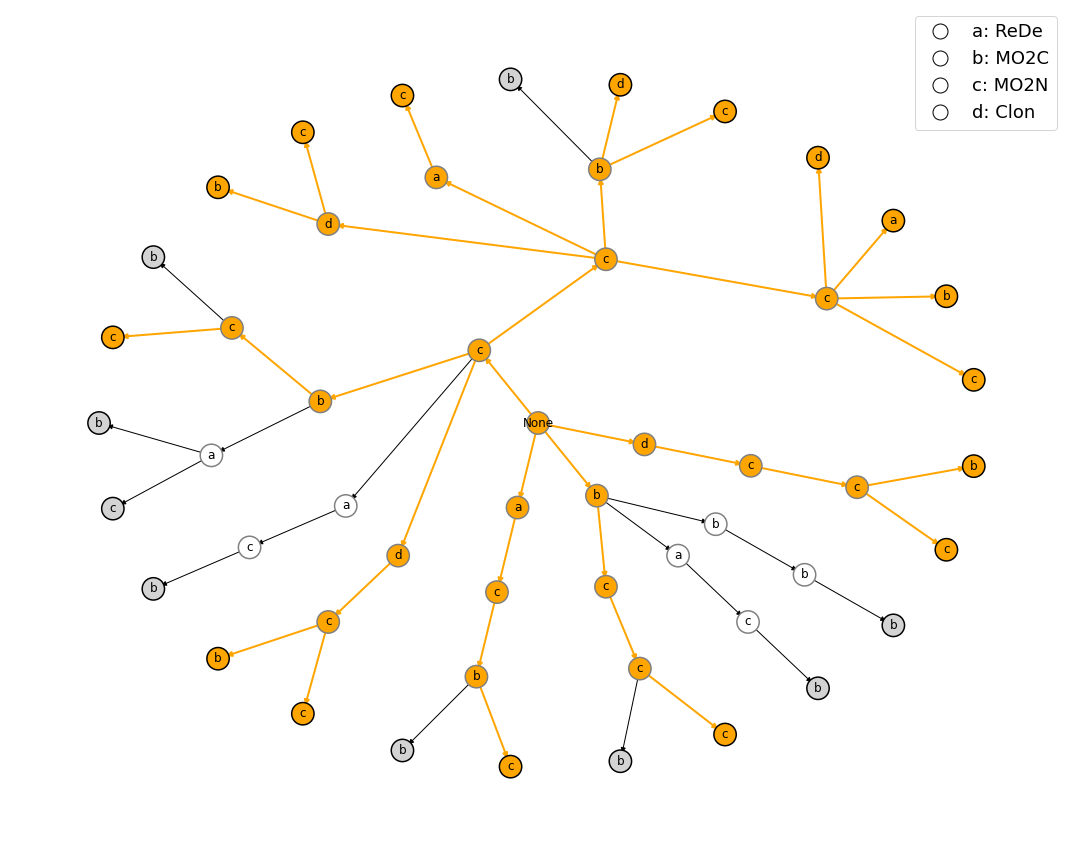}
	    \caption{baseline (\referenceP) - \ttbs\label{fig:trie-baseline-trainticket}}
	\end{subfigure}\hfill
	\begin{subfigure}{.5\textwidth}
		\includegraphics[width=0.95\textwidth]{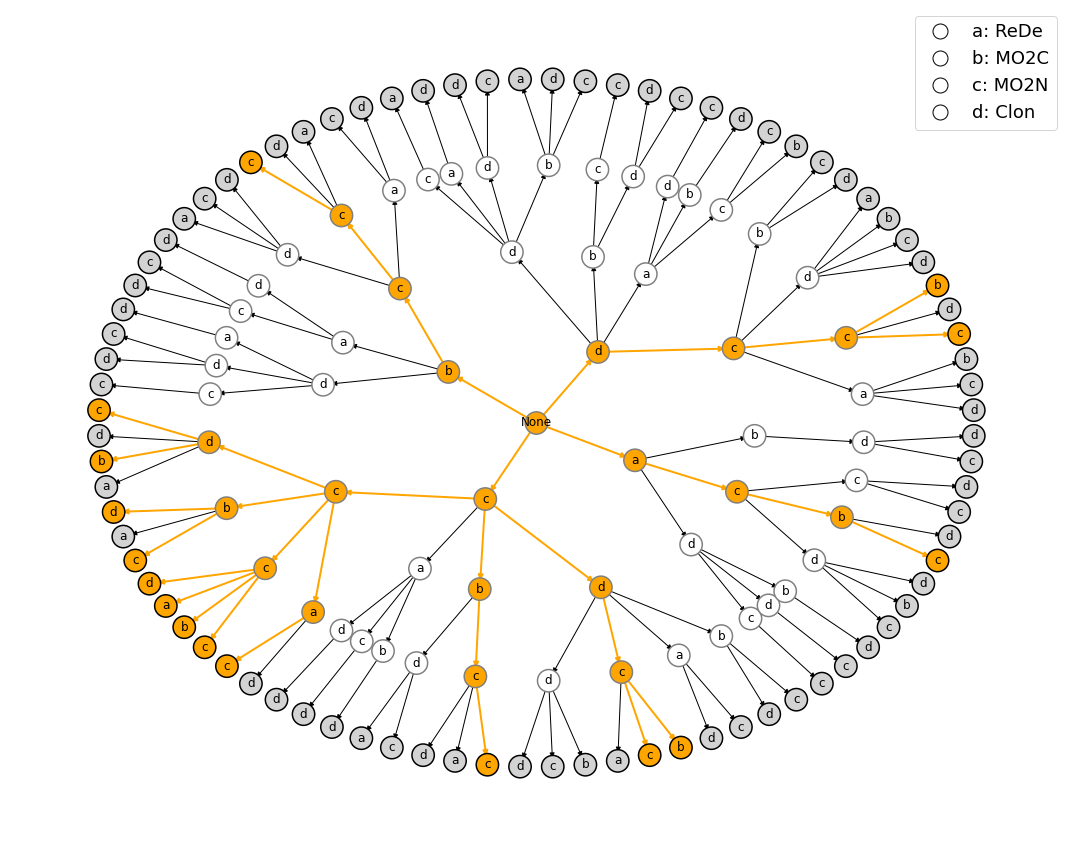}
		\caption{30 min budget - \ttbs\label{fig:trie-trainticket}}
	\end{subfigure}
    \caption{Examples of search spaces for \ttbs represented as trees, as generated by \nsga. The orange nodes and edges are sequences of refactoring actions shared by both trees (\ie intersections). Each node maps to an individual refactoring action as indicated in the legend.}
    \label{fig:search-trees-trainticket}
\end{figure*}

\begin{figure*}[t]
    \centering
	\begin{subfigure}{.5\textwidth}
	    \includegraphics[width=0.95\textwidth]{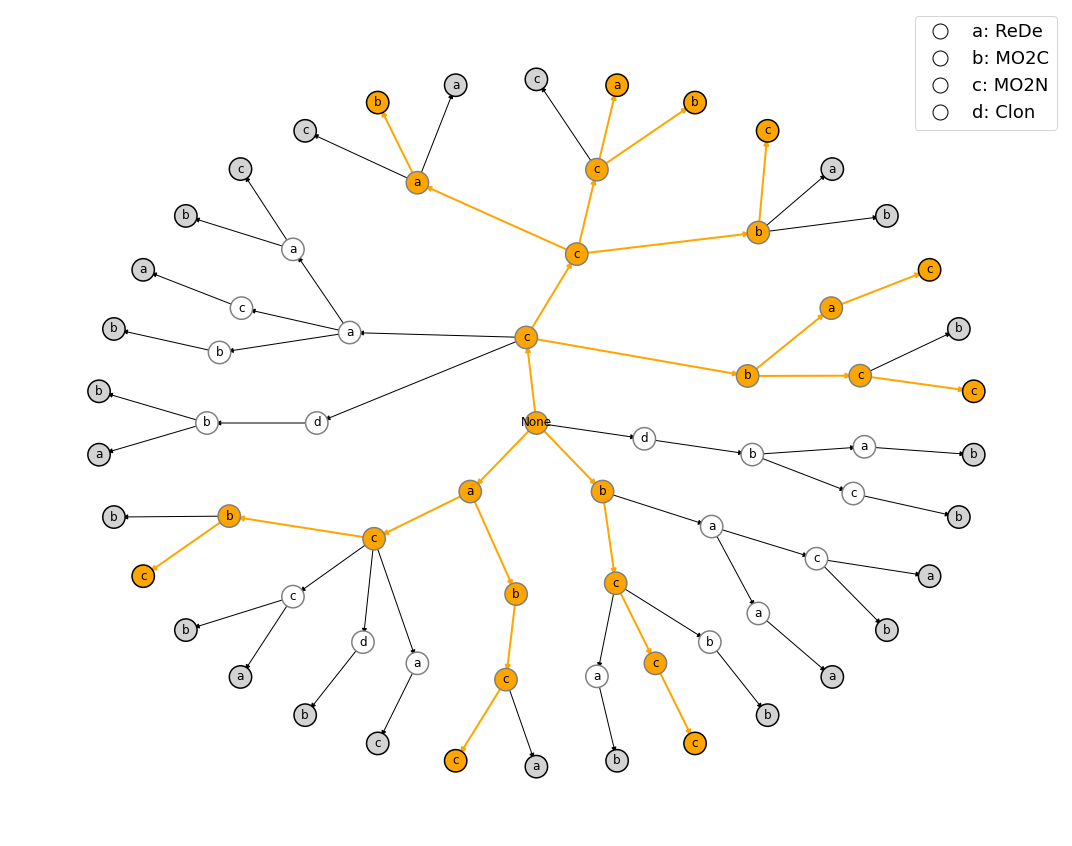}
	    \caption{baseline (\referenceP) - \ccm}
        \label{fig:trie-baseline-cocome}
	\end{subfigure}\hfill
	\begin{subfigure}{.5\textwidth}
		\includegraphics[width=0.95\textwidth]{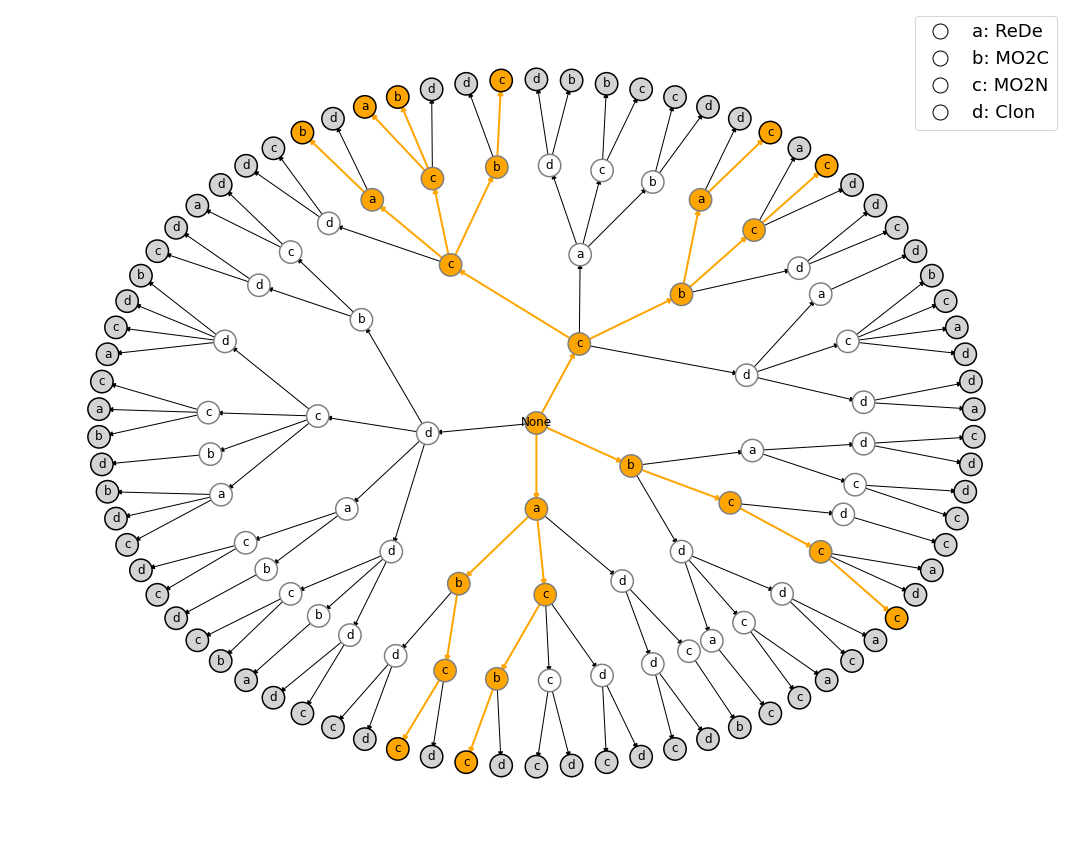}
		\caption{30 min budget - \ccm}
        \label{fig:trie-cocome}
	\end{subfigure}
    \caption{Examples of search spaces for \ccm represented as trees, as generated by \nsga. 
    The orange nodes and edges are sequences of refactoring actions shared by  both trees (\ie intersections). Each node maps to an individual refactoring action, as indicated in the legend.}
    \label{fig:search-trees-cocome}
\end{figure*}

For instance, \Cref{fig:search-trees-trainticket} and \Cref{fig:search-trees-cocome} show a pair of trees for \ttbs and \ccm experiments, respectively. 
Each path from the root to a leaf represents a unique sequence of refactoring actions, which can produce one or more models. 
All sequences involve exactly four refactoring actions. 
The colored paths correspond to common sequences (\ie an intersection) between both trees, while the remaining paths are particular to each tree. 
In this way, we can (approximately) determine that using a $30$ min time budget (either for \ttbs or \ccm) generates a subset of models that are structurally different from those generated by the optimization without any budget (\referenceP). 
Note also that the number of unique sequences in \referenceP is smaller (\ie less diverse) than that of the space explored with a budget. 
These trees help to establish a ``profile'' of refactoring actions for a given experiment and then make comparisons with other profiles. 
In general, the representation and analysis of search spaces have received less attention in the architecture optimization literature, since most works have focused on the objective space.

\begin{figure*}[t]
\begin{revfloatenv}
    \centering
	\begin{subfigure}{.5\textwidth}
	    \includegraphics[width=0.95\textwidth]{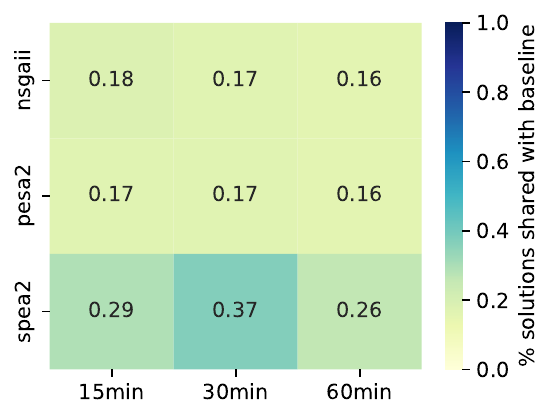}
	    \caption{\ttbs}
        \label{fig:heatmap-ttbs}
	\end{subfigure}\hfill
	\begin{subfigure}{.5\textwidth}
		\includegraphics[width=0.95\textwidth]{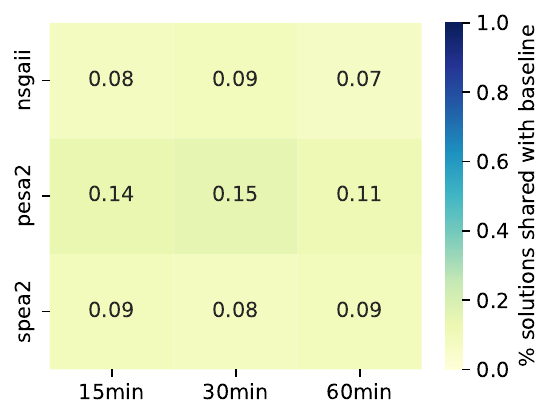}
		\caption{\ccm}
        \label{fig:heatmap-cocome}
	\end{subfigure}
    \caption{Percentages of shared solutions with the baseline (\referenceP) for the different combinations of time budgets and algorithms.}
    \label{fig:heatmaps-shared-solutions}
\end{revfloatenv}
\end{figure*}

\begin{figure*}
\begin{revfloatenv}
   \centering
   \begin{subfigure}{\dimexpr0.8\textwidth+20pt\relax}\centering
     \includegraphics[width=\textwidth]{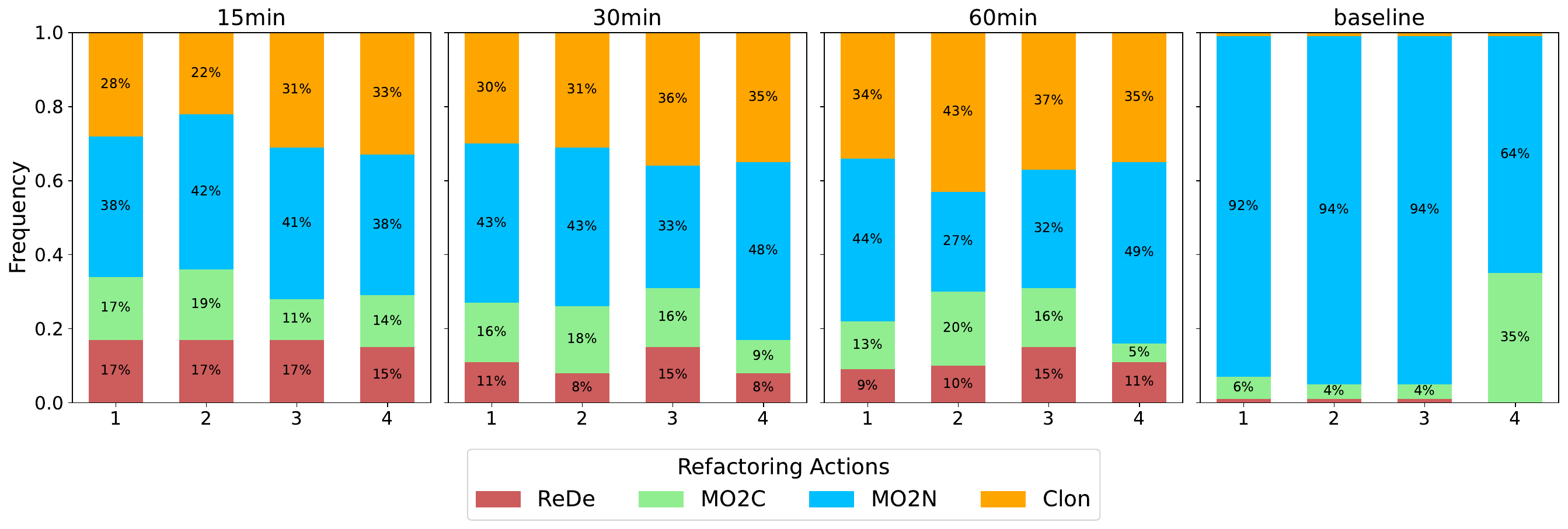}
     \caption{\nsga}
     \label{fig:stacked-nsga-ttbs}
   \end{subfigure}\vfill \begin{subfigure}{\dimexpr0.8\textwidth+17pt\relax}\centering
     \includegraphics[width=\textwidth]{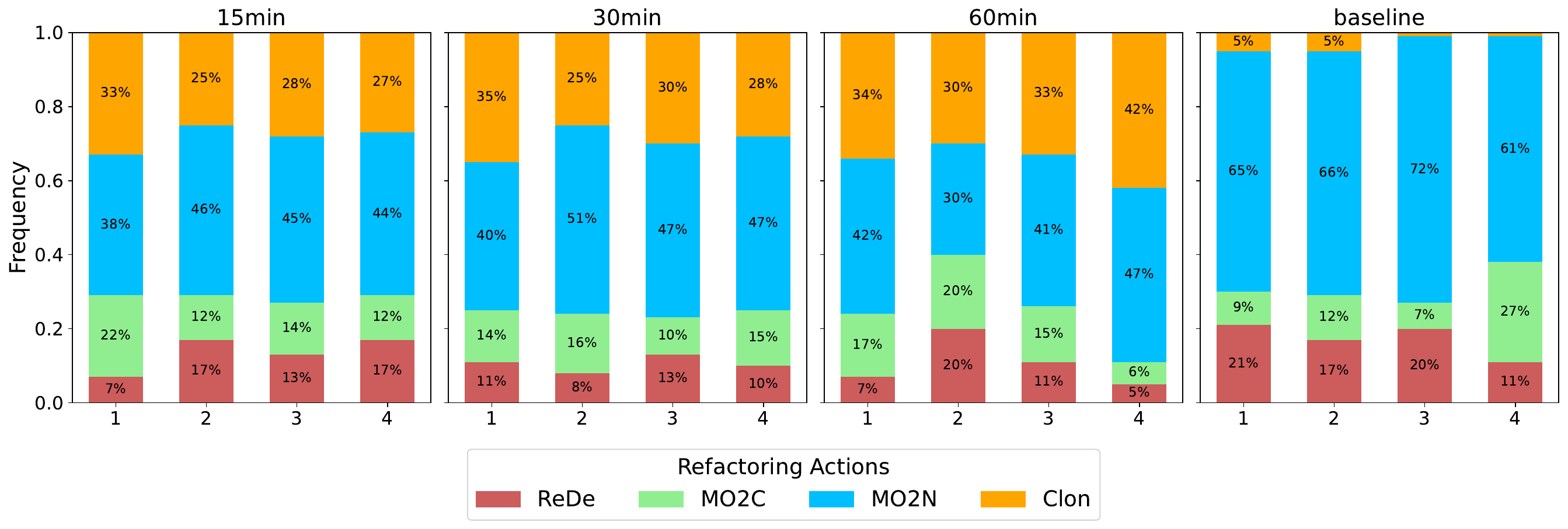}
     \caption{\pesa}
     \label{fig:stacked-pesa-ttbs}
   \end{subfigure}\vfill \begin{subfigure}{\dimexpr0.8\textwidth+17pt\relax}\centering
     \includegraphics[width=\textwidth]{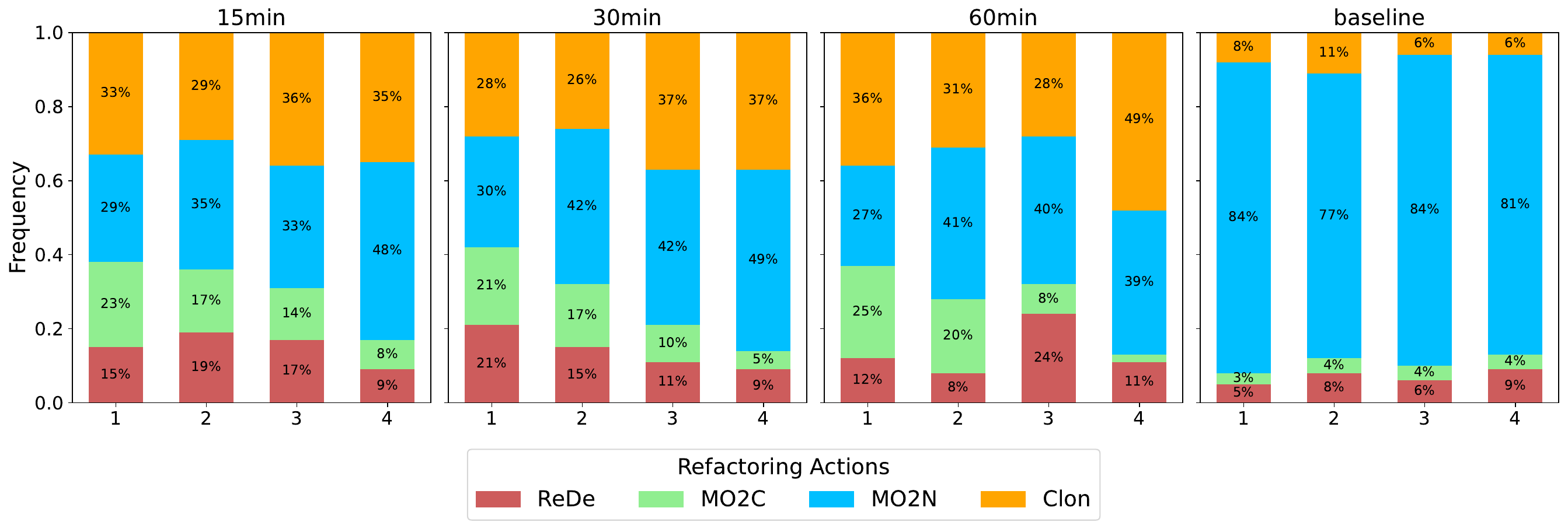}
     \caption{\spea}
     \label{fig:stacked-spea-ttbs}
   \end{subfigure}\vfill \caption{Frequency of refactoring actions used at each sequence position for \ttbs across the budgets and the baseline (\referenceP).
  \label{fig:stacked-refactions-ttbs}}
\end{revfloatenv}
\end{figure*}

\begin{figure*}
\begin{revfloatenv}
   \centering
   \begin{subfigure}{\dimexpr0.8\textwidth+20pt\relax}\centering
     \includegraphics[width=\textwidth]{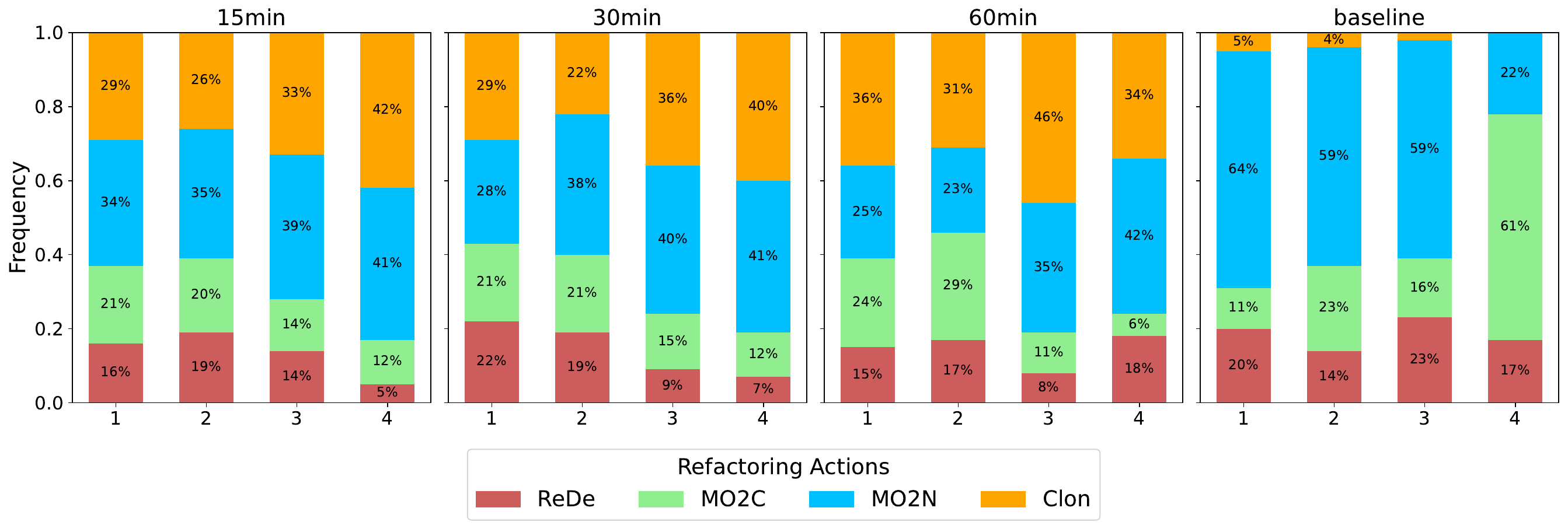}
     \caption{\nsga}
     \label{fig:stacked-nsga-ccm}
   \end{subfigure}\vfill \begin{subfigure}{\dimexpr0.8\textwidth+17pt\relax}\centering
     \includegraphics[width=\textwidth]{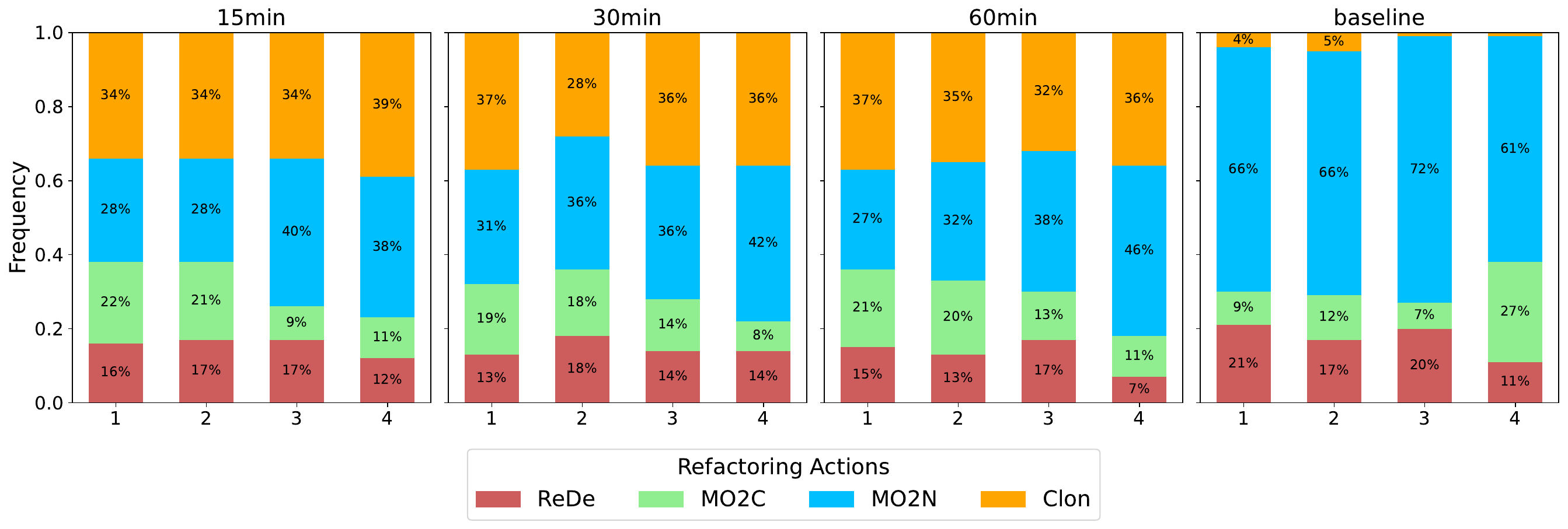}
     \caption{\pesa}
     \label{fig:stacked-pesa-ccm}
   \end{subfigure}\vfill \begin{subfigure}{\dimexpr0.8\textwidth+17pt\relax}\centering
     \includegraphics[width=\textwidth]{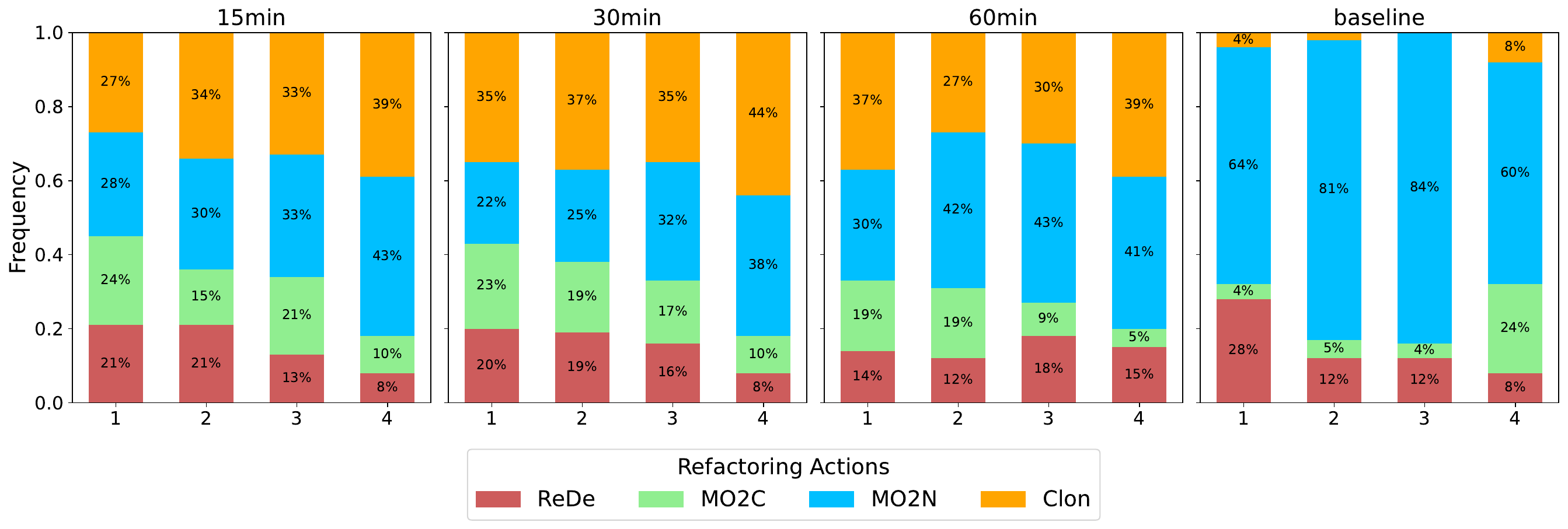}
     \caption{\spea}
     \label{fig:stacked-spea-cocm}
   \end{subfigure}\vfill \caption{Frequency of refactoring actions used at each sequence position for \ccm across the budgets and the baseline (\referenceP).
  \label{fig:stacked-refactions-ccm}}
\end{revfloatenv}
\end{figure*}

We compared the sequence trees obtained with different budgets among themselves and also compared each tree against the tree for \referenceP (baseline). The intersections of the trees resulting from the algorithms and budgets with the baseline are shown in \Cref{fig:heatmaps-shared-solutions}. For \ttbs, we found that between $18-37\%$ of the sequences obtained with time budgets were shared by the \referenceP. 
These percentages were in the range $8-14\%$ for \ccm. As an exception, we noticed a higher number of shared sequences in \spea for \ttbs.
Overall, these numbers indicate that more than half of the models generated when using budgets differ from those found in the \referenceP. 
Furthermore, regarding the \referenceP, the trees resulting from the budgets included many more sequences (in terms of types of refactoring actions) than the baseline trees. 
This was a common trend for both case studies, as hinted by \Cref{fig:search-trees-trainticket} and \Cref{fig:search-trees-cocome}, in which the trees on the right are denser than the trees on the left. 
We argue that this situation is due to the convergence of the solutions (and the corresponding sequences thereof) near the Pareto front, after a considerable number of evolutions.

To further analyze the model differences, we computed the frequency of the refactoring actions used at each position of the sequences explored in the experiments. 
The intuition here is that the repeated usage of certain actions is driven by the optimization objectives, which might originate the model differences within a given search space. 
The (normalized) frequency for the four available actions for \ccm and \ttbs is summarized in \Cref{fig:stacked-refactions-ttbs} and \Cref{fig:stacked-refactions-ccm}.

For \ccm, we can see that the frequency of actions resulting from imposing budgets are more or less similar in their composition, with \textit{MO2N} being on par with \textit{Clon} as the most prevalent actions. 
This pattern contrasts with the very high frequency of \textit{MO2N} observed in the baselines and the very low contributions of the remaining actions.
We conjecture that \textit{MO2N} could play a key role in the solutions in the \referenceP, and this could explain why some solutions in the experiments using time budgets did not reach the Pareto front. 
The frequencies for the baselines achieved by the three algorithms showed equivalent trends for the first three positions, except for the prevalence of \textit{MO2C} for \nsga in the last position of the sequence.
When it comes to \ttbs, the patterns were similar to those for \ccm, but the prevalence of \textit{MO2N} was even higher in the baselines, except for \pesa where other actions such \textit{ReDe} and \textit{MO2C} were sometimes applied. In addition, \textit{MO2C} became more relevant in the fourth sequence position for \pesa and \nsga, as in the case of \ccm. 
This could hint at a particular behavior in the last  position mainly for \nsga.
In general, the fact that less types of refactoring actions were used in the baselines for \ttbs and \ccm, and a greater variety of actions were used when imposing budgets is correlated with the results observed in \Cref{fig:search-trees-trainticket} and \Cref{fig:search-trees-cocome}. That is, baselines tend to consolidate the software models to a reduced set, presumably due to the optimization convergence, while the use of time budgets do not show that phenomenon. The prevalence of \textit{MO2N} over the other actions, particularly at the sequence end, seemed independent of the complexity of the \ttbs and \ccm models.

\begin{rqbox}
Overall, we can answer \textbf{RQ3} by saying that using time budgets generates different models in terms of sequences of refactoring actions, while sharing a small fraction of those models with the \referenceP. 
This fraction did not appear to be affected by budget increases, although the fraction was higher for \spea. 
The profiles of refactoring actions were similar, regardless of the budgets or algorithms being used, but differed from the baseline profiles, notably for \textit{MO2N}. 
\end{rqbox}
 
 \section{Implications}\label{sec:implications}

The findings of this study provide actionable insights for practitioners applying multi-objective optimization to software architecture refactoring. This section translates our results into recommendations for algorithm selection, budget configuration, and interpretation of optimization behavior. These implications aim to support informed decision-making in practical settings.

\paragraph*{Consider \pesa as a first default choice for reliable and high-quality trade-offs}
Across all experiments, \pesa consistently delivered strong results not only in solution quality but also in convergence and diversity. Its performance was stable across varying budget lengths and different case studies, particularly excelling with longer budgets in complex models like \ccm. This confirms that \pesa is a strong default algorithm for architecture refactoring with multiple objectives. It ensures not only good trade-offs among objectives but also a well-distributed Pareto front and a robust convergence pattern, making it suitable for production use where dependability is critical.

\paragraph*{Use \nsga for a quick exploration, not for quality}
\nsga demonstrated the shortest execution time across most runs, confirming its utility in fast-paced optimization scenarios. However, this speed comes at the cost of both diversity and convergence quality, with the algorithm frequently producing suboptimal and less varied solutions. Practitioners might use \nsga during early phases of exploration, especially when testing hypotheses or quickly scanning the design space. But if the quality of trade-offs matters, \nsga should be avoided in favor of more robust alternatives like \pesa.

\paragraph*{Calibrate budget length per case study}
Our results show that longer time budgets do not guarantee better performance across all cases. While \pesa benefits from longer runs, the same cannot be said for other algorithms or case studies. For instance, improvements plateaued for \ttbs, and in some settings, longer budgets did not significantly affect convergence or quality. This suggests that practitioners should not blindly increase computation time expecting better results. Instead, they should analyze budget sensitivity for their specific context and algorithm, conducting small-scale validations before committing to large runs.

\paragraph*{Watch for convergence issues through tree density or refactoring action profiles}
The refactoring trees generated during optimization runs were noticeably denser under budgeted configurations than in baseline settings. This density often reflects a lack of convergence, as many actions are being proposed, but the algorithm is not settling on optimal trade-offs. Monitoring tree density thus provides a lightweight heuristic to detect inefficiencies in the search process. If the trees remain dense over time without significant quality improvement, it may be a sign that the parameters of the algorithms need some adjustments. A similar heuristic can be followed with respect to the frequency of refactoring actions per sequence position.

\paragraph*{Leverage budgeted configurations for action diversity}
Budgeted configurations were shown to reduce over-reliance on certain refactoring actions, such as \texttt{MO2N}, which dominated in baseline scenarios, and promote a more balanced use of alternatives like \texttt{Clon}. This shift leads to structurally different solution candidates and broadens the design space, which can be beneficial for exploration of unforeseen alternatives. In practice, using budgeted settings may allow architects to discover novel or unexpected improvements, rather than reinforcing familiar patterns. This can be particularly useful when the goal is to explore creative refactoring strategies or to generate a portfolio of diverse architectural options.

\paragraph*{Select algorithm based on the most relevant QI}
No single algorithm outperformed the others across all quality indicators. While \pesa generally led in Hypervolume, SPREAD, and \igdp, other algorithms occasionally excelled in specific contexts or metrics. This reinforces the idea that algorithm selection should be driven by the specific priorities of the task at hand. For instance, if maximizing solution diversity is crucial, then SPREAD results may guide the decision. If the goal is to minimize deviation from a reference set, then \igdp might be more relevant.
 \section{Threats to validity}\label{sec:t2v}

In this section, we discuss threats that might affect our results.

\paragraph*{Construct validity}

An aspect that might affect our results is the estimation of the reference Pareto front (\referenceP), which is used to extract the quality indicators, as described in \Cref{sec:approach}. 
We mitigate this threat by building the \referenceP from a run without a search budget for each case study.
Therefore, \referenceP should contain all the non-dominated solutions across all configurations, and it should also represent a good Pareto front for computing the quality indicators.

Another important aspect that might threaten our experimentation concerns the parameters of the initial model. 
For example, \ccm showed higher initial reliability that might affect the search. 
However, in our experiments, it seems that \ttbs and \ccm initial configurations did not threaten the optimization process. 
We will further investigate how different initial parameters for the models could change the optimization results. 
We remark that changing a single model parameter means starting the optimization process at a different point of the solution space that might produce different results.

\paragraph*{External validity}
Our results might be affected by \emph{external validity} threats, as their generalization might be limited to some of the assumptions behind our approach.

In the first place, a threat might be represented by the use of a single modeling notation (\ie UML).
We cannot generalize our results to other modeling notations, which could imply using a different portfolio of refactoring actions. 
The syntax and semantics of the modeling notation determine the amount and nature of refactoring actions that can be performed. 
However, we have adopted UML, which is the de facto standard in the software modeling domain~\cite{Ozkaya_2019}. 
In general terms, this threat can be mitigated by porting the whole approach to a different modeling notation, but this is out of the scope of this paper. 

Another threat might be found in the fact that we have validated our approach on two case studies.
While the two case studies were selected from the available literature, they might not expose all the possible challenges that our approach could face in practice.

Another external validity threat lies in the limited refactoring space explored in this study. The four refactoring actions we implemented were selected for their simplicity and their ability to impact performance and reliability properties.
While these actions are representative of commonly applied model-level refactorings, they do not exhaust the full range of refactorings used in industrial scenarios. More complex or domain-specific refactoring catalogs might exhibit different runtime behaviors or trade-offs. Future work could extend the refactoring set to improve generalizability and capture a broader spectrum of architectural design concerns.

\paragraph*{Internal validity}
Our optimization approach might be affected by \emph{internal validity} threats. 
There are high degrees of freedom in our settings. 
For example, the variations of genetic configurations, such as the $P_{crossover}$ probability, may produce \computedP with different quality solutions. 
Also, problem configuration variations may also change our results.
For example, the population size, which is set to 16 individuals --- smaller than the standard values typically used in SBSE approaches. Each individual in our population is a UML model, \ie a complex object. Moreover, we did not observe any significant limitations in search space exploration in our previous studies~\cite{Di-Pompeo-Tucci-2022,Cortellessa-Di-Pompeo-Stoico-Tucci-2021,Cortellessa-Di-Pompeo-Stoico-Tucci-2023}. 
Another configuration that might affect our results, is the length of the chromosome might limit the space exploration.
In this study, we fixed it to 4 refactoring actions.
We decided this number on the basis of prior studies~\cite{cortellessaModeldrivenApproachContinuous2022a,arcelliExploitingArchitectureRuntime2019} and expertice~\cite{cortellessaIntroducingInteractionsMultiObjective2025} in the domain.
The degrees of freedom in our experimentation generate an unfeasible brute-force investigation of each suitable combination. 
For this reason, we limit the variability to subsets of problem configurations, as shown in \Cref{tab:config_params}. 
We also mitigate this threat by involving two different case studies derived from the literature, thus reducing biases in their construction.

Another aspect that might affect our findings is a misleading interpretation of the outcome due to the random nature of genetic algorithms.
In order to mitigate this threat, we performed \independentRun executions for each configuration~\cite{Zitzler-Deb-Thiele-2000}.

\paragraph*{Conclusion validity}
The  observations made by this study might change with different, better-tuned parameters for each algorithm.
For scoping reasons, we did not perform an extensive tuning phase for each algorithm.
Instead, we rely on common parameters to set up the algorithms, which should mitigate the threat~\cite{Arcuri-Fraser-2013}.
Wherever possible, we used appropriate statistical procedures with p-value and effect size measures to test the significance of the differences and their magnitude.

Another aspect that might affect our results is the estimation of the reference Pareto frontier (\referenceP). 
\referenceP is used for extracting the quality indicators as described in~\Cref{sec:rqs}. 
We soften this threat by building the \referenceP overall our \computedP for each case study. 
Therefore, the reference Pareto should optimistically contain all non-dominated solutions across all configurations.

Moreover, while the selected time budgets were inspired by prior findings and practical constraints, finer-grained or adaptive budget levels could reveal more nuanced behaviors, and we plan to explore this in future studies.
 \section{Conclusion and Future Work}\label{sec:conclusion}

In this study, we presented an investigation of the impact of the time budget for multi-objective refactoring optimization of software models. 
Imposing a time budget to limit the search within a solution space has been explored in various contexts~\cite{Arcuri-Briand-2014,Luong-Nguyen-Gupta-Rana-Venkatesh-2021,Zitzler_Künzli_2004}. 
This well-established practice helps reduce computational times in search-based processes that often as a black-box.

In this direction, the study aims at helping designers to select the best algorithm with respect to the time budget. 
We performed the study on two model benchmarks, \texttt{Train Ticket Booking Service}, and \ccm, and on three genetic algorithms, \nsga, \spea, and \pesa.

We assessed the quality of the results obtained by each algorithm through the six quality indicators (QIs).
From our results (see \Cref{sec:rq1}), \nsga emerged as the fastest algorithm because it performed the highest number of genetic evolutions within the search budget. 
\pesa, in turn, was the algorithm that in most of the cases generated the best quality indicator values.
\spea was the slowest algorithm and generated the results with the worst quality. 

Regarding the different budgets, they seem to produce similar models both in terms of structure and objective values (see \Cref{sec:rq3}).
In terms of their sequences of refactoring actions, the sets of models derived from the time budgets tend to have many models in common, despite some variations attributed to the policies of each algorithm. 
Moreover, a small fraction of these models was shared with those models in \referenceP, which indicate that the budgets might find optimal models. 

As future work, also with the goal of saving optimization time, we intend to analyze the Pareto front at each evolution in order to detect situations in which the quality is not having enough improvement, and thus one could decide to stop the algorithm. 
Additionally, future work could explore alternative strategies for reducing the computational cost of the evaluation phase. These include incorporating surrogate fitness functions, enforcing evaluation-time budgets per individual, and leveraging parallel evaluation frameworks. Specifically, we started to investigate the usage of machine learning techniques as surrogates for predicting quality objectives \cite{Diazpace:2025}. For instance, these surrogates can be useful for estimating the number of performance antipatterns in an architecture, which is one of the most expensive objectives to compute. We envision that surrogates can be combined with a Bayesian optimization algorithm \cite{Ozaki-et-al:2020} (\eg MO-TPE), which has been shown good results in benchmarks under limited time budgets, provided that we can adapt the refactoring actions to this context.

Furthermore, we would like to get more insights from the tree representation of the search spaces, which can enable the discovery of particular refactoring actions being correlated with the satisfaction of certain objectives by the optimization algorithms. 
Finally, we plan to experiment with additional case studies and further investigate the impact of the case study structure (\ie size and complexity) on the quality of the optimization results.
 
\backmatter

\bmhead{Funding}
\noindent Daniele Di Pompeo and Michele Tucci are supported by \SoBigDataITHack. J. Andres Diaz-Pace was supported by project PICT-2021-00757, Argentina.

\bmhead{Competing interests} The authors declare no competing interests.

\printbibliography

\end{document}